\documentclass[11pt,onecolumn]{article}
\usepackage[letterpaper,margin=1in]{geometry}
\usepackage{amsmath,amssymb,amsthm}
\usepackage{thmtools,thm-restate}
\usepackage{graphicx,subcaption,booktabs,multirow}
\usepackage[dvipsnames]{xcolor}
\usepackage{tikz}
\usetikzlibrary{patterns,quotes,angles,arrows.meta,calc}
\usepackage[ruled,vlined]{algorithm2e}
\usepackage{enumitem,cite,url}
\usepackage[hidelinks]{hyperref}
\newtheorem{theorem}{Theorem}[section]

\theoremstyle{definition}
\newtheorem{definition}[theorem]{Definition}

\theoremstyle{remark}

\newcommand{\manuscriptfigure}[1]{%
  \IfFileExists{#1}{\input{#1}}{%
    \IfFileExists{#1.tex}{\input{#1.tex}}{%
      \parbox{\linewidth}{\centering\scriptsize Original figure source unavailable:\\
        \resizebox{0.95\linewidth}{!}{\texttt{\detokenize{#1}}}}\PackageWarning{main}{Missing figure: #1}%
    }%
  }%
}
\title{Hamiltonian Cycle Construction in Rectangular Grid Graphs\\
with Three-Stacked Triangular Holes}

\author{
  Cecilia Chen\\
  Department of Computer Science\\
  Virginia Tech\\
  \texttt{cecilia833@vt.edu}
  \and
  Dianna Xu\\
  Department of Computer Science\\
  Bryn Mawr College\\
  \texttt{dxu@brynmawr.edu}
}
\date{}
\begin{document}
\maketitle

\section{Introduction}
\label{sec:introduction}

Hamiltonian cycles are a fundamental concept in graph theory.
Given a graph, the Hamiltonian cycle problem asks whether
there exists a cycle that visits every vertex exactly once.
In general, determining the existence of a Hamiltonian cycle
is NP-complete, and this hardness persists even for grid
graphs~\cite{article}. However, certain structured classes
of grid graphs admit efficient algorithms. For example,
Hamiltonian cycles in solid grid graphs can be found in
polynomial time~\cite{solid}, while rectangular grid graphs
admit simple constructions based on their dimensions and
parity properties.

The problem becomes more challenging when holes are introduced
into the grid. Removing vertices changes the structure of the
graph and may create narrow regions or additional constraints
on how a Hamiltonian cycle can be constructed. Nevertheless,
several structured grid graphs with holes remain tractable.
Previous work has developed efficient algorithms for rectangular
grid graphs with rectangular holes~\cite{rec_hole_even,rec_hole},
as well as C-shaped and L-shaped grid graphs
~\cite{keshavarzkohjerdi2016hamiltonianpathscshapedgrid,
KESHAVARZKOHJERDI201637}.

A common approach in these studies is to decompose the original
graph into smaller, more structured subgraphs. Hamiltonian paths
or cycles are first constructed within individual subgraphs and
then connected to obtain a global solution. This approach takes
advantage of the known structural properties of rectangular grid
graphs and reduces the difficulty of constructing Hamiltonian
cycles in more complex geometries. However, the decomposition
must be chosen carefully: the existence of Hamiltonian cycles
within individual subgraphs does not necessarily guarantee that
they can be combined into a Hamiltonian cycle of the entire graph.

In this paper, we study Hamiltonian cycles in rectangular grid 
graphs with a triangular hole removed. 
%The hole is formed by three stacked rectangular layers (of increasing widths), producing an irregular boundary within the enclosing rectangular grid. 
Unlike rectangular holes, the triangular structure introduces
different geometric and parity constraints across its layers,
making the construction of a global Hamiltonian cycle more
complicated. The main contribution is a constructive algorithm for determining and building Hamiltonian cycles in rectangular grid graphs with a triangular hole, when such cycles exist.

\subsection{Our Approaches}

We investigate two approaches based on decomposing the graph into rectangular
subgraphs whose Hamiltonian properties can be analyzed
individually. By examining the dimensions and parity of these
subgraphs, we identify conditions under which local Hamiltonian
paths or cycles can be constructed and connected. Depending on
the geometry of the hole and the surrounding regions, different
decomposition strategies may be required. 

\paragraph{Geometry-driven construction.}

The first approach is based on the geometry of the
triangular hole.
Depending on the relative position of the hole and the
outer boundary, we distinguish between small-hole and
large-hole configurations.

For small holes, the construction proceeds by finding
a Hamiltonian path in a reversed triangular region,
extending the path through a surrounding rectangular
region, and combining the resulting structure with
Hamiltonian cycles in the remaining graph.

Large-hole configurations require additional constructions
because some of the surrounding rectangular regions
may disappear or become too narrow to support the
same operations.

\paragraph{Rectangular-decomposition construction.}

The second approach partitions the graph into a collection
of rectangular subgraphs.

If every rectangular component has an even number of
vertices, a Hamiltonian cycle can be constructed locally
within each component.

The local cycles can then be merged whenever suitable
pairs of adjacent cycle edges exist.

If some components contain an odd number of vertices,
the original partition cannot directly support a
collection of local Hamiltonian cycles.
We therefore investigate alternative partitions and
boundary adjustments that transform the components into
even-sized rectangles.

The two approaches provide different strategies for addressing the same problem. The geometry-driven approach constructs Hamiltonian paths and cycles based on the specific structure of the triangular hole, but requires extensive case analysis when the hole approaches the outer boundary. The rectangular-decomposition approach aims to provide a more systematic alternative by reducing the problem to constructing and merging Hamiltonian cycles in rectangular subgraphs. While the first approach focuses on explicit geometric constructions, the second investigates whether parity-based repartitioning can simplify the construction and reduce the need for geometry-specific cases. The completeness of both approaches remains under investigation.

\subsection{Organization}

Section~\ref{sec:related} discusses related work.
Section~\ref{sec:problem} defines the graph family.
Section~\ref{sec:preliminaries} introduces rectangular
Hamiltonian-cycle constructions and cycle-merging operations.
Section~\ref{sec:geometric} presents the geometry-driven
construction.
Section~\ref{sec:decomposition} develops the
rectangular-decomposition approach.
Section~\ref{sec:discussion} discusses the remaining
theoretical questions.

\section{Related Work}\label{sec:related}

The Hamiltonian path problem in grid graphs has been extensively studied due to both its theoretical significance and its applications in problems such as the traveling salesman problem (TSP)~\cite{application, important}, robotic coverage, and routing in discrete environments. In these settings, a Hamiltonian path corresponds to visiting every location exactly once, making efficient constructions particularly important.

In their seminal work, Itai, Papadimitriou, and Szwarcfiter~\cite{article} proved that the Hamiltonian path (and cycle) problem is NP-complete for general grid graphs. At the same time, they showed that for rectangular grid graphs $R(m,n)$, the problem admits a complete characterization together with a linear-time algorithm. Their approach is based on bipartite coloring and forbidden configurations, providing both necessary and sufficient conditions and a constructive method for $(s,t)$-Hamiltonian paths.

In addition, their analysis relies on the bipartite structure of grid graphs. 
In particular, for any grid graph (not necessarily rectangular), the graph is bipartite under the parity coloring of vertices. 
As a consequence, if the total number of vertices is even, any Hamiltonian path must start and end at vertices of different colors. 
This parity condition serves as a fundamental necessary condition and will be used implicitly throughout our constructions.

Following this, Umans and Lenhart~\cite{solid} studied solid grid graphs, i.e., grid graphs without holes, and showed that Hamiltonian cycles can be found in polynomial time. This result highlights that the presence of holes fundamentally changes the structure of the problem. From an application perspective, holes correspond to obstacles or unavailable regions, which significantly complicate path construction.

A major line of work focuses on grid graphs with structured holes, where algorithmic techniques are developed to preserve linear-time solvability. Keshavarz-Kohjerdi and Bagheri proposed linear-time algorithms for finding Hamiltonian $(s,t)$-paths in rectangular grid graphs with a rectangular hole, covering both even-sized and odd-sized cases in~\cite{rec_hole_even} and ~\cite{rec_hole}. Their algorithms extend the classical framework by introducing additional necessary conditions induced by the hole. The key technique is to decompose the graph into structured subgraphs, construct Hamiltonian paths within each component, and then combine them into a global solution through carefully designed connections.

Beyond rectangular holes, several works have explored other structured geometries with an emphasis on constructive algorithms. In~\cite{keshavarzkohjerdi2016hamiltonianpathscshapedgrid}, the authors studied C-shaped grid graphs and presented a linear-time algorithm based on identifying separable substructures and recursively merging Hamiltonian paths. Similarly, in~\cite{KESHAVARZKOHJERDI201637}, L-shaped grid graphs were analyzed, and a linear-time construction algorithm was given. These approaches share a common strategy: partition the graph into simpler components, construct Hamiltonian paths or cycles in each part, and combine them via well-defined merging operations.

More recent work continued this direction by considering increasingly complex hole configurations. For example, rectangular grid graphs with an L-shaped hole can be handled by partitioning the graph into multiple rectangular and C-shaped subgraphs, constructing Hamiltonian cycles in each component, and merging them via parallel edges~\cite{Lhole_recent}. Similarly, algorithms for Hamiltonian cycles in C-shaped and related grid graphs have been refined using decomposition and merge-based techniques~\cite{Ccycle_recent}. These divide-and-conquer algorithms demonstrate that efficient constructions are possible when the graph can be reduced to a bounded number of structured components.

In addition to standard grid graphs, related models such as supergrid graphs and other generalized grid structures have also been studied. For example, Hamiltonian properties of supergrid graphs and their subclasses have been investigated in~\cite{supergrid1, supergrid2}, further demonstrating how structural constraints enable efficient algorithms even in extended settings.

Existing work is largely tailored to specific global shapes of the graph or the hole. The algorithms are typically designed for particular configurations, and their correctness depends on detailed case analyses corresponding to those shapes. In particular, the structure of the Hamiltonian path itself is often implicit in the construction. Our work will also follow the same strategy.

Even in the classical rectangular grid graph $R(m,n)$, although the existence problem is fully solved, there has been limited investigation into finer structural patterns. In particular, the relationship between the parity of $m$ and $n$, the feasible positions of $s$ and $t$, and the resulting shape of the Hamiltonian path has not been systematically analyzed. Understanding these patterns can lead to more explicit constructions and enable techniques such as systematic rectangular cutting, which simplifies the problem and is particularly useful in large-scale applications.

Furthermore, prior work on grid graphs with holes primarily focuses on single, well-structured hole shapes, such as rectangular, C-shaped, or L-shaped holes. There has been relatively little analysis of more complex holes that can be viewed as compositions of simpler components. Existing algorithms rely on global decomposition tailored to specific shapes. A compositional framework that analyzes how local structures interact will be needed to handle more complex holes.

\section{Setup of the Triangular Hole}\label{sec:problem}

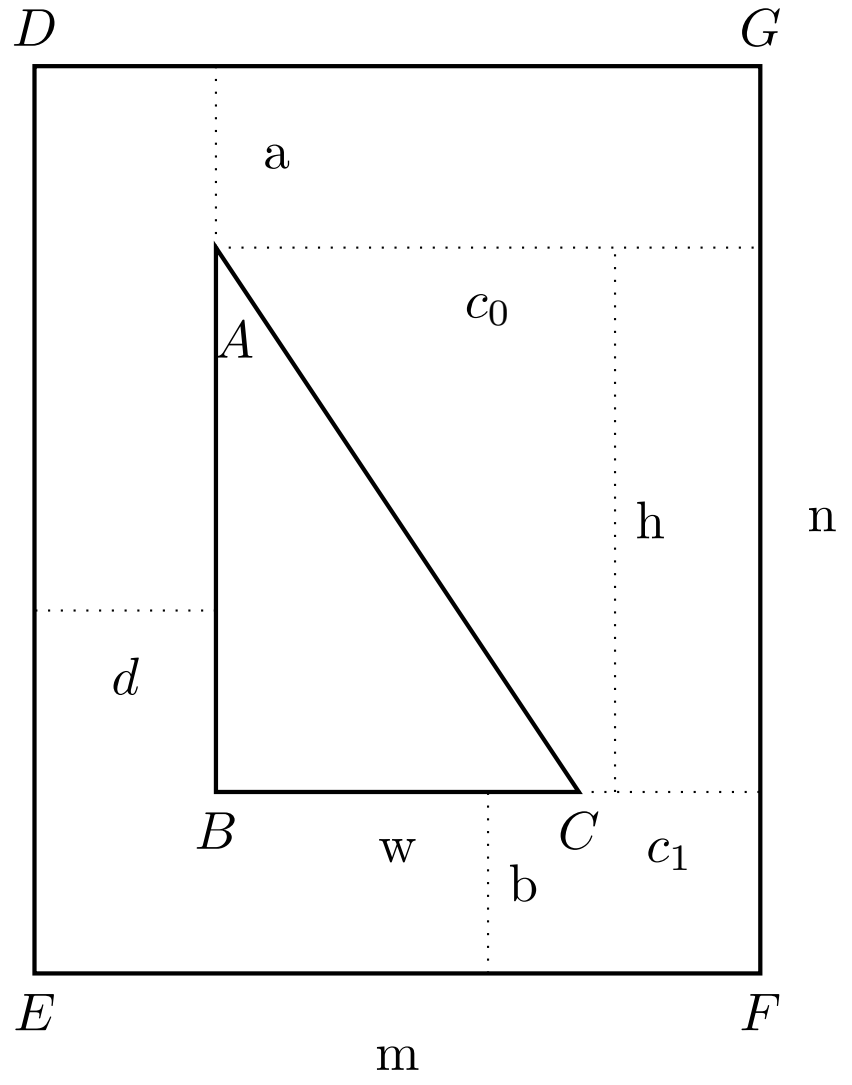
\begin{figure}[h]
    \centering
    \begin{tikzpicture}[scale=1.2, every node/.style={font=\small}]
  % Define points
  \coordinate (A) at (0, 0);
  \coordinate (B) at (4, 0);
  \coordinate (C) at (4, 5);
  \coordinate (D) at (0, 5);

  \node[below] at (A) {$E$};
  \node[below] at (B) {$F$};
  \node[above] at (C) {$G$};
  \node[above] at (D) {$D$};
  % Draw triangle
  \draw[thick] (A) -- (B) -- (C) -- (D) -- cycle;
  \coordinate (E'') at (1,1);
  \coordinate (E') at (0,1);
  \coordinate (E) at (1,1);
  \coordinate (F) at (3,1);
  \coordinate (F') at (4,1);
  \coordinate (F'') at (3,1);
  \coordinate (G) at (3,2);
  \coordinate (H) at (2.2,2);
  \coordinate (I) at (1,4);
  \coordinate (mid) at (1,4);
  \coordinate (I') at (4,4);
  \coordinate (J) at (1,2);
  \coordinate (J') at (0, 2);
  \coordinate (K) at (1.8,2);
  \coordinate (L) at (1,2);
  \coordinate (uppermid) at (1,5);
  \coordinate (uppermid') at (1, 4);
  \coordinate (lowermid) at (2,0);
  \coordinate (lowermid') at (2, 1);
  \draw[thick] (E) -- (F) -- (mid) -- cycle;

  % Label side lengths
  \draw [dotted](A) -- (B) 
    node[midway, below=10pt] {m};
  \draw[dotted] (I) -- (I') 
    node[midway, below=5pt] {$c_0$};  
    \draw[dotted] (I) -- (1,4) ;  
 \draw [dotted](J) -- (J') 
    node[midway, below=5pt] {$d$};
 \draw [dotted](F'') -- (F') 
    node[midway, below=5pt] {$c_1$};
     \draw [dotted](F'') -- (1,1);
  \draw [dotted](B) -- (C) 
    node[midway, right=5pt] {n};
\draw[dotted] (uppermid) -- (uppermid') 
    node[midway, right=5pt] {a};
\draw[dotted] (2.5,0) -- (2.5,1) 
    node[midway, right] {b};
\draw[dotted] (E) -- (F) 
    node[midway, below=5pt] {w};
    
\draw[dotted] (3.2,1) -- (3.2,4) 
    node[midway, right] {h};
    
  \node[below] at (E) {$B$};
  \node[below] at (F) {$C$};
  
  \node[below=10pt] at (1.1,4) {$A$};
\end{tikzpicture}

    \caption{A triangle hole inside a rectangle grid graph}

    \label{fig:converting}
\end{figure}

We will now formally introduce and define the setup for the triangular hole inside of the rectangular grid graph, which we construct by stacking L-shape holes, shown in Figure \ref{fig:converting} and \ref{fig:L shape triangle}. Note that we will work with right triangular hole, which reduces the analysis complexity. However, this doesn't mean that non-right triangular hole is completely unreachable from our work, instead, it can be treated as two right triangle placed side by side. Hence, it would just be more case analysis. 
The right triangle is placed so that its legs lie along grid lines parallel to the coordinate axes. In particular, one leg is parallel to the x-axis and the other is parallel to the y-axis of the enclosing rectangle. For example, in Figure \ref{fig:converting} for triangle $ABC$, edge $AB$ is parallel to the left edge of the outside rectangle and edge $BC$ is parallel to the bottom edge.

In a rectangle grid graph, $R(m,n)$, where, WLOG, $m$ is the width and $n$ is the height, we define a triangle hole $T(w, h)$, where $w$ is the base of the triangle (the edge that is parallel to the bottom edge of the outer rectangle), and $h$ is the height, as shown in Figure \ref{fig:converting}. We set the following variables to represent the distance between two edges or vertices, which is the number of vertices in between: $a$ as the distance between the top edge of the rectangle ($DC$) and the apex of the triangle ($A$); $b$ as the distance between the triangle's base ($BC$) and the rectangle's bottom edge ($EF$); $c_0$ is the distance between the triangle's apex ($A$) and the right edge of the rectangle ($GF$); $c_1$ is the distance between the right end point of the base ($C$) and right edge of the rectangle ($GF$); and $d$ is the distance between the vertical leg of the triangular hole ($AB$) and the left edge of the outer rectangle ($DE$). Note that triangle $T(w,h)$ is not necessarily centered in the rectangle grid graph, i.e., it is not necessary for $d$ to be equal to $c_1$ or for $a$ to be equal to $b$.

\subsection{The Tall Triangle}
Given a rectangle grid graph $R(m,n)$ with a triangle hole $T(w,h)$, if we add the requirement that: each layer of the triangle has constant width $w_0$, except for the bottom layer, which is wider with width $w_1$, as long as it fits inside the rectangle. We call the triangle represented by a grid graph holding the above property a \textbf{tall triangle}. A tall triangle is equivalent to an L-shape, shown in Figure \ref{fig:T shape}, where the height for the horizontal part is $h_1$, and the width for the vertical part is $w_0$.

\subsection{Tall Triangles Stacking}

\begin{figure}[h!]
    \centering
\begin{subfigure}[b]{0.3\textwidth}
    \centering
    \begin{tikzpicture}[scale=1.2, every node/.style={font=\small}]
  % Define points
  \coordinate (A) at (0, 0);
  \coordinate (B) at (4, 0);
  \coordinate (C) at (4, 5);
  \coordinate (D) at (0, 5);

  % Draw triangle
  \draw[thick] (A) -- (B) -- (C) -- (D) -- cycle;
  \coordinate (E'') at (0.5,1.5);
  \coordinate (E') at (0,1.5);
  \coordinate (E) at (0.5,1);
  \coordinate (F) at (3.5,1);
  \coordinate (F') at (4,1.2);
  \coordinate (F'') at (3.5,1.2);
  \coordinate (G) at (3.5,1.4);
  \coordinate (H) at (1,1.4);
  \coordinate (I) at (1,4);
  \coordinate (I') at (4,4);
  \coordinate (J) at (0.5,4);
  \coordinate (J') at (0, 4);
  \coordinate (K) at (0.5,1.4);
  \coordinate (L) at (0.5,1.4);
  \coordinate (uppermid) at (0.8,5);
  \coordinate (uppermid') at (0.8, 4);
  \coordinate (lowermid) at (2,0);
  \coordinate (lowermid') at (2, 1);
  \draw[thick] (E) -- (F) -- (G) -- (H) -- (I) -- (J) -- (K) -- (L) -- (E);
  % \draw[thick] (M) -- (N) -- (O);
  % Label points
  % \node[below left] at (A) {$A$};
  % \node[below right] at (B) {$B$};
  % \node[above right] at (C) {$C$};

  % Label side lengths
  \draw [dotted](A) -- (B) 
    node[midway, below=10pt] {m};
  \draw[dotted] (I) -- (I') 
    node[midway, below=5pt] {$c_0$};  
 % \draw [dotted](J) -- (J') 
    % node[midway, below=5pt] {$d_0$};
 \draw [dotted](F'') -- (F') 
    node[midway, below=5pt] {$c_1$};
 \draw [dotted](E'') -- (E') 
    node[midway, above=5pt] {$d$};
  \draw [dotted](B) -- (C) 
    node[midway, right=5pt] {n};
\draw[dotted] (uppermid) -- (uppermid') 
    node[midway, right=5pt] {a};
\draw[dotted] (lowermid) -- (lowermid') 
    node[midway, right=5pt] {b};
    \draw[dotted] (2.9,1) -- (2.9,1.4) 
    node[midway, right=1pt] {$h_1$};
    \draw[dotted] (2,1.4) -- (2,4) 
    node[midway, right=5pt] {$h_0$};
    \draw[dotted] (0.5, 3) -- (1, 3) 
    node[midway, below=5pt] {$w_0$};
    \draw[dotted] (0.5, 1.2) -- (4, 1.2) 
    node[midway, above=1pt, left=5pt] {$w_1$};
\end{tikzpicture}
    \caption{}
    \label{fig:T shape}
    \end{subfigure}
\hfill
    \begin{subfigure}[b]{0.3\textwidth}
        \centering
        \begin{tikzpicture}[scale=1.2, every node/.style={font=\small}]
  % Define points
  \coordinate (A) at (0, 0);
  \coordinate (B) at (4, 0);
  \coordinate (C) at (4, 5);
  \coordinate (D) at (0, 5);

  % Draw triangle
  \draw[thick] (A) -- (B) -- (C) -- (D) -- cycle;
  \coordinate (E'') at (0.5,1.2);
  \coordinate (E') at (0,1.2);
  \coordinate (E) at (0.8,1);
  \coordinate (F) at (3.5,1);
  \coordinate (F') at (4,1.2);
  \coordinate (F'') at (3.5,1.2);
  \coordinate (G) at (3.5,1.4);
  \coordinate (H') at (3.2,1.4);
  \coordinate (H) at (2.2,2.3);
  \coordinate (I) at (2.2,4);
  \coordinate (I') at (4,4);
  \coordinate (J) at (1.8,4);
  \coordinate (J') at (0, 4);
  \coordinate (K') at (0.8,1.4);
  \coordinate (K) at (1.8,2.3);
  \coordinate (L) at (0.8,4);
  \coordinate (M) at (0.8,1);
  \coordinate (N) at (0.8,1.8);
  \coordinate (O) at (3.2,1.8);
  \coordinate (P) at (3.2,1.4);
  \coordinate (uppermid) at (2,5);
  \coordinate (uppermid') at (2, 4);
  \coordinate (lowermid) at (2,0);
  \coordinate (lowermid') at (2, 1);
  \draw[thick] (E) -- (F) -- (G) -- (H');
  \draw[thick] (2.2, 2.3) -- (2.2,4) -- (L) -- (E) ;
  % \draw[thick] (H) -- (I) -- (J) -- (K);
  % \draw[thick] (M) -- (N) -- (1.2,1.8);
  % \draw[thick] (3, 1.8) -- (O) -- (P);
  % \draw[thick] (1.2, 1.8) -- (1.2, 2.3) -- (1.8, 2.3);
  \draw[thick] (2.2,2.3) -- (3, 2.3) -- (3, 1.4) -- (3.2, 1.4);
 % \draw[dotted] (3,2) -- (4,2) 
 %    node[midway, above=3pt] {$c_1$};
    
 % \draw[dotted] (0,2) -- (1.2,2) 
    % node[midway, above=3pt] {$d_1$};
    
  % Label side lengths
  \draw [dotted](A) -- (B) 
    node[midway, below=10pt] {m};
  % \draw[dotted] (I) -- (I') 
  %   node[midway, below=5pt] {$c_0$};  
 % \draw [dotted](J) -- (J') 
    % node[midway, below=5pt] {$d_0$};
 % \draw [dotted](3.5,1.1) -- (4,1.1) 
 %    node[midway, above=3pt] {$c_2$};
 % \draw [dotted](0,1.1) -- (0.5,1.1) 
 %    node[midway, above=3pt] {$d_3$};
  \draw [dotted](B) -- (C) 
    node[midway, right=5pt] {n};
% \draw[dotted] (uppermid) -- (uppermid') 
%     node[midway, right=5pt] {a};
% \draw[dotted] (lowermid) -- (lowermid') 
%     node[midway, right=5pt] {b};
%     \draw[dotted] (3.2,1) -- (3.2,1.4) 
%     node[midway, left=5pt] {$h_2$};
    % \draw[dotted] (3,1.4) -- (3.2,1.4) 
    % node[midway, right=5pt] {};
    % \draw[dotted] (2.7,1.4) -- (3,1.4) 
    % node[midway, right=5pt] {};
    % \draw[dotted] (2.5,2.3) -- (2.5,4) 
    % node[midway, right=5pt] {$h_0$};
    % \draw[dotted] (3,1.4) -- (3,1.8) 
    % node[midway, left=5pt] {$h_2$};
    % \draw[dotted] (2.7,1.4) -- (2.7,2.3) 
    % node[midway, left=3pt] {$h_1$};
    % \draw[dotted] (3,1.5) -- (0.8,1.5) 
    % node[midway, above=3pt] {$w_1$};
    % \draw[dotted] (0,2) -- (0.8,2) 
    % node[midway, above=3pt] {$d$};
    % \draw[dotted] (4,1.5) -- (3.2,1.5) 
    % node[midway, above=2pt] {$c_2$};
    % \draw[dotted] (0.8, 3) -- (2.2, 3) 
    % node[midway, below=5pt] {$w_0$};
    \draw[thin] (3,1.8) -- (2,1.8) -- (2,4);
    \draw[dotted] (2,3.5) -- (0,3.5) node[midway, above=5pt, left=7pt] {$d''$};
    \draw[dotted] (1.5,3) -- (0,3) node[midway, above=5pt, left=3pt] {$d'$};
    \draw[dotted] (0.8,2.5) -- (0,2.5) node[midway, above] {$d$};
    \draw[thin] (3,1.4) -- (1.5,1.4) -- (1.5,4);
    \draw[dotted] (0.8,2) -- (1.5,2) node[midway, above] {$w_0$};
    \draw[dotted] (2,2.3) -- (1.5,2.3) node[midway, above] {$w_0'$};
    \draw[dotted] (2,2.6) -- (2.2,2.6) node[midway, above=5pt, right=2pt] {$w_0''$};
    \draw[dotted] (2.6,0) -- (2.6,1.8) node[midway, right=6pt, below] {$b''$};
    \draw[dotted] (2.1,0) -- (2.1,1.4) node[midway, right=1pt] {$b'$};
    \draw[dotted] (1.6,0) -- (1.6,1) node[midway, right=1pt] {$b$};
    \draw[dotted] (3.2,1) -- (3.2,1.4) node[midway, left=1pt] {$h_1$};
    \draw[dotted] (2.3,1.8) -- (2.3,1.4) node[midway, left=1pt] {$h_1'$};
    \draw[dotted] (2.8,1.8) -- (2.8,2.3) node[midway, left=1pt] {$h_1''$};
\end{tikzpicture}
        \caption{}
    \label{fig:L shape triangle-1}
    \end{subfigure}
    \hfill
    \begin{subfigure}[b]{0.3\textwidth}
        \centering
        \begin{tikzpicture}[scale=1.2, every node/.style={font=\small}]
  % Define points
  \coordinate (A) at (0, 0);
  \coordinate (B) at (4, 0);
  \coordinate (C) at (4, 5);
  \coordinate (D) at (0, 5);

  % Draw triangle
  \draw[thick] (A) -- (B) -- (C) -- (D) -- cycle;
  \coordinate (E'') at (0.5,1.2);
  \coordinate (E') at (0,1.2);
  \coordinate (E) at (0.8,1);
  \coordinate (F) at (3.5,1);
  \coordinate (F') at (4,1.2);
  \coordinate (F'') at (3.5,1.2);
  \coordinate (G) at (3.5,1.4);
  \coordinate (H') at (3.2,1.4);
  \coordinate (H) at (2.2,2.3);
  \coordinate (I) at (2.2,4);
  \coordinate (I') at (4,4);
  \coordinate (J) at (1.8,4);
  \coordinate (J') at (0, 4);
  \coordinate (K') at (0.8,1.4);
  \coordinate (K) at (1.8,2.3);
  \coordinate (L) at (0.8,4);
  \coordinate (M) at (0.8,1);
  \coordinate (N) at (0.8,1.8);
  \coordinate (O) at (3.2,1.8);
  \coordinate (P) at (3.2,1.4);
  \coordinate (uppermid) at (2,5);
  \coordinate (uppermid') at (2, 4);
  \coordinate (lowermid) at (2,0);
  \coordinate (lowermid') at (2, 1);
  \draw[thick] (E) -- (F) -- (G) -- (H');
  \draw[thick] (2.2, 2.3) -- (2.2,4) -- (L) -- (E) ;
  % \draw[thick] (H) -- (I) -- (J) -- (K);
  % \draw[thick] (M) -- (N) -- (1.2,1.8);
  % \draw[thick] (3, 1.8) -- (O) -- (P);
  % \draw[thick] (1.2, 1.8) -- (1.2, 2.3) -- (1.8, 2.3);
  \draw[thick] (2.2,2.3) -- (3, 2.3) -- (3, 1.4) -- (3.2, 1.4);
 \draw[dotted] (3,2) -- (4,2) 
    node[midway, above=3pt] {$c_1$};
    
 % \draw[dotted] (0,2) -- (1.2,2) 
    % node[midway, above=3pt] {$d_1$};
    
  % Label side lengths
  \draw [dotted](A) -- (B) 
    node[midway, below=10pt] {m};
  \draw[dotted] (I) -- (I') 
    node[midway, below=5pt] {$c_0$};  
 % \draw [dotted](J) -- (J') 
    % node[midway, below=5pt] {$d_0$};
 \draw [dotted](3.5,1.1) -- (4,1.1) 
    node[midway, above=3pt] {$c_2$};
 % \draw [dotted](0,1.1) -- (0.5,1.1) 
 %    node[midway, above=3pt] {$d_3$};
  \draw [dotted](B) -- (C) 
    node[midway, right=5pt] {n};
\draw[dotted] (uppermid) -- (uppermid') 
    node[midway, right=5pt] {a};
\draw[dotted] (lowermid) -- (lowermid') 
    node[midway, right=5pt] {b};
    \draw[dotted] (3.2,1) -- (3.2,1.4) 
    node[midway, left=5pt] {$h_2$};
    % \draw[dotted] (3,1.4) -- (3.2,1.4) 
    % node[midway, right=5pt] {};
    \draw[dotted] (2.7,1.4) -- (3,1.4) 
    node[midway, right=5pt] {};
    \draw[dotted] (2.5,2.3) -- (2.5,4) 
    node[midway, right=5pt] {$h_0$};
    % \draw[dotted] (3,1.4) -- (3,1.8) 
    % node[midway, left=5pt] {$h_2$};
    \draw[dotted] (2.7,1.4) -- (2.7,2.3) 
    node[midway] {$h_1$};
    \draw[dotted] (3,2.1) -- (0.8,2.1) 
    node[midway] {$w_1$};
    \draw[dotted] (0,2) -- (0.8,2) 
    node[midway, above=3pt] {$d$};
    % \draw[dotted] (4,1.5) -- (3.2,1.5) 
    % node[midway, above=2pt] {$c_2$};
    \draw[dotted] (0.8, 3) -- (2.2, 3) 
    node[midway, below=5pt] {$w_0$};
    \draw[dotted] (0.8, 1.3) -- (4, 1.3) 
    node[midway] {$w_2$};
\end{tikzpicture}
        \caption{}
        \label{fig:L shape triangle right}
    \label{fig:L shape triangle-2}
    \end{subfigure}
    \caption{Convert several tall triangle holes into a right triangle hole}
    \label{fig:L shape triangle}
\end{figure}
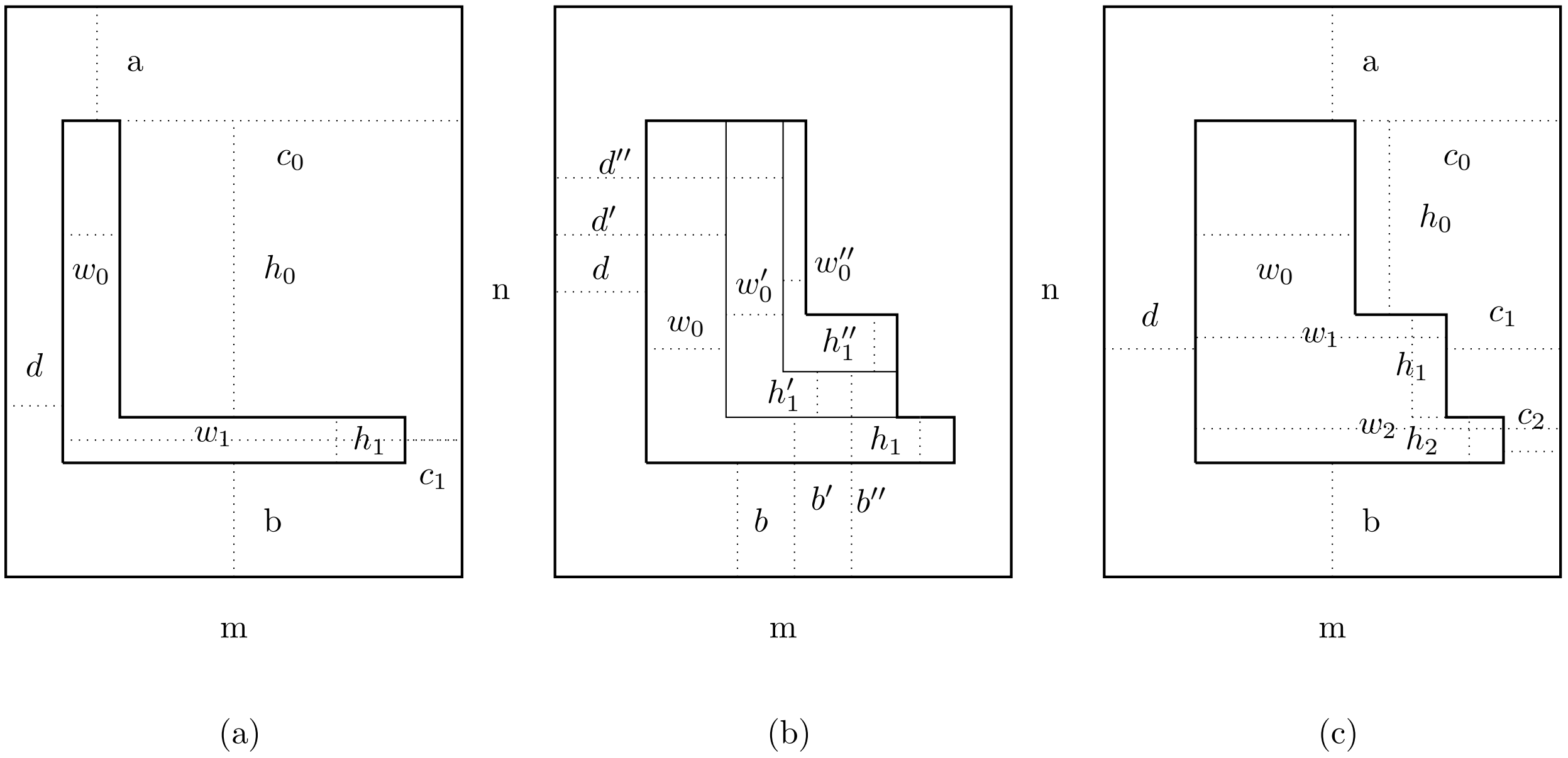

Now, we can stack several L-shape holes together by arranging the bottom rows layer by layer from the widest to the thinnest, as shown in Figure \ref{fig:L shape triangle-1}. In addition, we require that all L-shape holes have the vertical column (completely or partially) aligned with or right next to each other. This is equivalent to each L-shape hole $L(w,h)$ holding the following fact relatively to the hole right below it $L'(w',h')$: $d + w_0 = d'$ and $h_1 +b = b'$. Note that since the L-shapes are stacked from the widest to the thinnest, the spaces between the L-shapes stack and right edges decrease gradually from top to bottom, i.e, $c_0 \geq c_1\geq c_2\geq \cdots \geq c_n$, as shown in Figure \ref{fig:L shape triangle-2}.

It is not difficult to see that an arbitrary right triangle can be obtained by stacking a specific sequence of tall triangles.
Given this construction, we define $w_0$ as the width of the highest column, which is also equal to the total contribution of the corresponding widths from all L-shaped components, as shown in Figure \ref{fig:L shape triangle-2}. Let $d$ denote the horizontal distance between this column and the left boundary of the outer rectangle. We define $h_0$ as the height of the vertical column, which is also the height of the topmost L-shaped component.

The parameters $a$, $b$, $d$, and $c_0$ are defined analogously to the tall triangle case. Let $C = \{c_1, c_2, \cdots, c_n\}$ denote the set of horizontal distances from the base to the rightmost edge of each L-shaped component.
Note that the widths $w_1, w_2, \cdots, w_n$ are defined slightly differently. Specifically, $w_i$ is the base width of the $(i-1)^{\text{th}}$ L-shaped component (ordered from top to bottom), defined as the horizontal distance between the leftmost column of the bottom L-shaped hole and the right edge of the vertical segment of the $(i-1)^{\text{th}}$ L-shape. For example, $w_1$ corresponds to the width of the second L-shape shown in Figure~\ref{fig:L shape triangle right}.

\subsection{Three-stacked Right Triangular Hole}

We focus our work on right triangular holes, and specifically on three-stacked triangular hole, which is constructed by stacking three L-shaped holes from bottom to top.
Without loss of generality we assume that the two legs lie along the left and bottom boundaries of the rectangle, as shown in Figure~\ref{fig:L shape triangle}. 

The placement of the right triangular hole is determined by the parameters $c_0, c_1, c_2$, which describe the horizontal distances on the right, the offset $d$ from the left boundary, the vertical distances $a$ and $b$, and the width and height parameters for each L-shaped component, namely $w_0, w_1, w_2$ and $h_0, h_1, h_2$.

\section{Definitions and Preliminary Supporting Theorems}\label{sec:preliminaries}

In this section, we introduce several definitions and preliminary supporting results that will be useful for explaining the constructions and algorithms presented in later sections. These supporting results and the rectangular classification provide the building blocks for our constructions.

A grid graph is a finite, node-included subgraph of the infinite graph consisting of all points with integer coordinates in the plane, where two vertices are connected if and only if they are at Euclidean distance 1 from each other \cite{solid, article}. A (grid) graph has a Hamiltonian path between starting vertex $s$ and ending vertex $t$ if and only if there exists a path that visits each vertex of the graph exactly once from $s$ to $t$. A Hamiltonian cycle is such a path that returns to $s$ after visiting all vertices exactly once. A rectangular grid graph, $R(m,n)$, is a rectangle shape of grids, where $m$ is the number of columns and $n$ is the number of rows. 

We will also use a basic structural property of grid graphs that follows from their bipartite nature. 
By coloring each vertex $(x,y)$ according to the parity of $x+y$, every grid graph admits a bipartition in which each edge connects vertices of opposite colors \cite{article}. 
As a consequence, the endpoints of any Hamiltonian path are constrained by the parity of the graph.

\begin{restatable}{lemma}{turnandstraight}
\label{lem:turnandstraight}
Let $G$ be any grid graph with an even number of vertices. 
If there exists a Hamiltonian path between vertices $s$ and $t$, then $s$ and $t$ must lie in different bipartition classes, as $s$ and $t$ in Figure \ref{fig:keep upper-left and lower-left - second-1} and \ref{fig:turn upper-left and lower-right 0 second-1}, which lie in different classes.
\end{restatable}

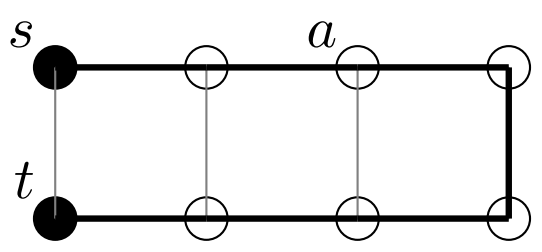
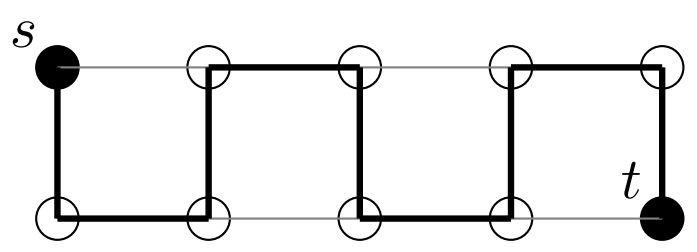
\begin{figure}[h!]
  \begin{subfigure}[b]{0.45\textwidth}
  \centering
  \begin{tikzpicture}
\fill (0,1) circle (4pt);
\fill (0,0) circle (4pt);
  % Draw grid points
  \foreach \x in {0,...,3} {
    \foreach \y in {0,1} {
        \draw (\x,\y) circle (4pt);
    }
  }

  % Draw optional grid lines
  \foreach \x in {0,...,3} {
    \draw[gray, thin] (\x,0) -- (\x,1);
  }
  \foreach \y in {0,1} {
    \draw[gray, thin] (0,\y) -- (3,\y);
  }

  % Bold vertical line from upper-left to lower-left
  \draw[very thick] (0,1) -- (1,1);
  \draw[very thick] (1,1) -- (2,1);
  \draw[very thick] (2,1) -- (3,1);
  \draw[very thick] (3,1) -- (3,0);
  \draw[very thick] (3,0) -- (2,0);
  \draw[very thick] (2,0) -- (1,0);
  \draw[very thick] (1,0) -- (0,0);
    \node[above left] at (0,1) {$s$};
    \node[above left] at (0,0) {$t$};
    \node[above left] at (2,1) {$a$};
\end{tikzpicture}\caption{}
  \label{fig:keep upper-left and lower-left - second-1}
  \end{subfigure}
  \hfill
  \begin{subfigure}[b]{0.45\textwidth}
  \centering
  \begin{tikzpicture}
\fill (0,1) circle (4pt);
\fill (4,0) circle (4pt);
  % Draw grid points
  \foreach \x in {0,...,4} {
    \foreach \y in {0,1} {
        \draw (\x,\y) circle (4pt);
    }
  }
  % Draw optional grid lines
  \foreach \x in {0,...,4} {
    \draw[gray, thin] (\x,0) -- (\x,1);
  }
  \foreach \y in {0,1} {
    \draw[gray, thin] (0,\y) -- (4,\y);
  }

  % Bold vertical line from upper-left to lower-left
  \draw[very thick] (0,0) -- (0,1);
  \draw[very thick] (0,0) -- (1,0);
  \draw[very thick] (1,0) -- (1,1);
  \draw[very thick] (1,1) -- (2,1);
  \draw[very thick] (2,1) -- (2,0);
  \draw[very thick] (2,0) -- (3,0);
  \draw[very thick] (3,0) -- (3,1);
  \draw[very thick] (3,1) -- (4,1);
  \draw[very thick] (4,1) -- (4,0);
  \node[above left] at (0,1) {$s$};
  \node[above left] at (4,0) {$t$};
\end{tikzpicture}\caption{}
  \label{fig:turn upper-left and lower-right 0 second-1}
  \end{subfigure}
  \caption{$s$ and $t$ are denoted by the solid black vertices}
  \label{fig:2-rectangle possible path - second}
\end{figure}

We use a Cartesian coordinate system to index the vertices of $R(m,n)$. The origin is placed at the leftmost vertex of the top row, so the top-left vertex has coordinate $(0,0)$. The $x$-coordinate increases from left to right, and the $y$-coordinate increases from top to bottom. Under this convention, the bottom-right vertex of $R(m,n)$ has coordinate $(m-1,n-1)$. For example, in the rectangle with width $4$ and height $2$ as shown in Figure \ref{fig:keep upper-left and lower-left - second-1}, the top left vertex is $(0,0)$, the bottom left vertex is $(0,1)$, and vertex $a$ is $(2,0)$.

We first consider the simplest nontrivial case, the $2$-rectangle grid graph, $R(m,2)$, i.e. a rectangle with 2 rows, which serves as a fundamental building block for understanding more general constructions. The following lemma describes the basic strategies for constructing Hamiltonian paths depending on the relative positions of $s$ and $t$.
We state the following supporting proposition.

\begin{restatable}{proposition}{rect}
For a 2-rectangle grid graph $R(m,2)$ to have a valid Hamiltonian path:
    \begin{enumerate}
        \item $s$ and $t$ are in the same column. Then $s_x = t_x$. A Hamiltonian path exists between $s$ and $t$ and we can find it by going horizontally and completing a single sideways U, 
        as shown in Figure \ref{fig:keep upper-left and lower-left - second-1}. 

        \item $s$ and $t$ are not in the same column. Then $s_x \neq t_x$, and we need to go vertically and perform successive U-turns, 
        as shown in Figure \ref{fig:turn upper-left and lower-right 0 second-1};
        
    \end{enumerate}
     \label{x values and valid strategy}
\end{restatable}

The previous lemma characterizes constructive strategies in the feasible cases. We now describe conditions under which such a construction is not possible.

\begin{restatable}{proposition}{nopath}

    The 2-rectangle grid graph $R(m,2)$ does not have a valid Hamiltonian path between $s$ and $t$, when $s$ and $t$ are in different columns and either:
    \begin{enumerate}
        \item $s$ and $t$ are not in the same row and $m$ is even, as shown in Figure \ref{fig:doesn't work even different y-1}.
        \item $s$ and $t$ are in the same row and $m$ is odd, as shown in Figure \ref{fig:doesn't work odd for same y-1}.
    \end{enumerate}
    \label{2 rectangle final-1}
\end{restatable}

    \begin{figure}[h!]
  \begin{subfigure}[b]{0.45\textwidth}
  \centering
\begin{tikzpicture}
\fill (0,1) circle (4pt);
\fill (3,0) circle (4pt);
\node[above left] at (0,1) {$s$};
\node[below right] at (3,0) {$t$};
  % Draw grid points
  \foreach \x in {0,...,3} {
    \foreach \y in {0,1} {
        \draw (\x,\y) circle (4pt);
    }
  }
  % Draw optional grid lines
  \foreach \x in {0,...,3} {
    \draw[gray, thin] (\x,0) -- (\x,1);
  }
  \foreach \y in {0,1} {
    \draw[gray, thin] (0,\y) -- (3,\y);
  }

  % Bold vertical line from upper-left to lower-left
  \draw[very thick] (0,0) -- (0,1);
  \draw[very thick] (0,0) -- (1,0);
  \draw[very thick] (1,0) -- (1,1);
  \draw[very thick] (1,1) -- (2,1);
  \draw[very thick] (2,1) -- (2,0);
  \draw[very thick] (2,0) -- (3,0);
\end{tikzpicture}
\caption{}
  \label{fig:doesn't work even different y-1}
    \end{subfigure}
    \hfill
  \begin{subfigure}[b]{0.45\textwidth}
  \centering
\begin{tikzpicture}
\fill (0,1) circle (4pt);
\node[above left] at (0,1) {$s$};
\node[above right] at (4,1) {$t$};
\fill (4,1) circle (4pt);
  % Draw grid points
  \foreach \x in {0,...,4} {
    \foreach \y in {0,1} {
        \draw (\x,\y) circle (4pt);
    }
  }

  % Draw optional grid lines
  \foreach \x in {0,...,4} {
    \draw[gray, thin] (\x,0) -- (\x,1);
  }
  \foreach \y in {0,1} {
    \draw[gray, thin] (0,\y) -- (4,\y);
  }

  % Bold vertical line from upper-left to lower-left
  \draw[very thick] (0,0) -- (0,1);
  \draw[very thick] (0,0) -- (1,0);
  \draw[very thick] (1,0) -- (1,1);
  \draw[very thick] (1,1) -- (2,1);
  \draw[very thick] (2,1) -- (2,0);
  \draw[very thick] (2,0) -- (3,0);
  \draw[very thick] (3,0) -- (3,1);
  \draw[very thick] (4,1) -- (3,1);
\end{tikzpicture}
\caption{}
  \label{fig:doesn't work odd for same y-1}
    \end{subfigure}
\caption{}
  \label{fig:show doesn't work}
\end{figure}

    \begin{figure}[h!]
\begin{subfigure}[b]{0.45\textwidth}
  \centering
\begin{tikzpicture}
\fill (0,3) circle (4pt);
\fill (0,0) circle (4pt);
  % Draw grid points
  \foreach \x in {0,...,3} {
    \foreach \y in {0,...,3} {
        \draw (\x,\y) circle (4pt);
    }
  }
  % Draw optional grid lines
  \foreach \x in {0,...,3} {
    \draw[gray, thin] (\x,0) -- (\x,3);
  }
  \foreach \y in {0,3} {
    \draw[gray, thin] (0,\y) -- (3,\y);
  }

  % Bold vertical line from upper-left to lower-left
  \draw[very thick] (0,3) -- (3,3) -- (3,2) -- (0,2) -- (0,1) -- (3,1) -- (3,0)--(0,0);
    \node[above left] at (0,3) {$s$};
    \node[above left] at (0,0) {$t$};
\end{tikzpicture}
\caption{}
  \label{same x - 1}
    \end{subfigure}
    \hfill  
    \begin{subfigure}[b]{0.45\textwidth}
  \centering
\begin{tikzpicture}
\fill (0,3) circle (4pt);
\fill (4,0) circle (4pt);
  % Draw grid points
  \foreach \x in {0,...,4} {
    \foreach \y in {0,...,3} {
        \draw (\x,\y) circle (4pt);
    }
  }
  % Draw optional grid lines
  \foreach \x in {0,...,4} {
    \draw[gray, thin] (\x,0) -- (\x,3);
  }
  \foreach \y in {0,3} {
    \draw[gray, thin] (0,\y) -- (4,\y);
  }

  % Bold vertical line from upper-left to lower-left
  \draw[very thick] (0,3) -- (0,0) -- (1,0) -- (1,3) -- (2,3) -- (2,0) -- (3,0) -- (3,3) -- (4,3) -- (4,0);
    \node[above left] at (0,3) {$s$};
    \node[above left] at (4,0) {$t$};
\end{tikzpicture}
\caption{}
  \label{different x-1}
    \end{subfigure}
  
\caption{}
  \label{even-rectangle-1}
\end{figure}

We now extend these observations from $2$-rectangles to more general rectangular grid graphs. The following lemma illustrates standard construction strategies when the height $n$ is even, which will be used as a basis for later constructions.

\begin{restatable}{lemma}{even}
    For any rectangle $R(m,n)$, if $n$ is even, the Hamiltonian Path can be created by:
    \begin{enumerate}
        \item if $s$ is the top left vertex and $t$ is bottom left vertex, then $s=(0,0)$, $t=(0,n-1)$, and $m$ is even, we can construct a Hamiltonian path by using strategy 1, go horizontally, then turn when necessary, as shown in Figure \ref{same x - 1}.
        \item if $s=(0,0)$ and $t=(m-1,n-1)$, we can construct one by using strategy 2, go vertically, then turn when necessary, as shown in Figure \ref{different x-1}.
    \end{enumerate}
    \label{even}
\end{restatable}

We observe that for a rectangular grid graph $R(m,n)$, different parity combinations of $m$ and $n$ may lead to different acceptable starting and ending vertex combinations. Hence, given an arbitrary rectangular grid graph, $R(m,n)$, we proposed a classification method into 10 categories based on the parity of its width and height, as shown in Equation \ref{eq:Ci-def-1}. Note that, apart from those purely based on parity, there are also size-induced cases, i.e. $C_1, C_2, C_3, C_5, C_7,$ and $C_9$. These special cases occur when $m$ or $n$ is $2$ or $3$, and there may not be enough space to turn with so few vertices. 

For example, Figure \ref{same x - 1} is a rectangle with $4$ columns and $4$ rows, $R(4,4)$, $m$ and $n$ are both even and greater than $2$. Hence, Figure \ref{same x - 1} is $C_4$. As shown in Figure \ref{fig:doesn't work odd for same y-1}, $R(2, 5)$, is a rectangle with $5$ columns and $2$ rows, which makes it fall into $C_7$.

\begin{equation}\label{eq:Ci-def-1}
\begin{aligned}
C_1&:\ m=2,\\
C_3&:\ m \text{ even},\ m>2,\ n=2,\\
C_4&:\ m \text{ even},\ m>2,\ n \text{ even},\ n>2,\\
C_5&:\ m \text{ even},\ m>2,\ n=3,\\
C_6&:\ m \text{ even},\ m>2,\ n \text{ odd},\ n>3,\\
C_2&:\ m=3,\\
C_7&:\ m \text{ odd},\ m>3,\ n=2,\\
C_8&:\ m \text{ odd},\ m>3,\ n \text{ even},\ n>2,\\
C_9&:\ m \text{ odd},\ m>3,\ n = 3\\
C_{10}&:\ m \text{ odd},\ m>3,\ n \text{ odd}, n > 3.
\end{aligned}
\end{equation}

\subsection{Constructing a Hamiltonian Cycle}

\subsubsection{Even-sized Rectangles}
Inspired by the snake-based construction of Lewis et al.~\cite{lewis2023methods}, we describe an alternative explicit procedure for constructing Hamiltonian cycles in even-sized rectangular grid graphs. We first introduce a basic construction that will be used repeatedly in
later sections. Consider an even-sized rectangular grid graph $R(m,n)$,
where $m$ denotes the number of columns and $n$ denotes the number of rows.
Since $R(m,n)$ is even-sized, at least one of $m$ and $n$ is even.
Therefore, if necessary, we rotate the rectangle by $90^\circ$ so that
the number of vertices in each row is even. After this rotation, we assume
that $m$ is even.

We use the coordinate system in which $(0,0)$ is the upper-left vertex,
the $x$-coordinate increases from left to right, and the $y$-coordinate
increases from top to bottom. Thus, the upper-right vertex is
$(m-1,0)$ and the bottom-right vertex is $(m-1,n-1)$.

\begin{figure}[htbp]
\centering
\begin{tikzpicture}[x=0.7cm,y=-0.7cm]
\foreach \x in {0,...,5}{\foreach \y in {0,...,4}{
 \ifnum\x<5 \draw[gray!35] (\x,\y)--++(1,0);\fi
 \ifnum\y<4 \draw[gray!35] (\x,\y)--++(0,1);\fi
 \fill (\x,\y) circle (1.3pt);
}}
\draw[blue!70!black,very thick,->] (0,0)--(3,0);
\draw[blue!70!black,very thick]
 (0,0)--(5,0)--(5,4)--(4,4)--(4,1)--(3,1)--(3,4)
 --(2,4)--(2,1)--(1,1)--(1,4)--(0,4)--cycle;
\node[above left] at (0,0) {$(0,0)$};
\node[above right] at (5,0) {$(m-1,0)$};
\node[below left] at (0,4) {$(0,n-1)$};
\node[below right] at (5,4) {$(m-1,n-1)$};
\end{tikzpicture}
\caption{The rectangular cycle construction for $R(6,5)$. The top
row is covered first; alternating vertical traversals stop below it,
and the reserved leftmost column closes the cycle.}
\label{even rectangle constructing hc}
\end{figure}
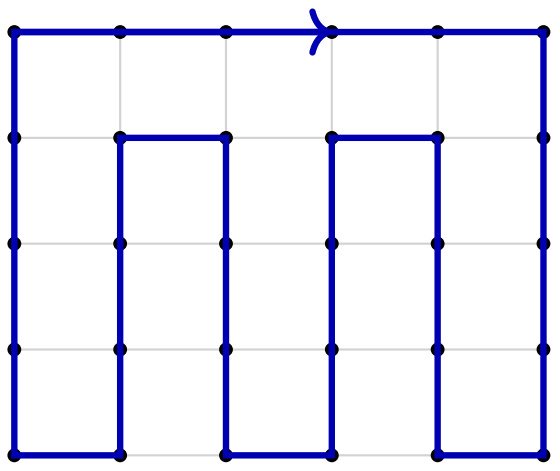

As shown in Fig. \ref{even rectangle constructing hc}, a Hamiltonian Cycle can be constructed in the following way: Starting from $(0,0)$, we first traverse the entire top row from left to
right and then move downward along the rightmost column. After reaching
the bottom boundary, we repeatedly move one column to the left and
traverse that column in the opposite vertical direction. In this way,
the path forms a sequence of alternating vertical traversals connected
by single horizontal edges. The process continues until every vertex is
covered, after which the path returns to $(0,0)$ through the leftmost
column. 

The construction is formally described in Algorithm~\ref{alg:even-rectangle-cycle}.

\begin{algorithm}[htbp]
\caption{Hamiltonian Cycle in an Even-Sized Rectangular Grid Graph}
\label{alg:even-rectangle-cycle}
\KwIn{An even-sized rectangular grid graph $R(m,n)$, with $m,n\geq2$}
\KwOut{A Hamiltonian cycle of $R(m,n)$}
Rotate the rectangle by $90^\circ$, if necessary, so that $m$ is even\;
Start at $(0,0)$ and traverse the top row to $(m-1,0)$\;
Move down the rightmost column to $(m-1,n-1)$\;
\For{$j=m-2,m-3,\ldots,1$}{
 Move left to column $j$\;
 \eIf{$j$ is even}{
  Traverse upward to $(j,1)$, stopping below the visited top row\;
 }{
  Traverse downward to $(j,n-1)$\;
 }
}
Move left to $(0,n-1)$\;
Traverse the leftmost column upward and close the cycle at $(0,0)$\;
Return the cycle, rotating it back if the input was rotated\;
\end{algorithm}

\begin{theorem}
\label{thm:even-rectangle-cycle}
For every even-sized rectangular grid graph $R(m,n)$ with $m,n\geq 2$,
Algorithm~\ref{alg:even-rectangle-cycle} constructs a Hamiltonian cycle.
\end{theorem}

\begin{proof}
Since $R(m,n)$ is even-sized, at least one of $m$ and $n$ is even.
Therefore, after rotating the rectangle if necessary, we may assume that
$m$ is even.
Algorithm~\ref{alg:even-rectangle-cycle} first covers the entire top row.
It then traverses the remaining columns alternately downward and upward,
moving one column to the left after each vertical traversal. Each upward
traversal stops at row $1$, so the already visited top row is not revisited.
The leftmost column is reserved for closing the cycle. Since $m$ is even,
the traversal of column $1$ ends at its bottom vertex, adjacent to $(0,n-1)$.

Each column is traversed only once, and the already-covered top-row
vertices are not revisited. Therefore, every vertex of $R(m,n)$ is
visited exactly once.
Finally, the algorithm moves upward along the leftmost column and returns
to the starting vertex $(0,0)$. Hence, the resulting path is a cycle that
contains every vertex exactly once, and is therefore a Hamiltonian cycle.
\end{proof}

\subsubsection{Odd-sized Rectangles}

The previous theorem shows that Algorithm~\ref{alg:even-rectangle-cycle}
constructs a Hamiltonian cycle whenever the rectangular grid graph is
even-sized. We now consider the remaining case, when the rectangular grid
graph has an odd number of vertices. In this case, a Hamiltonian cycle is
impossible due to the bipartite structure of grid graphs \cite{article}.

\begin{theorem}
\label{thm:odd-no-cycle}
Let $G$ be a finite grid graph. If $G$ has an odd number of
vertices, then $G$ does not admit a Hamiltonian cycle. 
\end{theorem}

% \begin{proof}
% Color each vertex $(x,y)$ of $G$ according to the parity of
% $x+y$. Since every edge joins two vertices of opposite
% colors, $G$ is bipartite. Every cycle in a bipartite graph
% has even length. A Hamiltonian cycle must visit every vertex
% exactly once, so its length equals $|V(G)|$. Therefore,
% if $|V(G)|$ is odd, $G$ cannot admit a Hamiltonian cycle.
% \end{proof}

Together, Theorems \ref{thm:even-rectangle-cycle} and \ref{thm:odd-no-cycle} imply that a rectangular grid graph
$R(m,n)$, with $m,n \geq 2$, has a Hamiltonian cycle if and only if
$mn$ is even. We will need to focus on odd-sized rectangular grid graphs for later larger grid's Hamiltonian Cycle construction.

\subsection{Connecting Two Hamiltonian Cycles}

The previous subsection gives a fixed construction of a Hamiltonian cycle
for every even-sized rectangular grid graph. We now consider how two such
cycles can be combined when the corresponding rectangular grid graphs are
adjacent. This operation will later allow us to construct Hamiltonian
cycles in larger grid graphs by decomposing them into rectangular
subgraphs.

\begin{definition}
Let $R_1$ and $R_2$ be two vertex-disjoint rectangular grid graphs. We say
that $R_1$ and $R_2$ are \emph{connected} if there exist at least two
consecutive boundary vertices of $R_1$ that are respectively adjacent to
two consecutive boundary vertices of $R_2$.
\end{definition}

The portions of the boundaries of $R_1$ and $R_2$ that face each other
will be called their \emph{connecting boundaries}.

For a Hamiltonian cycle constructed by
Algorithm~\ref{alg:even-rectangle-cycle}, the edges along a boundary have
one of two forms. On three of the four boundaries, all consecutive
boundary vertices are joined by the Hamiltonian cycle; we call such a
boundary a \emph{straight boundary}. On the remaining boundary, the cycle
contains alternating boundary edges: the first two vertices are joined,
the next pair is not, the following pair is joined, and so on. We call
this an \emph{alternating boundary}.

Thus, when two rectangular grid graphs are connected, their connecting
boundaries fall into one of the following three cases:
\begin{enumerate}
    \item a straight boundary is adjacent to an alternating boundary;
    \item a straight boundary is adjacent to a straight boundary;
    \item an alternating boundary is adjacent to an alternating boundary.
\end{enumerate}

\begin{theorem}
\label{thm:connect-two-cycles}
Let $R_1$ and $R_2$ be two connected even-sized rectangular grid graphs,
and let $C_1$ and $C_2$ be the Hamiltonian cycles constructed in $R_1$
and $R_2$, respectively, by
Algorithm~\ref{alg:even-rectangle-cycle}. Assume that the common interface contains an edge of each alternating
connecting boundary involved. In the alternating--alternating case,
also assume that the two alternating patterns are aligned as described
below. Then there
exist edges
$
u_1v_1\in C_1
\text{and }
u_2v_2\in C_2
$
on the two connecting boundaries such that
$
u_1u_2, v_1v_2
$
are edges of the full grid graph. In other words, the two cycle edges are
parallel and adjacent across the connecting boundary.
\end{theorem}

\begin{proof}
Let the connecting boundaries of $R_1$ and $R_2$ be considered.

If either connecting boundary is a straight boundary, then every pair of
consecutive vertices on that boundary is joined by an edge of its
Hamiltonian cycle. The connecting interface contains consecutive vertices that are adjacent
across the two rectangles. If the other boundary is alternating, the
interface assumption ensures that at least one of its used boundary
edges lies within this interface. The straight boundary contains the
corresponding parallel edge.

If the other boundary is also straight, those two vertices are joined by
a cycle edge immediately. If the other boundary is alternating, we choose
the corresponding consecutive pair for which the alternating boundary
edge belongs to the Hamiltonian cycle. Hence, in both the
straight--straight and straight--alternating cases, a pair of parallel
cycle edges can be found.

It remains to consider the alternating--alternating case. For this case,
we restrict our attention to two rectangles whose connecting boundaries
are aligned; equivalently, the rectangles share the same left, right,
top, or bottom outer boundary. Without loss of generality, assume that
they share the same left boundary. Because Algorithm~\ref{alg:even-rectangle-cycle}
produces the same alternating pattern from the same aligned endpoint, the
alternating cycle edges on the two connecting boundaries occur in
matching positions. Thus, there exists a pair of adjacent parallel cycle
edges, one belonging to $C_1$ and the other to $C_2$.

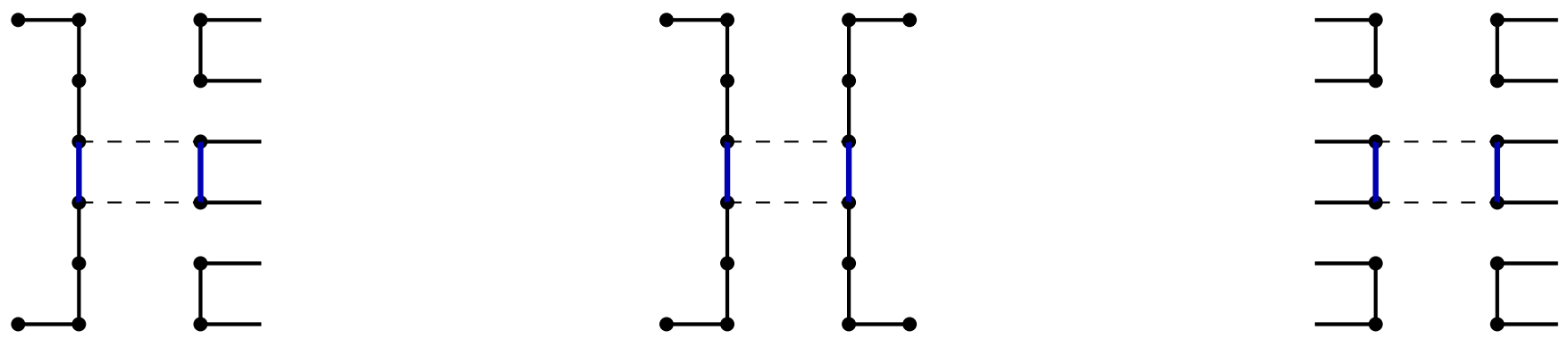
\begin{figure}[htbp]
\centering
\begin{tikzpicture}[x=0.45cm,y=0.45cm,every node/.style={font=\small}]
\foreach \k in {0,1,2}{
 \begin{scope}[xshift=\k*4.8cm]
 \foreach \y in {0,...,5}{
  \fill (0,\y) circle (1.5pt);\fill (2,\y) circle (1.5pt);
 }
 \ifnum\k<2
 \fill (-1,0) circle (1.5pt);
 \fill (-1,5) circle (1.5pt);
  \draw[thick] (-1,0)--(0,0)--(0,5)--(-1,5);
 \else
  \draw[thick] (-1,0)--(0,0)--(0,1)--(-1,1) (-1,2)--(0,2)--(0,3)--(-1,3) (-1,4)--(0,4)--(0,5)--(-1,5);
 \fi
 \ifnum\k=1
 \fill (3,0) circle (1.5pt);
 \fill (3,5) circle (1.5pt);
  \draw[thick] (3,0)--(2,0)--(2,5)--(3,5);
 \else
  \draw[thick] (2,0)--(3,0) (2,0)--(2,1) (2,1)--(3,1) (2,2)--(3,2) (2,2)--(2,3) (2,3)--(3,3) (2,4)--(3,4) (2,4)--(2,5) (2,5)--(3,5);
 \fi
 \draw[blue!70!black,very thick] (0,2)--(0,3) (2,2)--(2,3);
 \draw[dashed] (0,2)--(2,2) (0,3)--(2,3);
 \ifcase\k
  \node[below,align=center] at (1,-0.5) {Straight--alternating};
 \or
  \node[below,align=center] at (1,-0.5) {Straight--straight};
 \or
  \node[below,align=center] at (1,-0.5) {Alternating--alternating};
 \fi
 \end{scope}
}
\end{tikzpicture}
\caption{The three connecting-boundary cases. Blue segments are a
compatible pair of cycle edges; dashed segments are grid edges across
the interface. Alternating patterns must contain matching edges within
the common interface.}
\label{fig:two cycle connected}
\end{figure}

Therefore, under the stated interface and alignment conditions, such a
pair of parallel cycle edges can be found in all three cases.
\end{proof}

Once such a pair of parallel cycle edges has been found, the two
Hamiltonian cycles can be merged into a single Hamiltonian cycle
using the merge operation described in~\cite{Lhole_recent}.

Suppose the two parallel edges are
$
u_1v_1 \in C_1
\text{and }
u_2v_2 \in C_2,
$
where $u_1$ is adjacent to $u_2$ and $v_1$ is adjacent to $v_2$ in the
original grid graph. We remove the two edges $u_1v_1$ and $u_2v_2$, and
instead add the two connecting edges
$
u_1u_2
\text{ and }
v_1v_2.
$

Removing one edge from each Hamiltonian cycle turns each cycle into a
Hamiltonian path. The two newly added edges connect the endpoints of these
two paths and therefore form a single cycle containing all vertices of
both rectangles exactly once. Hence, the resulting cycle is a Hamiltonian
cycle of the combined grid graph.

\begin{figure}[ht]
    \centering

    % ---------------- Before ----------------
    \begin{tikzpicture}[scale=0.73,
        vertex/.style={circle, fill=black, inner sep=2.2pt},
        every node/.style={font=\small}
    ]

    % First cycle
    \draw[thick]
        (0,0) -- (0,2) -- (2,2) -- (2,1.3);
    \draw[thick]
        (2,0.7) -- (2,0) -- (0,0);

    % Second cycle
    \draw[thick]
        (3,0) -- (3,0.7);
    \draw[thick]
        (3,1.3) -- (3,2) -- (5,2) -- (5,0) -- (3,0);

    % Edges to remove
    \draw[very thick] (2,0.7) -- (2,1.3);
    \draw[very thick] (3,0.7) -- (3,1.3);

    % Vertices
    \node[vertex,label=left:$u_1$] at (2,1.3) {};
    \node[vertex,label=left:$v_1$] at (2,0.7) {};

    \node[vertex,label=right:$u_2$] at (3,1.3) {};
    \node[vertex,label=right:$v_2$] at (3,0.7) {};

    \node at (1,1) {$C_1$};
    \node at (4,1) {$C_2$};

    % adjacency across rectangles
    \draw[dashed] (2,1.3) -- (3,1.3);
    \draw[dashed] (2,0.7) -- (3,0.7);

    \node at (2.5,-0.55) {(a) Before merging};

    \end{tikzpicture}
    \hspace{0.4cm}
    % ---------------- After ----------------
    \begin{tikzpicture}[scale=0.73,
        vertex/.style={circle, fill=black, inner sep=2.2pt},
        every node/.style={font=\small}
    ]

    % First former cycle, with u1-v1 removed
    \draw[thick]
        (0,0) -- (0,2) -- (2,2) -- (2,1.3);

    \draw[thick]
        (2,0.7) -- (2,0) -- (0,0);

    % Second former cycle, with u2-v2 removed
    \draw[thick]
        (3,1.3) -- (3,2) -- (5,2) -- (5,0) -- (3,0) -- (3,0.7);

    % New connecting edges
    \draw[very thick] (2,1.3) -- (3,1.3);
    \draw[very thick] (2,0.7) -- (3,0.7);

    % Vertices
    \node[vertex,label=left:$u_1$] at (2,1.3) {};
    \node[vertex,label=left:$v_1$] at (2,0.7) {};

    \node[vertex,label=right:$u_2$] at (3,1.3) {};
    \node[vertex,label=right:$v_2$] at (3,0.7) {};

    \node at (2.5,-0.55) {(b) After merging};

    \end{tikzpicture}

    \caption{Merging two Hamiltonian cycles. In (a), $u_1v_1$ and
    $u_2v_2$ are parallel cycle edges, while the dashed edges indicate
    adjacency between the two rectangular grid graphs. In (b), the edges
    $u_1v_1$ and $u_2v_2$ are removed and replaced by $u_1u_2$ and
    $v_1v_2$, producing one larger Hamiltonian cycle.}
    \label{fig:merge-cycles}
\end{figure}
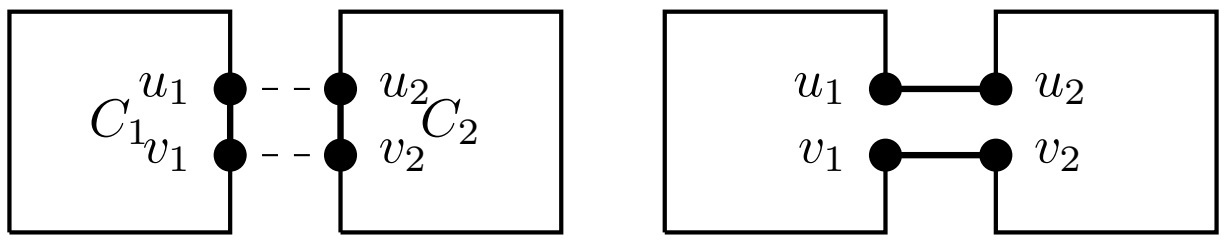

This merging operation is illustrated in Figure~\ref{fig:merge-cycles}.

\paragraph{Conclusion.}
We have now shown how to construct a Hamiltonian cycle in any even-sized
rectangular grid graph, and how to merge two rectangular Hamiltonian cycles when compatible
boundary edges are available.
Therefore, whenever a larger grid graph can be decomposed into connected
even-sized rectangular subgraphs, we may construct a Hamiltonian cycle
within each rectangle and repeatedly apply the merging operation, provided
that a compatible edge pair remains available at every step, to obtain
one Hamiltonian cycle for the entire graph.

This observation allows us to return to our original setting of a
rectangular grid graph with a triangular hole. Rather than constructing
the Hamiltonian cycle globally, we will separate the remaining grid into
a collection of rectangular subgraphs and view the full graph as a set of
rectangles connected along their boundaries. For every even-sized
rectangle, the construction from Algorithm~\ref{alg:even-rectangle-cycle}
can be applied directly, and the resulting cycles can be combined using
the merging operation established above.

The only rectangles that require additional treatment are odd-by-odd
rectangles, since these subgraphs are odd-sized and therefore do not admit
Hamiltonian cycles by themselves. In Section~\ref{sec:decomposition}, we
define how the larger grid graph is separated into rectangular subgraphs.
We then analyze the possible appearances of odd-by-odd rectangles and
develop separate constructions for handling them.

\section{Geometry-Driven Hamiltonian Cycle Construction}\label{sec:geometric}

We present an algorithm for constructing the Hamiltonian cycle in a rectangular Grid Graph with a right triangular hole if such a cycle exists. In general, there are two cases: if there is enough space around the hole, we construct the Hamiltonian cycle by generating a Hamiltonian Path that covers the complementary reversed triangle in the smallest rectangle that contains the triangular hole, and then connect it to the Hamiltonian cycle in the outer grid; if the hole is tightly bounded by the rectangular grid graph, i.e. the path of the Hamiltonian Cycle is constrained because the space is too narrow,  then a Hamiltonian cycle's construction is determined by each boundary variables' value, i.e. $a, b, h_0, \cdots$. 

\subsection{Case Determination}

\begin{figure}[h!]
    \centering
    
    \begin{subfigure}[b]{0.45\textwidth}
        \centering
        \begin{tikzpicture}[scale=1.2, every node/.style={font=\small}]
  % Define points
  \coordinate (A) at (0, 0);
  \coordinate (B) at (4, 0);
  \coordinate (C) at (4, 5);
  \coordinate (D) at (0, 5);

  % Draw triangle
  \draw[thick] (A) -- (B) -- (C) -- (D) -- cycle;
  \coordinate (E'') at (0.5,1.2);
  \coordinate (E') at (0,1.2);
  \coordinate (E) at (0.8,1);
  \coordinate (F) at (3.5,1);
  \coordinate (F') at (4,1.2);
  \coordinate (F'') at (3.5,1.2);
  \coordinate (G) at (3.5,1.4);
  \coordinate (H') at (3.2,1.4);
  \coordinate (H) at (2.2,2.3);
  \coordinate (I) at (2.2,4);
  \coordinate (I') at (4,4);
  \coordinate (J) at (1.8,4);
  \coordinate (J') at (0, 4);
  \coordinate (K') at (0.8,1.4);
  \coordinate (K) at (1.8,2.3);
  \coordinate (L) at (0.8,4);
  \coordinate (M) at (0.8,1);
  \coordinate (N) at (0.8,1.8);
  \coordinate (O) at (3.2,1.8);
  \coordinate (P) at (3.2,1.4);
  \coordinate (uppermid) at (2,5);
  \coordinate (uppermid') at (2, 4);
  \coordinate (lowermid) at (2,0);
  \coordinate (lowermid') at (2, 1);
  \draw[thick] (E) -- (F) -- (G) -- (H');
  \draw[thick] (2.2, 2.3) -- (2.2,4) -- (L) -- (E) ;
  % \draw[thick] (H) -- (I) -- (J) -- (K);
  % \draw[thick] (M) -- (N) -- (1.2,1.8);
  % \draw[thick] (3, 1.8) -- (O) -- (P);
  % \draw[thick] (1.2, 1.8) -- (1.2, 2.3) -- (1.8, 2.3);
  \draw[thick] (2.2,2.3) -- (3, 2.3) -- (3, 1.4) -- (3.2, 1.4);
 \draw[dotted] (3,2) -- (4,2) 
    node[midway, above=3pt] {$c_1$};
    
 % \draw[dotted] (0,2) -- (1.2,2) 
    % node[midway, above=3pt] {$d_1$};
    
  % Label side lengths
  \draw [dotted](A) -- (B) 
    node[midway, below=10pt] {m};
  \draw[dotted] (I) -- (I') 
    node[midway, below=5pt] {$c_0$};  
 % \draw [dotted](J) -- (J') 
    % node[midway, below=5pt] {$d_0$};
 \draw [dotted](3.5,1.1) -- (4,1.1) 
    node[midway, above=3pt] {$c_2$};
 % \draw [dotted](0,1.1) -- (0.5,1.1) 
 %    node[midway, above=3pt] {$d_3$};
  \draw [dotted](B) -- (C) 
    node[midway, right=5pt] {n};
\draw[dotted] (uppermid) -- (uppermid') 
    node[midway, right=5pt] {a};
\draw[dotted] (lowermid) -- (lowermid') 
    node[midway, right=5pt] {b};
    \draw[dotted] (3.2,1) -- (3.2,1.4) 
    node[midway] {$h_2$};
    % \draw[dotted] (3,1.4) -- (3.2,1.4) 
    % node[midway, right=5pt] {};
    \draw[dotted] (2.7,1.4) -- (3,1.4) 
    node[midway, right=5pt] {};
    \draw[dotted] (2.5,2.3) -- (2.5,4) 
    node[midway, right=5pt] {$h_0$};
    % \draw[dotted] (3,1.4) -- (3,1.8) 
    % node[midway, left=5pt] {$h_2$};
    \draw[dotted] (2.7,1.4) -- (2.7,2.3) 
    node[midway] {$h_1$};
    % \draw[dotted] (3,2.1) -- (0.8,2.1) 
    % node[midway] {$w_1$};
    \draw[dotted] (0,2) -- (0.8,2) 
    node[midway, above=3pt] {$d$};
    % \draw[dotted] (4,1.5) -- (3.2,1.5) 
    % node[midway, above=2pt] {$c_2$};
    % \draw[dotted] (0.8, 3) -- (2.2, 3) 
    % node[midway, below=5pt] {$w_0$};
    % \draw[dotted] (0.8, 1.3) -- (4, 1.3) 
    % node[midway] {$w_2$};
      \fill[pattern=north east lines, pattern color=gray] (2.2,5) -- (0.8,5) -- (0.8,4) -- (2.2,4) -- (2.2,2.3) -- (3,2.3) -- (3,1.4) -- (3.5,1.4) -- (4,1.4) -- (4,5);
\end{tikzpicture}
        \caption{Rectangular Grid Graph with 3-stacked Right Triangular Hole}
        \label{fig:general-setting}
    \end{subfigure}
    \hfill
    \begin{subfigure}[b]{0.45\textwidth}
        \centering
         \begin{tikzpicture}
  % Background gray dots for all grid points
  \foreach \x in {0,...,7}
    \foreach \y in {0,...,9}
      \fill[gray] (\x,\y) circle(0.04);

  % ── Parity-coloured nodes ──────────────────────────────────────
  % Rule: black if (x+y) even, white if (x+y) odd

  % All nodes using parity colouring

    \foreach \x in {0,...,7} {
    \foreach \y in {0,1,8,9} {
      \pgfmathparse{int(mod(\x+\y,2))==0 ? 1 : 0}
      \ifnum\pgfmathresult=1
        \fill[black] (\x,\y) circle(4pt);
      \else
        \draw[fill=white] (\x,\y) circle(4pt);
      \fi
    }
  }
  
  \foreach \x in {0,1} {
    \foreach \y in {2,...,7} {
      \pgfmathparse{int(mod(\x+\y,2))==0 ? 1 : 0}
      \ifnum\pgfmathresult=1
        \fill[black] (\x,\y) circle(4pt);
      \else
        \draw[fill=white] (\x,\y) circle(4pt);
      \fi
    }
  } 
  \foreach \x in {6,7} {
    \foreach \y in {4,5,6,7} {
      \pgfmathparse{int(mod(\x+\y,2))==0 ? 1 : 0}
      \ifnum\pgfmathresult=1
        \fill[black] (\x,\y) circle(4pt);
      \else
        \draw[fill=white] (\x,\y) circle(4pt);
      \fi
    }
  } 
  \foreach \x in {5} {
    \foreach \y in {6,7} {
      \pgfmathparse{int(mod(\x+\y,2))==0 ? 1 : 0}
      \ifnum\pgfmathresult=1
        \fill[black] (\x,\y) circle(4pt);
      \else
        \draw[fill=white] (\x,\y) circle(4pt);
      \fi
    }
  }
  % \foreach \x in {3,4} {
  %   \foreach \y in {2,3,4,5} {
  %     \pgfmathparse{int(mod(\x+\y,2))==0 ? 1 : 0}
  %     \ifnum\pgfmathresult=1
  %       \fill[black] (\x,\y) circle(4pt);
  %     \else
  %       \draw[fill=white] (\x,\y) circle(4pt);
  %     \fi
  %   }
  % } 
  % \foreach \x in {2} {
  %   \foreach \y in {4,5} {
  %     \pgfmathparse{int(mod(\x+\y,2))==0 ? 1 : 0}
  %     \ifnum\pgfmathresult=1
  %       \fill[black] (\x,\y) circle(4pt);
  %     \else
  %       \draw[fill=white] (\x,\y) circle(4pt);
  %     \fi
  %   }
  % }

  % ── Start label (col 1, row 6 = top-left of path) ─────────────
  % \node[above left] at (3,4) {$s$};

  % ── End label (col 4, row 1 = bottom-right of staircase) ──────
  % \node[above left] at (1,5) {$t$};

  % ── Path edges (thick blue staircase) ─────────────────────────
  % Top row: col 1 → col 5, row 6
  % \draw[blue, line width=1.5pt]
  %   (1,6)--(2,6)--(3,6)--(4,6)--(5,6);

  % Right side col 5: row 6 → row 5
  \draw[green, line width=1.5pt]
   (7,9)--(0,9)--(0,0)--(7,0)--(7,1)--(1,1)--(1,8)--(5,8)--(5,6)--(6,6)--(6,4)--(7,4)--(7,7)--(6,7)--(6,8)--(7,8)--(7,9);

  % Step 1 box: row 5, col 2 → col 5
  % \draw[blue, line width=1.5pt]
  %   (2,5)--(3,5)--(4,5)--(5,5);

  % Left wall col 2: row 5 → row 4
  % \draw[blue, line width=1.5pt]
  %   (2,5)--(2,4);

  % Right side col 5: row 5 → row 4
  % \draw[blue, line width=1.5pt]
  %   (5,5)--(5,4);

  % Step 2 box: row 4, col 2 → col 5
  % \draw[blue, line width=1.5pt]
  %   (2,4)--(3,4)--(4,4)--(5,4);

  % Left wall col 2: row 4 → row 3
  % \draw[blue, line width=1.5pt]
  %   (2,4)--(2,3);

  % Right side col 5: row 4 → row 3
  % \draw[blue, line width=1.5pt]
  %   (5,4)--(5,3);

  % Step 3 box: row 3, col 3 → col 5  (staircase steps in)
  % \draw[blue, line width=1.5pt]
  %   (3,3)--(4,3)--(5,3);

  % Left wall col 3: row 3 → row 2
  % \draw[blue, line width=1.5pt]
  %   (3,3)--(3,2);

  % Right side col 5: row 3 → row 2 (but path ends at col 4)
  % col 4 down: row 3 → row 2
  % \draw[blue, line width=1.5pt]
  %   (4,3)--(4,2);

  % Step 4: row 2, col 3 → col 4 (innermost box, 2 wide)
  % \draw[blue, line width=1.5pt]
  %   (3,2)--(4,2);

  % col 4 down: row 2 → row 1 (final descent to t)
  % \draw[blue, line width=1.5pt]
  %   (4,2)--(4,1);

\end{tikzpicture}
        
        \caption{Large Hole Case When $c_2 = 0$}
        \label{fig:large hole}
    \end{subfigure}
    \caption{}
\end{figure}

Given a rectangular Grid Graph with a 3-stacked right triangular hole, shown in Figure \ref{fig:general-setting}, to construct a Hamiltonian cycle if one exists, we want to construct the path for the irregular part of the graph first (the shaded area in Figure \ref{fig:general-setting}), then find a way to connect it to the rest of the graph, if there is enough space to do so. Therefore, given such a Grid Graph, we need to first determine if there is enough space to do so. We define the following criteria:  if $c_2>0$, $a\geq 2$, $d>0$, and $b>0$, then there is enough space to build the irregular part first and then connect it to the whole grid graph. We call this the small hole case. Otherwise, if $c_2 =0$, $a<2$, $d=0$, or $b=0$, as long as any of these holds, we will need to construct the Hamiltonian cycle globally, which we will call the large hole case, as shown in Figure \ref{fig:large hole}. We next present the constructions for these two cases.

\subsection{The Small Hole Case}\label{sec:small-hole}
\begin{figure}[h!]
    \centering
    \begin{subfigure}[b]{0.3\textwidth}
        \centering
        \begin{tikzpicture}[scale=1.2, every node/.style={font=\small}]
  % Define points
  \coordinate (A) at (0, 0);
  \coordinate (B) at (4, 0);
  \coordinate (C) at (4, 5);
  \coordinate (D) at (0, 5);

  \draw[thick] (A) -- (B) -- (C) -- (D) -- cycle;
  \coordinate (E'') at (0.5,1.2);
  \coordinate (E') at (0,1.2);
  \coordinate (E) at (0.8,1);
  \coordinate (F) at (3.5,1);
  \coordinate (F') at (4,1.2);
  \coordinate (F'') at (3.5,1.2);
  \coordinate (G) at (3.5,1.4);
  \coordinate (H') at (3.2,1.4);
  \coordinate (H) at (2.2,2.3);
  \coordinate (I) at (2.2,4);
  \coordinate (I') at (4,4);
  \coordinate (J) at (1.8,4);
  \coordinate (J') at (0, 4);
  \coordinate (K') at (0.8,1.4);
  \coordinate (K) at (1.8,2.3);
  \coordinate (L) at (0.8,4);
  \coordinate (M) at (0.8,1);
  \coordinate (N) at (0.8,1.8);
  \coordinate (O) at (3.2,1.8);
  \coordinate (P) at (3.2,1.4);
  \coordinate (uppermid) at (2,5);
  \coordinate (uppermid') at (2,4);
  \coordinate (lowermid) at (2,0);
  \coordinate (lowermid') at (2,1);

  \draw[thick] (E) -- (F) -- (G) -- (H');
  \draw[thick] (2.2, 2.3) -- (2.2,4) -- (L) -- (E);
  \draw[thick] (2.2,2.3) -- (3, 2.3) -- (3, 1.4) -- (3.2, 1.4);

  % \draw[dotted] (3,2) -- (4,2)
  %   node[midway, above=3pt] {$c_1$};

  \draw[dotted](A) -- (B)
    node[midway, below=10pt] {$m$};

  % \draw[dotted] (2.2,3.5) -- (4,3.5)
  %   node[midway, below=5pt] {$c_0$};

  % \draw[dotted](3.5,1.1) -- (4,1.1)
  %   node[midway, above=3pt] {$c_2$};

  \draw[dotted](B) -- (C)
    node[midway, right=5pt] {$n$};

  % \draw[dotted] (1.8,4) -- (1.8,5)
  %   node[midway, right=5pt] {$a$};

  % \draw[dotted] (lowermid) -- (lowermid')
  %   node[midway, right=5pt] {$b$};

  % \draw[dotted] (0,2) -- (0.8,2)
  %   node[midway, above=3pt] {$d$};

    \draw[thick] (3.5,1.4) -- (3.6,1.4) -- (3.6,4.2) -- (0.8,4.2) -- (0.8,4);
    \draw[thick] (0.8,4.2) -- (0.8,4) node[midway, left] {$n_1$};
     \draw[dotted] (1.4,4) -- (3.6,4) node[midway, below] {$m_1$};
     \draw[thick] (2.2,4) -- (2.2,2.3) node[midway, left] {$n_2$};
     \draw[thick] (3,1.5) -- (3,2.3) node[midway, left] {$n_3$};
     \draw[] (3,1.4) -- (3.6,1.4) node[midway, below] {$m_3$};
     \draw[dotted] (2.2,2.3) -- (3.6,2.3) node[midway, above] {$m_2$};
    % \draw[->, thick] (4.2,3.2) -- (3.3,3.2) node[near start, right=5pt] {$R_2$};
    % \draw[->, thick] (4.2,1.7) -- (3.3,1.7) node[near start, right=5pt] {$R_3$};
    % \draw[->, thick] (4.2,4.1) -- (3.3,4.1) node[near start, right=5pt] {$R_1$};
    \fill[pattern=north east lines, pattern color=red] (0.8,4.2) -- (0.8, 4) -- (3.6,4) -- (3.6,4.2);
    \fill[pattern=north west lines, pattern color=green] (0,0)--(0,5)--(4,5)--(4,0)--(3.5,0)--(3.5,1.4)--(3.6,1.4)--(3.6,4.2)--(4,4.2)--(0.8,4.2)--(0.8,1)--(4,1)--(4,0)--(0,0);
    \fill[pattern=north east lines, pattern color=red] (2.2,4)-- (2.2,2.3) -- (3,2.3) -- (3.6,2.3) -- (3.6,4);
    \fill[pattern=north east lines, pattern color=red]   (3,2.3) -- (3,1.4) -- (3.6,1.4) -- (3.6,2.3);
\end{tikzpicture}
        \caption{}
        \label{inner&outer}
    \end{subfigure}
    \hfill
    \begin{subfigure}[b]{0.3\textwidth}
        \centering
        \begin{tikzpicture}[scale=1.2, every node/.style={font=\small}]
  % Define points
  \coordinate (A) at (0, 0);
  \coordinate (B) at (4, 0);
  \coordinate (C) at (4, 5);
  \coordinate (D) at (0, 5);

  \draw[thick] (A) -- (B) -- (C) -- (D) -- cycle;
  \coordinate (E'') at (0.5,1.2);
  \coordinate (E') at (0,1.2);
  \coordinate (E) at (0.8,1);
  \coordinate (F) at (3.5,1);
  \coordinate (F') at (4,1.2);
  \coordinate (F'') at (3.5,1.2);
  \coordinate (G) at (3.5,1.4);
  \coordinate (H') at (3.2,1.4);
  \coordinate (H) at (2.2,2.3);
  \coordinate (I) at (2.2,4);
  \coordinate (I') at (4,4);
  \coordinate (J) at (1.8,4);
  \coordinate (J') at (0, 4);
  \coordinate (K') at (0.8,1.4);
  \coordinate (K) at (1.8,2.3);
  \coordinate (L) at (0.8,4);
  \coordinate (M) at (0.8,1);
  \coordinate (N) at (0.8,1.8);
  \coordinate (O) at (3.2,1.8);
  \coordinate (P) at (3.2,1.4);
  \coordinate (uppermid) at (2,5);
  \coordinate (uppermid') at (2,4);
  \coordinate (lowermid) at (2,0);
  \coordinate (lowermid') at (2,1);

  \draw[thick] (E) -- (F) -- (G) -- (H');
  \draw[thick] (2.2, 2.3) -- (2.2,4) -- (L) -- (E);
  \draw[thick] (2.2,2.3) -- (3, 2.3) -- (3, 1.4) -- (3.2, 1.4);

  % \draw[dotted] (3,2) -- (4,2)
  %   node[midway, above=3pt] {$c_1$};

  \draw[dotted](A) -- (B)
    node[midway, below=10pt] {$m$};

  % \draw[dotted] (2.2,3.5) -- (4,3.5)
  %   node[midway, below=5pt] {$c_0$};

  % \draw[dotted](3.5,1.1) -- (4,1.1)
  %   node[midway, above=3pt] {$c_2$};

  \draw[dotted](B) -- (C)
    node[midway, right=5pt] {$n$};

  % \draw[dotted] (1.8,4) -- (1.8,5)
  %   node[midway, right=5pt] {$a$};

  % \draw[dotted] (lowermid) -- (lowermid')
  %   node[midway, right=5pt] {$b$};

  % \draw[dotted] (0,2) -- (0.8,2)
  %   node[midway, above=3pt] {$d$};

    \draw[thick] (3.5,1.4) -- (3.6,1.4) -- (3.6,4.2) -- (0.8,4.2) -- (0.8,4);
    \draw[thick] (0.8,4.2) -- (0.8,4) node[midway, left] {$n_1=2$};
     \draw[dotted] (1.4,4) -- (3.6,4) node[midway, below] {$m_1$};
     \draw[thick] (2.2,4) -- (2.2,2.3) node[midway, left] {$n_2$};
     \draw[thick] (3,1.5) -- (3,2.3) node[midway, left] {$n_3$};
     \draw[] (3,1.4) -- (3.6,1.4) node[midway, above] {$m_3$};
     \draw[dotted] (2.2,2.3) -- (3.6,2.3) node[midway, above] {$m_2$};
    \draw[thick] (3.6,1.4) -- (3.6,4.2) node[midway, above=5pt, left=4pt] {$G_1$};
      \fill[pattern=north east lines, pattern color=gray] (0.8,4.2) -- (0.8, 4) -- (2.2,4)-- (2.2,2.3) -- (3,2.3) -- (3,1.4) -- (3.6,1.4) -- (3.6,4.2);
    % \draw[thick] (3.5,1.4) -- (3.6,1.4) node[midway, right=1pt, below=1pt] {1};
    \draw[->, thick] (3.55,0.8) -- (3.55,1.4) node[near start, below=2pt] {$1$};
\end{tikzpicture}
        \caption{}
        \label{grid graph with rectangular cover}
    \end{subfigure}
    \hfill
    \begin{subfigure}[b]{0.3\textwidth}
        \centering
        \begin{tikzpicture}[scale=1.2, every node/.style={font=\small}]
  % Define points
  \coordinate (A) at (0, 0);
  \coordinate (B) at (4, 0);
  \coordinate (C) at (4, 5);
  \coordinate (D) at (0, 5);

  \draw[thick] (A) -- (B) -- (C) -- (D) -- cycle;
  \coordinate (E'') at (0.5,1.2);
  \coordinate (E') at (0,1.2);
  \coordinate (E) at (0.8,1);
  \coordinate (F) at (3.5,1);
  \coordinate (F') at (4,1.2);
  \coordinate (F'') at (3.5,1.2);
  \coordinate (G) at (3.5,1.4);
  \coordinate (H') at (3.2,1.4);
  \coordinate (H) at (2.2,2.3);
  \coordinate (I) at (2.2,4);
  \coordinate (I') at (4,4);
  \coordinate (J) at (1.8,4);
  \coordinate (J') at (0, 4);
  \coordinate (K') at (0.8,1.4);
  \coordinate (K) at (1.8,2.3);
  \coordinate (L) at (0.8,4);
  \coordinate (M) at (0.8,1);
  \coordinate (N) at (0.8,1.8);
  \coordinate (O) at (3.2,1.8);
  \coordinate (P) at (3.2,1.4);
  \coordinate (uppermid) at (2,5);
  \coordinate (uppermid') at (2,4);
  \coordinate (lowermid) at (2,0);
  \coordinate (lowermid') at (2,1);

  \draw[thick] (E) -- (F) -- (G) -- (H');
  \draw[thick] (2.2, 2.3) -- (2.2,4) -- (L) -- (E);
  \draw[thick] (2.2,2.3) -- (3, 2.3) -- (3, 1.4) -- (3.2, 1.4);

  % \draw[dotted] (3,2) -- (4,2)
  %   node[midway, above=3pt] {$c_1$};

  \draw[dotted](A) -- (B)
    node[midway, below=10pt] {$m$};

  % \draw[dotted] (2.2,3.5) -- (4,3.5)
  %   node[midway, below=5pt] {$c_0$};

  % \draw[dotted](3.5,1.1) -- (4,1.1)
  %   node[midway, above=3pt] {$c_2$};

  \draw[dotted](B) -- (C)
    node[midway, right=5pt] {$n$};

  % \draw[dotted] (1.8,4) -- (1.8,5)
  %   node[midway, right=5pt] {$a$};

  % \draw[dotted] (lowermid) -- (lowermid')
  %   node[midway, right=5pt] {$b$};

  % \draw[dotted] (0,2) -- (0.8,2)
  %   node[midway, above=3pt] {$d$};

    \draw[thick] (3.5,1.4) -- (3.6,1.4) -- (3.6,4.2) -- (0.8,4.2) -- (0.8,4);
    \draw[thick] (0.8,4.2) -- (0.8,4) node[midway, left] {$n_1$};
     \draw[dotted] (1.4,4) -- (3.6,4) node[midway, below] {$m_1$};
     \draw[thick] (2.2,4) -- (2.2,2.3) node[midway, left] {$n_2$};
     \draw[thick] (3,1.5) -- (3,2.3) node[midway, left] {$n_3$};
     \draw[] (3,1.4) -- (3.6,1.4) node[midway, below] {$m_3$};
     \draw[dotted] (2.2,2.3) -- (3.6,2.3) node[midway, above] {$m_2$};
    \draw[->, thick] (4.2,3.2) -- (3.3,3.2) node[near start, right=5pt] {$R_2$};
    \draw[->, thick] (4.2,1.7) -- (3.3,1.7) node[near start, right=5pt] {$R_3$};
    \draw[->, thick] (4.2,4.1) -- (3.3,4.1) node[near start, right=5pt] {$R_1$};
    \fill[pattern=north east lines, pattern color=gray] (0.8,4.2) -- (0.8, 4) -- (3.6,4) -- (3.6,4.2);
    \fill[pattern=north west lines, pattern color=blue] (2.2,4)-- (2.2,2.3) -- (3,2.3) -- (3.6,2.3) -- (3.6,4);
    \fill[pattern=north east lines, pattern color=red]   (3,2.3) -- (3,1.4) -- (3.6,1.4) -- (3.6,2.3);
\end{tikzpicture}
        \caption{}
        \label{grid graph with rectangular sep}
    \end{subfigure}
        \caption{Rectangular Grid Graph with 3-stacked Right Triangular Hole}
\end{figure}
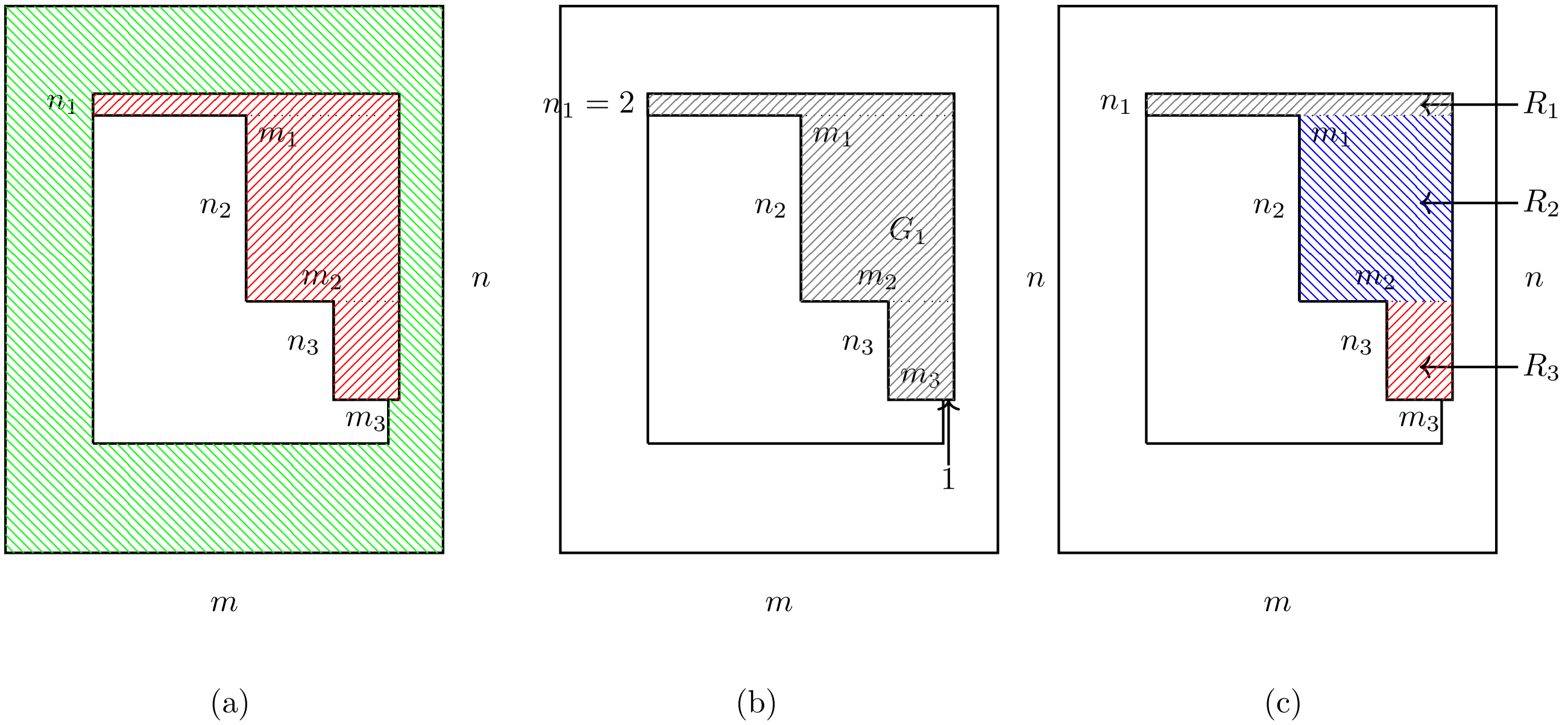

For the small hole case, as mentioned above, we construct a Hamiltonian Cycle for the inner irregular part, denoted as the red shadowed area in Figure \ref{inner&outer}, and a Hamiltonian Cycle for the outer part, denoted as the green shadowed area in Figure \ref{inner&outer}, and then connect those two together. We first construct the Hamiltonian Path for the irregular part, i.e., the reversed right rectangle that forms the minimal axis-aligned rectangle determined by the legs of the triangular hole, but with two vertices up and one column extended to the right, labeled as $G_1$ in Figure \ref{grid graph with rectangular cover}. 

We will construct the Hamiltonian Path of $G_1$ depending on the cases of the three-stacked subgraph between $s$ and $t$, which is the
leftmost vertex in the second row of $G_1$ and the rightmost vertex in the last row of $G_1$, respectively, and then connect it to the outer grids.
In order to better explain the algorithm for constructing a Hamiltonian Cycle under the small hole case, we will use an example to show how this works step by step in each subsection as we present the algorithm.

\subsubsection{Reversed Right Triangle}

In this section, we will explain the construction of the Hamiltonian Path in the complementary reversed right triangle, i.e., $G_1$ in Figure \ref{grid graph with rectangular cover} (the shaded subgraph).
We can also view this $G_1$ as three rectangles stacked together as shown in Figure \ref{grid graph with rectangular sep}, where from top to bottom, $R_1, R_2, $ and $R_3$ have $m_i$ and $n_i$ as their widths and heights. Since $G_1$ is a reversed three-stacked triangle, we require that the width strictly decreases from top to bottom, i.e $m_1> m_2>m_3$.

Note that in order to explain the structure of this three-stacking grid, we will describe each vertex using local coordinates, where $(0,0)$ represent the leftmost vertex on the top row of the subgraph $G_1$, and $(m_1,n_3+n_2+n_1-1)$ represents the rightmost vertex of the bottom row of $G_1$. We may also use a more local representation that use the information of each rectangle, i.e., $R_1[right]$ represents the rightmost column of the topmost rectangle, $(m_1-1, y)$, and $R_2[left]$ represents the leftmost column of the middle rectangle, $(m_1-m_2-1, y)$.

The main difficulty of constructing a Hamiltonian path in $G_1$ lies in coordinating the local paths across the stacked rectangles. While each rectangle individually admits Hamiltonian paths under appropriate boundary conditions, the vertical stacking imposes additional constraints: the endpoint of one rectangle must coincide with the starting point of the next. Thus, feasibility is determined not only by local structure, but also by the compatibility of endpoints across layers.

From the case classification of the rectangular grid graphs, the existence of a Hamiltonian path between boundary vertices is governed primarily by parity conditions, which partition the rectangles into two classes $S_1$ and $S_2$. Consequently, the problem reduces to selecting starting vertices whose $x$-coordinates satisfy these parity constraints while ensuring consistent alignment between adjacent rectangles. As what is described in algorithm \ref{algorithm_imp}, it systematically determines such starting positions based on the case type of each rectangle, thereby guaranteeing that the resulting local paths can be composed into a global Hamiltonian path.

Given the placement and property of $G_1$ in the grid graph, we have the following restriction for our Hamiltonian path construction in addition to the three-stacked setup: the starting vertex $s$ must lie on the bottom left of $R_1$, $(0,1)$, and the ending vertex $t$ is $(m_1-1, n_3+n_2+n_1-1)$, which is the rightmost vertex on the bottom row of $G_1$. Given those three rectangles, the following should hold: $n_1 =2$ by design, $m_1=w_2+1$, $n_2 = h_0$, $m_2 =w_2+1-w_0$, $n_3 = h_1$, and $m_3 = w_2 - w_1 + 1$. 

In order to construct a Hamiltonian Path between the desired $s$ and $t$ in the grid graph $G_1$, we proposed the following algorithm, which clearly identifies a way to construct a Hamiltonian Path. 
As shown in the algorithm \ref{algorithm_imp}, $s$, the starting vertex's x-coordinate is assigned clearly for all possible cases and y-coordinate can be computed simply, since the starting vertex of each rectangle stacked in this reversed right triangle always lies on the first row of each rectangle, which allows us to always find the starting vertex for each rectangle. The ending vertex $t$ will just be the vertex that connects to the starting vertex of the rectangle below, i.e., $t_1$ is right above $s_2$, $t_2$ is right above $s_3$, and $t_3$ is the rightmost vertex in the last row of the $R_3$. Note that the algorithm is only concerned with assigning the x-coordinates of the starting vertices because the y-coordinates can be easily inferred, which is simply the top most row and bottom most row shown in each rectangle.
Therefore, since the three starting vertices can always be found for all possible cases with some provable exceptions, i.e., all different stacking combinations of the three rectangles, a Hamiltonian path can always be constructed within $G_1$.

\begin{algorithm}[H]
\label{three stack}
\caption{Construct Hamiltonian Path in Reversed Right Triangle}
\KwIn{$R_1, R_2, \text{ and} R_3$}
\KwOut{the x-coordinates of the starting vertices}

$S_1 = \{C_1, C_2, C_3, C_4, C_7, C_8, C_9, C_{10}\}$\;

$S_2 = \{C_5, C_6\}$\;

\If{$R_3 \in S_1$}{
    \If{$m_2=3, m_3=2, \text{ and } n_3 \text{ is odd}$}{
        \tcp{Not exist I}
    }
    \ElseIf{$m_3=2, n_3\text{ is even}, m_2=3, m_1 \text{ is odd, and }n_1=2$}{
        \tcp{Not exist II}
    }
    \ElseIf{$m_3=2, n_3 \text{ is odd, and } m_2\text{ is even }\geq 4$}{
        \tcp{Subcase I}
        
        $s_{3x} \gets 0$\;
        
        \If{$R_2 \in S_1$}{
            $s_{1x} \gets m - 1\;$ 
            $s_{2x} \gets 0\;$
        }
        \Else{
            $s_{1x} \gets m-1\;$
            
            $s_{2x} \gets m-1\;$
        }
    }

    \ElseIf{$m_3 \text{ is even, }n_3 \text{ is odd, }m_2 \text{ is odd, and }n_2 \text{ is odd}$}{
        \tcp{Not exist II}
    }
    
    \ElseIf{$m_2 \text{ is odd }>3, m_3 = 2, n_3 \text{ is odd}$}{
        \tcp{Subcase II}
    
        $s_{1x} \gets m-1\;$
        
        $s_{2x} \gets 1\; $
        
        $s_{3x} \gets 0\;$
    }
    \Else{      

        $s_{3x} \gets m\;$
    
        \If{$R_2 \in S_1$}{
            $s_{1x} \gets m\;$
            
            $s_{2x} \gets m\;$
        
        }
        \Else{
            $s_{1x}\gets m-3\; $
            
            $s_{2x}\gets m-1\;$
        }
    }
}
\Else{
    \If{$m_2 \text{ is odd, }n_2\text{ is odd, }m_3\text{ is even, and }n_3\text{ is odd}$}{
        \tcp{Not exist II}
    }
    \Else{
        \If{$R_2\in S_1$}{
            $s_{1x}\gets m-1\;$
            
            $s_{2x} \gets m-3\;$
        }
        \Else{
            $s_{1x}\gets m\;$ 
            
            $s_{2x} \gets m-2\;$
        }
    }
    $s_{3x} \gets m-1\;$
}
\label{algorithm_imp}
\end{algorithm}

\paragraph{Example}

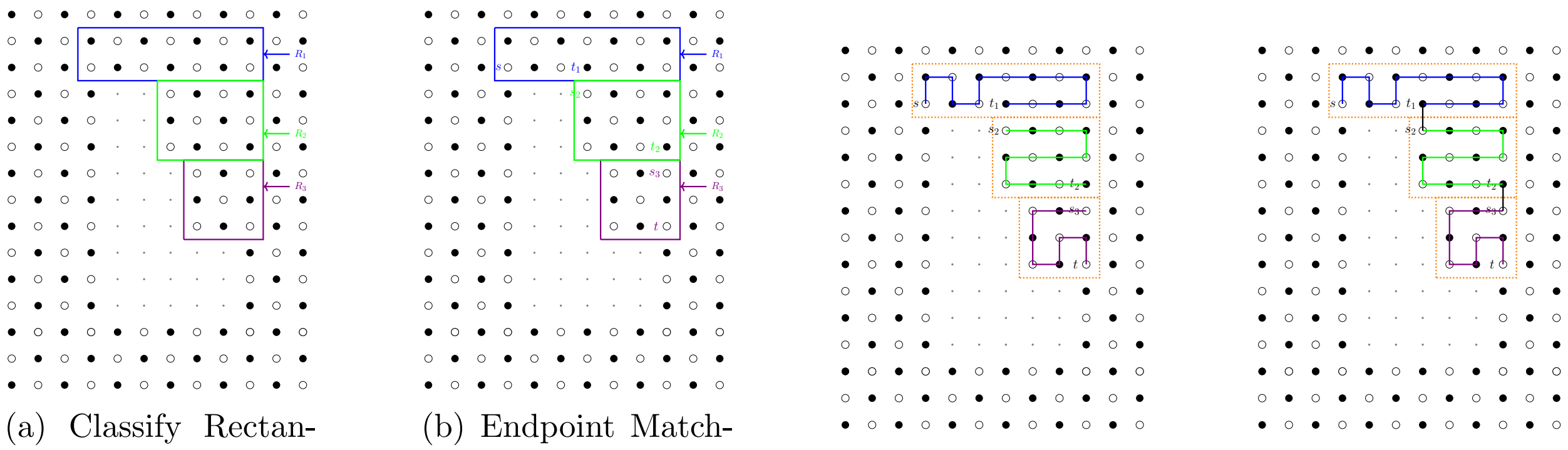
\begin{figure}[h!]
    \centering
    \begin{subfigure}[b]{0.20\textwidth}
        \centering
        \resizebox{\linewidth}{!}{
            \begin{tikzpicture}
  % Background gray dots for all grid points
  \foreach \x in {0,...,11}
    \foreach \y in {0,...,14}
      \fill[gray] (\x,\y) circle(0.04);

  % ── Parity-coloured nodes ──────────────────────────────────────
  % Rule: black if (x+y) even, white if (x+y) odd

  % All nodes using parity colouring

    \foreach \x in {0,...,11} {
    \foreach \y in {0,1,2,12,13,14} {
      \pgfmathparse{int(mod(\x+\y,2))==0 ? 1 : 0}
      \ifnum\pgfmathresult=1
        \fill[black] (\x,\y) circle(4pt);
      \else
        \draw[fill=white] (\x,\y) circle(4pt);
      \fi
    }
  }
  
  \foreach \x in {0,1,2,3,9,10,11} {
    \foreach \y in {2,...,11} {
      \pgfmathparse{int(mod(\x+\y,2))==0 ? 1 : 0}
      \ifnum\pgfmathresult=1
        \fill[black] (\x,\y) circle(4pt);
      \else
        \draw[fill=white] (\x,\y) circle(4pt);
      \fi
    }
  }
   \foreach \x in {7,8} {
    \foreach \y in {6,7,8,9,10,11} {
      \pgfmathparse{int(mod(\x+\y,2))==0 ? 1 : 0}
      \ifnum\pgfmathresult=1
        \fill[black] (\x,\y) circle(4pt);
      \else
        \draw[fill=white] (\x,\y) circle(4pt);
      \fi
    }
  }
  \foreach \x in {6} {
    \foreach \y in {9,10,11} {
      \pgfmathparse{int(mod(\x+\y,2))==0 ? 1 : 0}
      \ifnum\pgfmathresult=1
        \fill[black] (\x,\y) circle(4pt);
      \else
        \draw[fill=white] (\x,\y) circle(4pt);
      \fi
    }
  }

    \draw[blue, line width=1.5pt] (2.5,13.5)--(2.5,11.5)--(9.5,11.5)--(9.5,13.5)--(2.5,13.5);

    \draw[violet, line width=1.5pt] (6.5,5.5)--(6.5,8.5)--(9.5,8.5)--(9.5,5.5)--(6.5,5.5);

    \draw[green, line width=1.5pt] (5.5,11.5)--(5.5,8.5)--(9.5,8.5)--(9.5,11.5)--(5.5,11.5);

    \draw[->, thin, blue, line width=1.5pt] (10.5,12.5) -- (9.5,12.5) node[near start, right=8pt] {$R_1$};

    \draw[->, thin, green, line width=1.5pt] (10.5,9.5) -- (9.5,9.5) node[near start, right=8pt] {$R_2$};

    \draw[->, thin, violet, line width=1.5pt] (10.5,7.5) -- (9.5,7.5) node[near start, right=8pt] {$R_3$};

\end{tikzpicture}
          }
        \caption{Classify Rectangle}
        \label{small_example_1_1}
    \end{subfigure}
    \hfill
    \begin{subfigure}[b]{0.20\textwidth}
        \centering
        \resizebox{\linewidth}{!}{
            \begin{tikzpicture}
  % Background gray dots for all grid points
  \foreach \x in {0,...,11}
    \foreach \y in {0,...,14}
      \fill[gray] (\x,\y) circle(0.04);

  % ── Parity-coloured nodes ──────────────────────────────────────
  % Rule: black if (x+y) even, white if (x+y) odd

  % All nodes using parity colouring

    \foreach \x in {0,...,11} {
    \foreach \y in {0,1,2,12,13,14} {
      \pgfmathparse{int(mod(\x+\y,2))==0 ? 1 : 0}
      \ifnum\pgfmathresult=1
        \fill[black] (\x,\y) circle(4pt);
      \else
        \draw[fill=white] (\x,\y) circle(4pt);
      \fi
    }
  }
  
  \foreach \x in {0,1,2,3,9,10,11} {
    \foreach \y in {2,...,11} {
      \pgfmathparse{int(mod(\x+\y,2))==0 ? 1 : 0}
      \ifnum\pgfmathresult=1
        \fill[black] (\x,\y) circle(4pt);
      \else
        \draw[fill=white] (\x,\y) circle(4pt);
      \fi
    }
  }
   \foreach \x in {7,8} {
    \foreach \y in {6,7,8,9,10,11} {
      \pgfmathparse{int(mod(\x+\y,2))==0 ? 1 : 0}
      \ifnum\pgfmathresult=1
        \fill[black] (\x,\y) circle(4pt);
      \else
        \draw[fill=white] (\x,\y) circle(4pt);
      \fi
    }
  }
  \foreach \x in {6} {
    \foreach \y in {9,10,11} {
      \pgfmathparse{int(mod(\x+\y,2))==0 ? 1 : 0}
      \ifnum\pgfmathresult=1
        \fill[black] (\x,\y) circle(4pt);
      \else
        \draw[fill=white] (\x,\y) circle(4pt);
      \fi
    }
  }
    \draw[blue, line width=1.5pt] (2.5,13.5)--(2.5,11.5)--(9.5,11.5)--(9.5,13.5)--(2.5,13.5);

    \draw[violet, line width=1.5pt] (6.5,5.5)--(6.5,8.5)--(9.5,8.5)--(9.5,5.5)--(6.5,5.5);

    \draw[green, line width=1.5pt] (5.5,11.5)--(5.5,8.5)--(9.5,8.5)--(9.5,11.5)--(5.5,11.5);

    \draw[->, thin, blue, line width=1.5pt] (10.5,12.5) -- (9.5,12.5) node[near start, right=8pt] {$R_1$};

    \draw[->, thin, green, line width=1.5pt] (10.5,9.5) -- (9.5,9.5) node[near start, right=8pt] {$R_2$};

    \draw[->, thin, violet, line width=1.5pt] (10.5,7.5) -- (9.5,7.5) node[near start, right=8pt] {$R_3$};
    
    \draw[dotted] (3,12)
    node[left=3pt,font=\Large,color=blue] {$s$};
    \draw[dotted] (6,12)
    node[left=3pt,font=\Large,color=blue] {$t_1$};
    \draw[dotted] (6,11)
    node[left=3pt,font=\Large,color=green] {$s_2$};
    \draw[dotted] (9,9)
    node[left=3pt,font=\Large,color=green] {$t_2$};
    \draw[dotted] (9,8)
    node[left=3pt,font=\Large,color=violet] {$s_3$};
    \draw[dotted] (9,6)
    node[left=5pt,font=\Large,color=violet] {$t$};
    
    % \draw[black, line width=1.5pt]
   % (3,12)--(3,13);
   % \draw[blue,line width=1.5pt] (3,13)--(4,13)--(4,12)--(5,12)--(5,13)--(9,13)--(9,12)--(6,12);
   % \draw[green, line width=1.5pt]
   % (6,11)--(9,11)--(9,10)--(6,10)--(6,9)--(9,9);
   % \draw[violet, line width=1.5pt]
   % (9,8)--(7,8)--(7,6)--(8,6)--(8,7)--(9,7)--(9,6);
   % \draw[black,line width=1.5pt](6,12)--(6,11);
   % \draw[black,line width=1.5pt](9,9)--(9,8);

   % \draw[black,line width = 1.5pt](9,6)--(9,2)--(3,2)--(3,12);
   % \draw[orange,dotted,line width=1.5pt](9.5,5.5)--(9.5,1.5)--(8.5,1.5)--(8.5,5.5)--(9.5,5.5);
   % \draw[orange,dotted,line width=1.5pt](8.5,1.5)--(8.5,2.5)--(2.5,2.5)--(2.5,1.5)--(8.5,1.5);
   % \draw[orange,dotted,line width=1.5pt](2.5,2.5)--(3.5,2.5)--(3.5,11.5)--(2.5,11.5)--(2.5,2.5);

   % \draw[black,line width=1.5pt] (2,2)--(0,2)--(0,0)--(1,0)--(1,1)--(2,1)--(2,0)--(3,0)--(3,1)--(4,1)--(4,0)--(5,0)--(5,1)--(6,1)--(6,0)--(7,0)--(7,1)--(8,1)--(8,0)--(9,0)--(9,1)--(10,1)--(10,0)--(11,0)--(11,2)--(10,2)--(10,3)--(11,3)--(11,4)--(10,4)--(10,5)--(11,5)--(11,6)--(10,6)--(10,7)--(11,7)--(11,8)--(10,8)--(10,9)--(11,9)--(11,10)--(10,10)--(10,11)--(11,11)--(11,12)--(10,12)--(10,13)--(11,13)--(11,14)--(4,14)
   
   % \draw[black,line width=1.5pt](3,14)--(0,14)--(0s,13)--(2,13)--(2,12)--(0,12)--(0,11)--(2,11)--(2,10);
   % \draw[black,line width=1.5pt](2,10)--(0,10)--(0,9)--(2,9)--(2,8)--(0,8)--(0,7)--(2,7)--(2,6)--(0,6)--(0,5)--(2,5)--(2,4)--(0,4)--(0,3)--(2,3)--(2,2);
   
% \draw[red,line width=2.5pt] (3,14)--(3,13);\draw[red,line width=2.5pt] (4,14)--(4,13);
   % \draw[red,line width=2pt](3,14)--(4,14);
   % \draw[red,line width=2pt](3,13)--(4,13);
\end{tikzpicture}
          }
        \caption{Endpoint Matching}
        \label{small_example_1_2}
    \end{subfigure}
    \hfill
    \begin{subfigure}[b]{0.20\textwidth}
        \centering
        \resizebox{\linewidth}{!}{
            \begin{tikzpicture}
  % Background gray dots for all grid points
  \foreach \x in {0,...,11}
    \foreach \y in {0,...,14}
      \fill[gray] (\x,\y) circle(0.04);

  % ── Parity-coloured nodes ──────────────────────────────────────
  % Rule: black if (x+y) even, white if (x+y) odd

  % All nodes using parity colouring

    \foreach \x in {0,...,11} {
    \foreach \y in {0,1,2,12,13,14} {
      \pgfmathparse{int(mod(\x+\y,2))==0 ? 1 : 0}
      \ifnum\pgfmathresult=1
        \fill[black] (\x,\y) circle(4pt);
      \else
        \draw[fill=white] (\x,\y) circle(4pt);
      \fi
    }
  }
  
  \foreach \x in {0,1,2,3,9,10,11} {
    \foreach \y in {2,...,11} {
      \pgfmathparse{int(mod(\x+\y,2))==0 ? 1 : 0}
      \ifnum\pgfmathresult=1
        \fill[black] (\x,\y) circle(4pt);
      \else
        \draw[fill=white] (\x,\y) circle(4pt);
      \fi
    }
  }
   \foreach \x in {7,8} {
    \foreach \y in {6,7,8,9,10,11} {
      \pgfmathparse{int(mod(\x+\y,2))==0 ? 1 : 0}
      \ifnum\pgfmathresult=1
        \fill[black] (\x,\y) circle(4pt);
      \else
        \draw[fill=white] (\x,\y) circle(4pt);
      \fi
    }
  }
  \foreach \x in {6} {
    \foreach \y in {9,10,11} {
      \pgfmathparse{int(mod(\x+\y,2))==0 ? 1 : 0}
      \ifnum\pgfmathresult=1
        \fill[black] (\x,\y) circle(4pt);
      \else
        \draw[fill=white] (\x,\y) circle(4pt);
      \fi
    }
  }
    \draw[orange, dotted, line width=1.5pt]
    (2.5,13.5)--(2.5,11.5)--(9.5,11.5)--(9.5,13.5)--(2.5,13.5);
    \draw[orange, dotted, line width=1.5pt]
    (6.5,5.5)--(6.5,8.5)--(9.5,8.5)--(9.5,5.5)--(6.5,5.5);
    \draw[orange, dotted, line width=1.5pt]
    (5.5,11.5)--(5.5,8.5)--(9.5,8.5)--(9.5,11.5)--(5.5,11.5);
    \draw[dotted] (3,12)
    node[left=3pt,font=\Large] {$s$};
    \draw[dotted] (6,12)
    node[left=3pt,font=\Large] {$t_1$};
    \draw[dotted] (6,11)
    node[left=3pt,font=\Large] {$s_2$};
    \draw[dotted] (9,9)
    node[left=3pt,font=\Large] {$t_2$};
    \draw[dotted] (9,8)
    node[left=3pt,font=\Large] {$s_3$};
    \draw[dotted] (9,6)
    node[left=5pt,font=\Large] {$t$};
    
    % \draw[black, line width=1.5pt]
   % (3,12)--(3,13);
   \draw[blue,line width=1.5pt] (3,12)--(3,13)--(4,13)--(4,12)--(5,12)--(5,13)--(9,13)--(9,12)--(6,12);
   \draw[green, line width=1.5pt]
   (6,11)--(9,11)--(9,10)--(6,10)--(6,9)--(9,9);
   \draw[violet, line width=1.5pt]
   (9,8)--(7,8)--(7,6)--(8,6)--(8,7)--(9,7)--(9,6);
   % \draw[black,line width=1.5pt](6,12)--(6,11);
   % \draw[black,line width=1.5pt](9,9)--(9,8);

   % \draw[black,line width = 1.5pt](9,6)--(9,2)--(3,2)--(3,12);
   % \draw[orange,dotted,line width=1.5pt](9.5,5.5)--(9.5,1.5)--(8.5,1.5)--(8.5,5.5)--(9.5,5.5);
   % \draw[orange,dotted,line width=1.5pt](8.5,1.5)--(8.5,2.5)--(2.5,2.5)--(2.5,1.5)--(8.5,1.5);
   % \draw[orange,dotted,line width=1.5pt](2.5,2.5)--(3.5,2.5)--(3.5,11.5)--(2.5,11.5)--(2.5,2.5);

   % \draw[black,line width=1.5pt] (2,2)--(0,2)--(0,0)--(1,0)--(1,1)--(2,1)--(2,0)--(3,0)--(3,1)--(4,1)--(4,0)--(5,0)--(5,1)--(6,1)--(6,0)--(7,0)--(7,1)--(8,1)--(8,0)--(9,0)--(9,1)--(10,1)--(10,0)--(11,0)--(11,2)--(10,2)--(10,3)--(11,3)--(11,4)--(10,4)--(10,5)--(11,5)--(11,6)--(10,6)--(10,7)--(11,7)--(11,8)--(10,8)--(10,9)--(11,9)--(11,10)--(10,10)--(10,11)--(11,11)--(11,12)--(10,12)--(10,13)--(11,13)--(11,14)--(4,14)
   
   % \draw[black,line width=1.5pt](3,14)--(0,14)--(0s,13)--(2,13)--(2,12)--(0,12)--(0,11)--(2,11)--(2,10);
   % \draw[black,line width=1.5pt](2,10)--(0,10)--(0,9)--(2,9)--(2,8)--(0,8)--(0,7)--(2,7)--(2,6)--(0,6)--(0,5)--(2,5)--(2,4)--(0,4)--(0,3)--(2,3)--(2,2);
   
% \draw[red,line width=2.5pt] (3,14)--(3,13);\draw[red,line width=2.5pt] (4,14)--(4,13);
   % \draw[red,line width=2pt](3,14)--(4,14);
   % \draw[red,line width=2pt](3,13)--(4,13);
\end{tikzpicture}
          }
        \caption{Generate Paths}
        \label{small_example_1_3}
    \end{subfigure}
    \hfill
    \begin{subfigure}[b]{0.20\textwidth}
        \centering
        \resizebox{\linewidth}{!}{
            \begin{tikzpicture}
  % Background gray dots for all grid points
  \foreach \x in {0,...,11}
    \foreach \y in {0,...,14}
      \fill[gray] (\x,\y) circle(0.04);

  % ── Parity-coloured nodes ──────────────────────────────────────
  % Rule: black if (x+y) even, white if (x+y) odd

  % All nodes using parity colouring

    \foreach \x in {0,...,11} {
    \foreach \y in {0,1,2,12,13,14} {
      \pgfmathparse{int(mod(\x+\y,2))==0 ? 1 : 0}
      \ifnum\pgfmathresult=1
        \fill[black] (\x,\y) circle(4pt);
      \else
        \draw[fill=white] (\x,\y) circle(4pt);
      \fi
    }
  }
  
  \foreach \x in {0,1,2,3,9,10,11} {
    \foreach \y in {2,...,11} {
      \pgfmathparse{int(mod(\x+\y,2))==0 ? 1 : 0}
      \ifnum\pgfmathresult=1
        \fill[black] (\x,\y) circle(4pt);
      \else
        \draw[fill=white] (\x,\y) circle(4pt);
      \fi
    }
  }
   \foreach \x in {7,8} {
    \foreach \y in {6,7,8,9,10,11} {
      \pgfmathparse{int(mod(\x+\y,2))==0 ? 1 : 0}
      \ifnum\pgfmathresult=1
        \fill[black] (\x,\y) circle(4pt);
      \else
        \draw[fill=white] (\x,\y) circle(4pt);
      \fi
    }
  }
  \foreach \x in {6} {
    \foreach \y in {9,10,11} {
      \pgfmathparse{int(mod(\x+\y,2))==0 ? 1 : 0}
      \ifnum\pgfmathresult=1
        \fill[black] (\x,\y) circle(4pt);
      \else
        \draw[fill=white] (\x,\y) circle(4pt);
      \fi
    }
  }
    \draw[orange, dotted, line width=1.5pt]
    (2.5,13.5)--(2.5,11.5)--(9.5,11.5)--(9.5,13.5)--(2.5,13.5);
    \draw[orange, dotted, line width=1.5pt]
    (6.5,5.5)--(6.5,8.5)--(9.5,8.5)--(9.5,5.5)--(6.5,5.5);
    \draw[orange, dotted, line width=1.5pt]
    (5.5,11.5)--(5.5,8.5)--(9.5,8.5)--(9.5,11.5)--(5.5,11.5);
    \draw[dotted] (3,12)
    node[left=3pt,font=\Large] {$s$};
    \draw[dotted] (6,12)
    node[left=3pt,font=\Large] {$t_1$};
    \draw[dotted] (6,11)
    node[left=3pt,font=\Large] {$s_2$};
    \draw[dotted] (9,9)
    node[left=3pt,font=\Large] {$t_2$};
    \draw[dotted] (9,8)
    node[left=3pt,font=\Large] {$s_3$};
    \draw[dotted] (9,6)
    node[left=5pt,font=\Large] {$t$};
    
    % \draw[black, line width=1.5pt]
   % (3,12)--(3,13);
   \draw[blue,line width=1.5pt] (3,12)--(3,13)--(4,13)--(4,12)--(5,12)--(5,13)--(9,13)--(9,12)--(6,12);
   \draw[green, line width=1.5pt]
   (6,11)--(9,11)--(9,10)--(6,10)--(6,9)--(9,9);
   \draw[violet, line width=1.5pt]
   (9,8)--(7,8)--(7,6)--(8,6)--(8,7)--(9,7)--(9,6);
   \draw[black,line width=1.5pt](6,12)--(6,11);
   \draw[black,line width=1.5pt](9,9)--(9,8);

   % \draw[black,line width = 1.5pt](9,6)--(9,2)--(3,2)--(3,12);
   % \draw[orange,dotted,line width=1.5pt](9.5,5.5)--(9.5,1.5)--(8.5,1.5)--(8.5,5.5)--(9.5,5.5);
   % \draw[orange,dotted,line width=1.5pt](8.5,1.5)--(8.5,2.5)--(2.5,2.5)--(2.5,1.5)--(8.5,1.5);
   % \draw[orange,dotted,line width=1.5pt](2.5,2.5)--(3.5,2.5)--(3.5,11.5)--(2.5,11.5)--(2.5,2.5);

   % \draw[black,line width=1.5pt] (2,2)--(0,2)--(0,0)--(1,0)--(1,1)--(2,1)--(2,0)--(3,0)--(3,1)--(4,1)--(4,0)--(5,0)--(5,1)--(6,1)--(6,0)--(7,0)--(7,1)--(8,1)--(8,0)--(9,0)--(9,1)--(10,1)--(10,0)--(11,0)--(11,2)--(10,2)--(10,3)--(11,3)--(11,4)--(10,4)--(10,5)--(11,5)--(11,6)--(10,6)--(10,7)--(11,7)--(11,8)--(10,8)--(10,9)--(11,9)--(11,10)--(10,10)--(10,11)--(11,11)--(11,12)--(10,12)--(10,13)--(11,13)--(11,14)--(4,14)
   
   % \draw[black,line width=1.5pt](3,14)--(0,14)--(0s,13)--(2,13)--(2,12)--(0,12)--(0,11)--(2,11)--(2,10);
   % \draw[black,line width=1.5pt](2,10)--(0,10)--(0,9)--(2,9)--(2,8)--(0,8)--(0,7)--(2,7)--(2,6)--(0,6)--(0,5)--(2,5)--(2,4)--(0,4)--(0,3)--(2,3)--(2,2);
   
% \draw[red,line width=2.5pt] (3,14)--(3,13);\draw[red,line width=2.5pt] (4,14)--(4,13);
   % \draw[red,line width=2pt](3,14)--(4,14);
   % \draw[red,line width=2pt](3,13)--(4,13);
\end{tikzpicture}
          }
        \caption{Connect Paths}
        \label{small_example_1_4}
    \end{subfigure}
    \caption{An example of a rectangular grid graph with a small three-stacked right-triangular hole}
    \label{small_example_1}
\end{figure}

We will use the example shown in Figure \ref{small_example_1} to explain this algorithm, where $c_2=3, c_1=5, c_0=6, d=4, a=3$ and $b=3$. As shown in Figure \ref{small_example_1_1}, the reversed triangle part, $G_1$ is denoted via the orange box, which is a combination of three stacked rectangular subgraphs, $R_1, R_2,$ and $R_3$ as denoted. For those three rectangles, $R_1(7,2), R_2(4,3),$ and $R_3(3,3)$, based on the parity of width and height, $R_1$ is categorized as $C_7$, $R_2$ is categorized as $C_5$, and $R_3$ is categorized as $C_2$. Then, according to Algorithm \ref{algorithm_imp}, as shown in Figure \ref{small_example_1_2}, we can find the positioning for $t_1, s_2,t_2,$
and $s_3$. Between each pair of starting vertex and ending vertex, a Hamiltonian path can be found within each rectangle as shown in Figure \ref{small_example_1_3}. 
Then complete the paths by connecting the ending vertex of the rectangle above to the starting vertex of rectangle below, i.e. $t_1$ to $s_2$, and $t_2$ to $s_3$, as shown in Figure \ref{small_example_1_4}.
Therefore, we found a Hamiltonian path within the reversed right triangle for this example grid graph.

\subsubsection{Bounding Rectangle}

We first define a bounding rectangle that contains the entire three-stacked triangular hole. Specifically, this rectangle extends the hole by one column to the left and right, one row downward, and two rows upward. This bounding region is illustrated as the red rectangle in Figure~\ref{grid graph with rectangular cover}.

\begin{figure}[h!]
    \centering
    \begin{subfigure}[b]{0.45\textwidth}
        \centering
        \begin{tikzpicture}[scale=1.2, every node/.style={font=\small}]
  % Define points
  \coordinate (A) at (0, 0);
  \coordinate (B) at (4, 0);
  \coordinate (C) at (4, 5);
  \coordinate (D) at (0, 5);

  \draw[thick] (A) -- (B) -- (C) -- (D) -- cycle;
  \coordinate (E'') at (0.5,1.2);
  \coordinate (E') at (0,1.2);
  \coordinate (E) at (0.8,1);
  \coordinate (F) at (3.5,1);
  \coordinate (F') at (4,1.2);
  \coordinate (F'') at (3.5,1.2);
  \coordinate (G) at (3.5,1.4);
  \coordinate (H') at (3.2,1.4);
  \coordinate (H) at (2.2,2.3);
  \coordinate (I) at (2.2,4);
  \coordinate (I') at (4,4);
  \coordinate (J) at (1.8,4);
  \coordinate (J') at (0, 4);
  \coordinate (K') at (0.8,1.4);
  \coordinate (K) at (1.8,2.3);
  \coordinate (L) at (0.8,4);
  \coordinate (M) at (0.8,1);
  \coordinate (N) at (0.8,1.8);
  \coordinate (O) at (3.2,1.8);
  \coordinate (P) at (3.2,1.4);
  \coordinate (uppermid) at (2,5);
  \coordinate (uppermid') at (2,4);
  \coordinate (lowermid) at (2,0);
  \coordinate (lowermid') at (2,1);

  \draw[thick] (E) -- (F) -- (G) -- (H');
  \draw[thick] (2.2, 2.3) -- (2.2,4) -- (L) -- (E);
  \draw[thick] (2.2,2.3) -- (3, 2.3) -- (3, 1.4) -- (3.2, 1.4);

  \draw[dotted](A) -- (B)
    node[midway, below=10pt] {$m$};

  \draw[dotted](B) -- (C)
    node[midway, right=5pt] {$n$};

    \draw[thick] (3.5,1.4) -- (3.6,1.4) -- (3.6,4.2) -- (0.8,4.2) ;
    \draw[thick] (0.8,4) -- (0.7,4);
    \draw[dotted] (0.8,4.2) -- (0.8,4) node[midway, right] {$n_1$};
     \draw[dotted] (1.4,4) -- (3.6,4) node[midway, below] {$m_1$};
     \draw[thick] (2.2,4) -- (2.2,2.3) node[midway, left] {$n2$};
     \draw[thick] (3,1.5) -- (3,2.3) node[midway, left] {$n_3$};
     \draw[] (3,1.4) -- (3.6,1.4) node[midway, above] {$m_3$};
     \draw[dotted] (2.2,2.3) -- (3.6,2.3) node[midway, above] {$m_2$};
    \draw[thick] (3.6,1.4) -- (3.6,4.2) node[midway, above=5pt, left=4pt] {$G_1$};
    \draw[thick] (3.5,0.9) -- (3.6,0.9) node[midway, right=1pt, below=1pt] {1};
    \draw[thick] (3.6,1.4) -- (3.6,0.9) -- (3.5, 0.9) -- (3.5, 1);
    \draw[thick] (0.7, 1) -- (0.8, 1);
    \draw[thick] (3.6, 0.9) -- (0.7, 0.9) -- (0.7, 4.2) -- (0.8, 4.2);
    \draw[->, thin] (4.2,1.2) -- (3.6,1.2) node[near start, right=2pt] {$G_2$};
    \draw[->, thin] (2.2,0.5) -- (2.2,0.9) node[near start, below=2pt] {$G_3$};
    \draw[->, thin] (0.4,2.5) -- (0.7,2.5) node[near start, left=0pt] {$G_4$};
    \draw[thick, red](3.6,0.9)--(0.7,0.9)--(0.7,4.2) -- (3.6,4.2) -- (3.6, 0.9);
\end{tikzpicture}
        \caption{$G_1, G_2, G_3, $ and $G4$ structure}
        \label{rec cover g1 g2}
    \end{subfigure}
    \hfill
    \begin{subfigure}[b]{0.45\textwidth}
        \centering
        \resizebox{0.7\linewidth}{!}{
            \begin{tikzpicture}
  % Background gray dots for all grid points
  \foreach \x in {0,...,11}
    \foreach \y in {0,...,14}
      \fill[gray] (\x,\y) circle(0.04);

  % ── Parity-coloured nodes ──────────────────────────────────────
  % Rule: black if (x+y) even, white if (x+y) odd

  % All nodes using parity colouring

    \foreach \x in {0,...,11} {
    \foreach \y in {0,1,2,12,13,14} {
      \pgfmathparse{int(mod(\x+\y,2))==0 ? 1 : 0}
      \ifnum\pgfmathresult=1
        \fill[black] (\x,\y) circle(4pt);
      \else
        \draw[fill=white] (\x,\y) circle(4pt);
      \fi
    }
  }
  
  \foreach \x in {0,1,2,3,9,10,11} {
    \foreach \y in {2,...,11} {
      \pgfmathparse{int(mod(\x+\y,2))==0 ? 1 : 0}
      \ifnum\pgfmathresult=1
        \fill[black] (\x,\y) circle(4pt);
      \else
        \draw[fill=white] (\x,\y) circle(4pt);
      \fi
    }
  }
   \foreach \x in {7,8} {
    \foreach \y in {6,7,8,9,10,11} {
      \pgfmathparse{int(mod(\x+\y,2))==0 ? 1 : 0}
      \ifnum\pgfmathresult=1
        \fill[black] (\x,\y) circle(4pt);
      \else
        \draw[fill=white] (\x,\y) circle(4pt);
      \fi
    }
  }
  \foreach \x in {6} {
    \foreach \y in {9,10,11} {
      \pgfmathparse{int(mod(\x+\y,2))==0 ? 1 : 0}
      \ifnum\pgfmathresult=1
        \fill[black] (\x,\y) circle(4pt);
      \else
        \draw[fill=white] (\x,\y) circle(4pt);
      \fi
    }
  }
    \draw[orange, dotted, line width=1.5pt]
    (2.5,13.5)--(2.5,11.5)--(9.5,11.5)--(9.5,13.5)--(2.5,13.5);
    \draw[orange, dotted, line width=1.5pt]
    (6.5,5.5)--(6.5,8.5)--(9.5,8.5)--(9.5,5.5)--(6.5,5.5);
    \draw[orange, dotted, line width=1.5pt]
    (5.5,11.5)--(5.5,8.5)--(9.5,8.5)--(9.5,11.5)--(5.5,11.5);
    \draw[dotted] (3,12)
    node[left=3pt,font=\Large] {$s$};
    \draw[dotted] (6,12)
    node[left=3pt,font=\Large] {$t_1$};
    \draw[dotted] (6,11)
    node[left=3pt,font=\Large] {$s_2$};
    \draw[dotted] (9,9)
    node[left=3pt,font=\Large] {$t_2$};
    \draw[dotted] (9,8)
    node[left=3pt,font=\Large] {$s_3$};
    \draw[dotted] (9,6)
    node[left=5pt,font=\Large] {$t$};
    
    \draw[black, line width=1.5pt]
   (3,12)--(3,13);
   \draw[black,line width=1.5pt] (3,13)--(4,13)--(4,12)--(5,12)--(5,13)--(9,13)--(9,12)--(6,12);
   \draw[black, line width=1.5pt] (6,11)--(9,11)--(9,10)--(6,10)--(6,9)--(9,9);
   \draw[black, line width=1.5pt] (9,8)--(7,8)--(7,6)--(8,6)--(8,7)--(9,7)--(9,6);
   \draw[black,line width=1.5pt](6,12)--(6,11);
   \draw[black,line width=1.5pt](9,9)--(9,8);

   \draw[black,line width = 1.5pt](9,6)--(9,2)--(3,2)--(3,12);
   
   \draw[orange,dotted,line width=1.5pt](9.5,5.5)--(9.5,1.5)--(8.5,1.5)--(8.5,5.5)--(9.5,5.5);
   \draw[orange,dotted,line width=1.5pt](8.5,1.5)--(8.5,2.5)--(2.5,2.5)--(2.5,1.5)--(8.5,1.5);
   \draw[orange,dotted,line width=1.5pt](2.5,2.5)--(3.5,2.5)--(3.5,11.5)--(2.5,11.5)--(2.5,2.5);

   % \draw[black,line width=1.5pt] (2,2)--(0,2)--(0,0)--(1,0)--(1,1)--(2,1)--(2,0)--(3,0)--(3,1)--(4,1)--(4,0)--(5,0)--(5,1)--(6,1)--(6,0)--(7,0)--(7,1)--(8,1)--(8,0)--(9,0)--(9,1)--(10,1)--(10,0)--(11,0)--(11,2)--(10,2)--(10,3)--(11,3)--(11,4)--(10,4)--(10,5)--(11,5)--(11,6)--(10,6)--(10,7)--(11,7)--(11,8)--(10,8)--(10,9)--(11,9)--(11,10)--(10,10)--(10,11)--(11,11)--(11,12)--(10,12)--(10,13)--(11,13)--(11,14)--(4,14)
   
   % \draw[black,line width=1.5pt](3,14)--(0,14)--(0,13)--(2,13)--(2,12)--(0,12)--(0,11)--(2,11)--(2,10);
   % \draw[black,line width=1.5pt](2,10)--(0,10)--(0,9)--(2,9)--(2,8)--(0,8)--(0,7)--(2,7)--(2,6)--(0,6)--(0,5)--(2,5)--(2,4)--(0,4)--(0,3)--(2,3)--(2,2);
   
% \draw[red,line width=2.5pt] (3,14)--(3,13);\draw[red,line width=2.5pt] (4,14)--(4,13);
   % \draw[red,line width=2pt](3,14)--(4,14);
   % \draw[red,line width=2pt](3,13)--(4,13);
\end{tikzpicture}
          }
        \caption{Example of HP in the Bounding Rectangle}
        \label{small_example_2_1}
    \end{subfigure}
    \caption{}
    \label{fig:rec_g1_g2}
\end{figure}

Since $m_3 = w_2 - w_0$, which is one column wider than the protruding part of the base L-shape as shown in Figure~\ref{grid graph with rectangular cover}, we can connect $t$ to the vertex at the lower-right corner of the rightmost vertex of the triangular hole, i.e., $(d+w_2, a+h_0+h_1+h_2)$, which covers $G_2$ in Figure~\ref{rec cover g1 g2}. Similarly, $G_3$ and $G_4$ can be covered in the same manner. 
It remains to connect the Hamiltonian path constructed within the inner region $G_1 \cup G_2 \cup G_3 \cup G_4$ to the outer portion of the grid.

\paragraph{Example}
Using the same grid graph example, as shown in \ref{small_example_2_1}: We complete the Hamiltonian Cycle for the minimal rectangular grid containing the three-stacked right triangular hole by connecting the starting vertex $s$ and ending vertex $t$ together via the thin U-shaped pathway.

\subsubsection{Outer Grids}
\begin{figure}[h!]
    \centering
  \begin{subfigure}[b]{0.23\textwidth}
  \centering
   \resizebox{\linewidth}{!}{%
    \begin{tikzpicture}
  % Background gray dots for all grid points
  \foreach \x in {0,...,11}
    \foreach \y in {0,...,14}
      \fill[gray] (\x,\y) circle(0.04);

  % ── Parity-coloured nodes ──────────────────────────────────────
  % Rule: black if (x+y) even, white if (x+y) odd

  % All nodes using parity colouring

    \foreach \x in {0,...,11} {
    \foreach \y in {0,1,2,12,13,14} {
      \pgfmathparse{int(mod(\x+\y,2))==0 ? 1 : 0}
      \ifnum\pgfmathresult=1
        \fill[black] (\x,\y) circle(4pt);
      \else
        \draw[fill=white] (\x,\y) circle(4pt);
      \fi
    }
  }
  
  \foreach \x in {0,1,2,3,9,10,11} {
    \foreach \y in {2,...,11} {
      \pgfmathparse{int(mod(\x+\y,2))==0 ? 1 : 0}
      \ifnum\pgfmathresult=1
        \fill[black] (\x,\y) circle(4pt);
      \else
        \draw[fill=white] (\x,\y) circle(4pt);
      \fi
    }
  }
   \foreach \x in {7,8} {
    \foreach \y in {6,7,8,9,10,11} {
      \pgfmathparse{int(mod(\x+\y,2))==0 ? 1 : 0}
      \ifnum\pgfmathresult=1
        \fill[black] (\x,\y) circle(4pt);
      \else
        \draw[fill=white] (\x,\y) circle(4pt);
      \fi
    }
  }
  \foreach \x in {6} {
    \foreach \y in {9,10,11} {
      \pgfmathparse{int(mod(\x+\y,2))==0 ? 1 : 0}
      \ifnum\pgfmathresult=1
        \fill[black] (\x,\y) circle(4pt);
      \else
        \draw[fill=white] (\x,\y) circle(4pt);
      \fi
    }
  }

   \draw[orange,dotted,line width=1.5pt](2.5,-0.5)--(2.5,14.5);
   \draw[orange,dotted,line width=1.5pt](9.5,-0.5)--(9.5,14.5);
   \draw[orange,dotted,line width=1.5pt](2.5,13.5)--(9.5,13.5);
   \draw[orange,dotted,line width=1.5pt](2.5,2.5)--(8.5,2.5);
   \draw[orange,dotted,line width=1.5pt](2.5,1.5)--(9.5,1.5);
   \draw[orange,dotted,line width=1.5pt](3.5,2.5)--(3.5,11.5);
   \draw[orange,dotted,line width=1.5pt](8.5,5.5)--(8.5,1.5);
   \draw[orange,dotted,line width=1.5pt](2.5,11.5)--(5.5,11.5)--(5.5,8.5)--(6.5,8.5)--(6.5,5.5)--(9.5,5.5);

   \node[font=\huge, text=orange] at (1,7.5) {$G_8$};
   \node[font=\huge, text=orange] at (3,7.5) {$G_4$};
   \node[font=\huge, text=orange] at (5.5,14) {$G_5$};
   \node[font=\huge, text=orange] at (7.5,10.5) {$G_1$};
   \node[font=\huge, text=orange] at (5.5,2) {$G_3$};
   \node[font=\huge, text=orange] at (6.5,0.5) {$G_7$};
   \node[font=\huge, text=orange] at (10.5,7.5) {$G_6$};
   \node[font=\huge, text=orange] at (9,3.5) {$G_2$};

    % \draw[blue, line width=1.5pt] (2.5,13.5)--(2.5,11.5)--(9.5,11.5)--(9.5,13.5)--(2.5,13.5);

    % \draw[violet, line width=1.5pt] (6.5,5.5)--(6.5,8.5)--(9.5,8.5)--(9.5,5.5)--(6.5,5.5);

    % \draw[green, line width=1.5pt] (5.5,11.5)--(5.5,8.5)--(9.5,8.5)--(9.5,11.5)--(5.5,11.5);

    % \draw[->, thin, blue, line width=1.5pt] (10.5,12.5) -- (9.5,12.5) node[near start, right=8pt] {$R_1$};

    % \draw[->, thin, green, line width=1.5pt] (10.5,9.5) -- (9.5,9.5) node[near start, right=8pt] {$R_2$};

    % \draw[->, thin, violet, line width=1.5pt] (10.5,7.5) -- (9.5,7.5) node[near start, right=8pt] {$R_3$};

\end{tikzpicture}
  }
  
    \caption{Subgraphs}
    \label{fig:outer}
  \end{subfigure}
    \hfill
    \begin{subfigure}[b]{0.23\textwidth}
        \centering
        \resizebox{\linewidth}{!}{
            \begin{tikzpicture}
  % Background gray dots for all grid points
  \foreach \x in {0,...,11}
    \foreach \y in {0,...,14}
      \fill[gray] (\x,\y) circle(0.04);

  % ── Parity-coloured nodes ──────────────────────────────────────
  % Rule: black if (x+y) even, white if (x+y) odd

  % All nodes using parity colouring

    \foreach \x in {0,...,11} {
    \foreach \y in {0,1,2,12,13,14} {
      \pgfmathparse{int(mod(\x+\y,2))==0 ? 1 : 0}
      \ifnum\pgfmathresult=1
        \fill[black] (\x,\y) circle(4pt);
      \else
        \draw[fill=white] (\x,\y) circle(4pt);
      \fi
    }
  }
  
  \foreach \x in {0,1,2,3,9,10,11} {
    \foreach \y in {2,...,11} {
      \pgfmathparse{int(mod(\x+\y,2))==0 ? 1 : 0}
      \ifnum\pgfmathresult=1
        \fill[black] (\x,\y) circle(4pt);
      \else
        \draw[fill=white] (\x,\y) circle(4pt);
      \fi
    }
  }
   \foreach \x in {7,8} {
    \foreach \y in {6,7,8,9,10,11} {
      \pgfmathparse{int(mod(\x+\y,2))==0 ? 1 : 0}
      \ifnum\pgfmathresult=1
        \fill[black] (\x,\y) circle(4pt);
      \else
        \draw[fill=white] (\x,\y) circle(4pt);
      \fi
    }
  }
  \foreach \x in {6} {
    \foreach \y in {9,10,11} {
      \pgfmathparse{int(mod(\x+\y,2))==0 ? 1 : 0}
      \ifnum\pgfmathresult=1
        \fill[black] (\x,\y) circle(4pt);
      \else
        \draw[fill=white] (\x,\y) circle(4pt);
      \fi
    }
  }
    \draw[orange, dotted, line width=1.5pt]
    (2.5,13.5)--(2.5,11.5)--(9.5,11.5)--(9.5,13.5)--(2.5,13.5);
    \draw[orange, dotted, line width=1.5pt]
    (6.5,5.5)--(6.5,8.5)--(9.5,8.5)--(9.5,5.5)--(6.5,5.5);
    \draw[orange, dotted, line width=1.5pt]
    (5.5,11.5)--(5.5,8.5)--(9.5,8.5)--(9.5,11.5)--(5.5,11.5);
    \draw[dotted] (3,12)
    node[left=3pt,font=\Large] {$s$};
    \draw[dotted] (6,12)
    node[left=3pt,font=\Large] {$t_1$};
    \draw[dotted] (6,11)
    node[left=3pt,font=\Large] {$s_2$};
    \draw[dotted] (9,9)
    node[left=3pt,font=\Large] {$t_2$};
    \draw[dotted] (9,8)
    node[left=3pt,font=\Large] {$s_3$};
    \draw[dotted] (9,6)
    node[left=5pt,font=\Large] {$t$};
    \draw[red, line width=2.5pt]
   (4,13)--(3,13);
    \draw[black, line width=1.5pt]
   (3,12)--(3,13);
   \draw[black,line width=1.5pt] (4,13)--(4,12)--(5,12)--(5,13)--(9,13)--(9,12)--(6,12);
   \draw[black, line width=1.5pt]
   (6,11)--(9,11)--(9,10)--(6,10)--(6,9)--(9,9);
   \draw[black, line width=1.5pt]
   (9,8)--(7,8)--(7,6)--(8,6)--(8,7)--(9,7)--(9,6);
   \draw[black,line width=1.5pt](6,12)--(6,11);
   \draw[black,line width=1.5pt](9,9)--(9,8);

   \draw[black,line width = 1.5pt](9,6)--(9,2)--(3,2)--(3,12);
   \draw[orange,dotted,line width=1.5pt](9.5,5.5)--(9.5,1.5)--(8.5,1.5)--(8.5,5.5)--(9.5,5.5);
   \draw[orange,dotted,line width=1.5pt](8.5,1.5)--(8.5,2.5)--(2.5,2.5)--(2.5,1.5)--(8.5,1.5);
   \draw[orange,dotted,line width=1.5pt](2.5,2.5)--(3.5,2.5)--(3.5,11.5)--(2.5,11.5)--(2.5,2.5);

   \draw[black,line width=1.5pt] (2,2)--(0,2)--(0,0)--(1,0)--(1,1)--(2,1)--(2,0)--(3,0)--(3,1)--(4,1)--(4,0)--(5,0)--(5,1)--(6,1)--(6,0)--(7,0)--(7,1)--(8,1)--(8,0)--(9,0)--(9,1)--(10,1)--(10,0)--(11,0)--(11,2)--(10,2)--(10,3)--(11,3)--(11,4)--(10,4)--(10,5)--(11,5)--(11,6)--(10,6)--(10,7)--(11,7)--(11,8)--(10,8)--(10,9)--(11,9)--(11,10)--(10,10)--(10,11)--(11,11)--(11,12)--(10,12)--(10,13)--(11,13)--(11,14)--(4,14);
   
   \draw[black,line width=1.5pt] (3,14)--(0,14)--(0,13)--(2,13)--(2,12)--(0,12)--(0,11)--(2,11)--(2,10);
   
   \draw[black,line width=1.5pt](2,10)--(0,10)--(0,9)--(2,9)--(2,8)--(0,8)--(0,7)--(2,7)--(2,6)--(0,6)--(0,5)--(2,5)--(2,4)--(0,4)--(0,3)--(2,3)--(2,2);
   
   \draw[red,line width=2.5pt] (3,14)--(4,14);
   % \draw[red,line width=2.5pt] (4,14)--(4,13);
   % \draw[red,line width=2pt](3,14)--(4,14);
   % \draw[red,line width=2pt](3,13)--(4,13);
\end{tikzpicture}
          }
        \caption{Two Parallel Edges}
        \label{small_example_3_1}
    \end{subfigure}
    \hfill
    \begin{subfigure}[b]{0.23\textwidth}
        \centering
        \resizebox{\linewidth}{!}{
            \begin{tikzpicture}
  % Background gray dots for all grid points
  \foreach \x in {0,...,11}
    \foreach \y in {0,...,14}
      \fill[gray] (\x,\y) circle(0.04);

  % ── Parity-coloured nodes ──────────────────────────────────────
  % Rule: black if (x+y) even, white if (x+y) odd

  % All nodes using parity colouring

    \foreach \x in {0,...,11} {
    \foreach \y in {0,1,2,12,13,14} {
      \pgfmathparse{int(mod(\x+\y,2))==0 ? 1 : 0}
      \ifnum\pgfmathresult=1
        \fill[black] (\x,\y) circle(4pt);
      \else
        \draw[fill=white] (\x,\y) circle(4pt);
      \fi
    }
  }
  
  \foreach \x in {0,1,2,3,9,10,11} {
    \foreach \y in {2,...,11} {
      \pgfmathparse{int(mod(\x+\y,2))==0 ? 1 : 0}
      \ifnum\pgfmathresult=1
        \fill[black] (\x,\y) circle(4pt);
      \else
        \draw[fill=white] (\x,\y) circle(4pt);
      \fi
    }
  }
   \foreach \x in {7,8} {
    \foreach \y in {6,7,8,9,10,11} {
      \pgfmathparse{int(mod(\x+\y,2))==0 ? 1 : 0}
      \ifnum\pgfmathresult=1
        \fill[black] (\x,\y) circle(4pt);
      \else
        \draw[fill=white] (\x,\y) circle(4pt);
      \fi
    }
  }
  \foreach \x in {6} {
    \foreach \y in {9,10,11} {
      \pgfmathparse{int(mod(\x+\y,2))==0 ? 1 : 0}
      \ifnum\pgfmathresult=1
        \fill[black] (\x,\y) circle(4pt);
      \else
        \draw[fill=white] (\x,\y) circle(4pt);
      \fi
    }
  }
    \draw[orange, dotted, line width=1.5pt]
    (2.5,13.5)--(2.5,11.5)--(9.5,11.5)--(9.5,13.5)--(2.5,13.5);
    \draw[orange, dotted, line width=1.5pt]
    (6.5,5.5)--(6.5,8.5)--(9.5,8.5)--(9.5,5.5)--(6.5,5.5);
    \draw[orange, dotted, line width=1.5pt]
    (5.5,11.5)--(5.5,8.5)--(9.5,8.5)--(9.5,11.5)--(5.5,11.5);
    \draw[dotted] (3,12)
    node[left=3pt,font=\Large] {$s$};
    \draw[dotted] (6,12)
    node[left=3pt,font=\Large] {$t_1$};
    \draw[dotted] (6,11)
    node[left=3pt,font=\Large] {$s_2$};
    \draw[dotted] (9,9)
    node[left=3pt,font=\Large] {$t_2$};
    \draw[dotted] (9,8)
    node[left=3pt,font=\Large] {$s_3$};
    \draw[dotted] (9,6)
    node[left=5pt,font=\Large] {$t$};
    % \draw[red, line width=1.5pt] (4,13)--(3,13);
    \draw[black, line width=1.5pt]
   (3,12)--(3,13);
   \draw[black,line width=1.5pt] (4,13)--(4,12)--(5,12)--(5,13)--(9,13)--(9,12)--(6,12);
   \draw[black, line width=1.5pt]
   (6,11)--(9,11)--(9,10)--(6,10)--(6,9)--(9,9);
   \draw[black, line width=1.5pt]
   (9,8)--(7,8)--(7,6)--(8,6)--(8,7)--(9,7)--(9,6);
   \draw[black,line width=1.5pt](6,12)--(6,11);
   \draw[black,line width=1.5pt](9,9)--(9,8);

   \draw[black,line width = 1.5pt](9,6)--(9,2)--(3,2)--(3,12);
   \draw[orange,dotted,line width=1.5pt](9.5,5.5)--(9.5,1.5)--(8.5,1.5)--(8.5,5.5)--(9.5,5.5);
   \draw[orange,dotted,line width=1.5pt](8.5,1.5)--(8.5,2.5)--(2.5,2.5)--(2.5,1.5)--(8.5,1.5);
   \draw[orange,dotted,line width=1.5pt](2.5,2.5)--(3.5,2.5)--(3.5,11.5)--(2.5,11.5)--(2.5,2.5);

   \draw[black,line width=1.5pt] (2,2)--(0,2)--(0,0)--(1,0)--(1,1)--(2,1)--(2,0)--(3,0)--(3,1)--(4,1)--(4,0)--(5,0)--(5,1)--(6,1)--(6,0)--(7,0)--(7,1)--(8,1)--(8,0)--(9,0)--(9,1)--(10,1)--(10,0)--(11,0)--(11,2)--(10,2)--(10,3)--(11,3)--(11,4)--(10,4)--(10,5)--(11,5)--(11,6)--(10,6)--(10,7)--(11,7)--(11,8)--(10,8)--(10,9)--(11,9)--(11,10)--(10,10)--(10,11)--(11,11)--(11,12)--(10,12)--(10,13)--(11,13)--(11,14)--(4,14);
   
   \draw[black,line width=1.5pt](3,14)--(0,14)--(0,13)--(2,13)--(2,12)--(0,12)--(0,11)--(2,11)--(2,10);
   \draw[black,line width=1.5pt](2,10)--(0,10)--(0,9)--(2,9)--(2,8)--(0,8)--(0,7)--(2,7)--(2,6)--(0,6)--(0,5)--(2,5)--(2,4)--(0,4)--(0,3)--(2,3)--(2,2);
   
   % \draw[red,line width=2.5pt] (3,14)--(4,14);
   % \draw[red,line width=2.5pt] (4,14)--(4,13);
   % \draw[red,line width=2pt](3,14)--(4,14);
   % \draw[red,line width=2pt](3,13)--(4,13);
\end{tikzpicture}
          }
        \caption{Disconnect}
        \label{small_example_3_2}
    \end{subfigure}
    \hfill
    \begin{subfigure}[b]{0.23\textwidth}
        \centering
        \resizebox{\linewidth}{!}{
            \begin{tikzpicture}
  % Background gray dots for all grid points
  \foreach \x in {0,...,11}
    \foreach \y in {0,...,14}
      \fill[gray] (\x,\y) circle(0.04);

  % ── Parity-coloured nodes ──────────────────────────────────────
  % Rule: black if (x+y) even, white if (x+y) odd

  % All nodes using parity colouring

    \foreach \x in {0,...,11} {
    \foreach \y in {0,1,2,12,13,14} {
      \pgfmathparse{int(mod(\x+\y,2))==0 ? 1 : 0}
      \ifnum\pgfmathresult=1
        \fill[black] (\x,\y) circle(4pt);
      \else
        \draw[fill=white] (\x,\y) circle(4pt);
      \fi
    }
  }
  
  \foreach \x in {0,1,2,3,9,10,11} {
    \foreach \y in {2,...,11} {
      \pgfmathparse{int(mod(\x+\y,2))==0 ? 1 : 0}
      \ifnum\pgfmathresult=1
        \fill[black] (\x,\y) circle(4pt);
      \else
        \draw[fill=white] (\x,\y) circle(4pt);
      \fi
    }
  }
   \foreach \x in {7,8} {
    \foreach \y in {6,7,8,9,10,11} {
      \pgfmathparse{int(mod(\x+\y,2))==0 ? 1 : 0}
      \ifnum\pgfmathresult=1
        \fill[black] (\x,\y) circle(4pt);
      \else
        \draw[fill=white] (\x,\y) circle(4pt);
      \fi
    }
  }
  \foreach \x in {6} {
    \foreach \y in {9,10,11} {
      \pgfmathparse{int(mod(\x+\y,2))==0 ? 1 : 0}
      \ifnum\pgfmathresult=1
        \fill[black] (\x,\y) circle(4pt);
      \else
        \draw[fill=white] (\x,\y) circle(4pt);
      \fi
    }
  }
    \draw[orange, dotted, line width=1.5pt]
    (2.5,13.5)--(2.5,11.5)--(9.5,11.5)--(9.5,13.5)--(2.5,13.5);
    \draw[orange, dotted, line width=1.5pt]
    (6.5,5.5)--(6.5,8.5)--(9.5,8.5)--(9.5,5.5)--(6.5,5.5);
    \draw[orange, dotted, line width=1.5pt]
    (5.5,11.5)--(5.5,8.5)--(9.5,8.5)--(9.5,11.5)--(5.5,11.5);
    \draw[dotted] (3,12)
    node[left=3pt,font=\Large] {$s$};
    \draw[dotted] (6,12)
    node[left=3pt,font=\Large] {$t_1$};
    \draw[dotted] (6,11)
    node[left=3pt,font=\Large] {$s_2$};
    \draw[dotted] (9,9)
    node[left=3pt,font=\Large] {$t_2$};
    \draw[dotted] (9,8)
    node[left=3pt,font=\Large] {$s_3$};
    \draw[dotted] (9,6)
    node[left=5pt,font=\Large] {$t$};
    
    \draw[black, line width=1.5pt]
   (3,12)--(3,13);
   \draw[black,line width=1.5pt] (4,13)--(4,12)--(5,12)--(5,13)--(9,13)--(9,12)--(6,12);
   \draw[black, line width=1.5pt]
   (6,11)--(9,11)--(9,10)--(6,10)--(6,9)--(9,9);
   \draw[black, line width=1.5pt]
   (9,8)--(7,8)--(7,6)--(8,6)--(8,7)--(9,7)--(9,6);
   \draw[black,line width=1.5pt](6,12)--(6,11);
   \draw[black,line width=1.5pt](9,9)--(9,8);

   \draw[black,line width = 1.5pt](9,6)--(9,2)--(3,2)--(3,12);
   \draw[orange,dotted,line width=1.5pt](9.5,5.5)--(9.5,1.5)--(8.5,1.5)--(8.5,5.5)--(9.5,5.5);
   \draw[orange,dotted,line width=1.5pt](8.5,1.5)--(8.5,2.5)--(2.5,2.5)--(2.5,1.5)--(8.5,1.5);
   \draw[orange,dotted,line width=1.5pt](2.5,2.5)--(3.5,2.5)--(3.5,11.5)--(2.5,11.5)--(2.5,2.5);

   \draw[black,line width=1.5pt] (2,2)--(0,2)--(0,0)--(1,0)--(1,1)--(2,1)--(2,0)--(3,0)--(3,1)--(4,1)--(4,0)--(5,0)--(5,1)--(6,1)--(6,0)--(7,0)--(7,1)--(8,1)--(8,0)--(9,0)--(9,1)--(10,1)--(10,0)--(11,0)--(11,2)--(10,2)--(10,3)--(11,3)--(11,4)--(10,4)--(10,5)--(11,5)--(11,6)--(10,6)--(10,7)--(11,7)--(11,8)--(10,8)--(10,9)--(11,9)--(11,10)--(10,10)--(10,11)--(11,11)--(11,12)--(10,12)--(10,13)--(11,13)--(11,14)--(4,14);
   
   \draw[black,line width=1.5pt](3,14)--(0,14)--(0,13)--(2,13)--(2,12)--(0,12)--(0,11)--(2,11)--(2,10);
   \draw[black,line width=1.5pt](2,10)--(0,10)--(0,9)--(2,9)--(2,8)--(0,8)--(0,7)--(2,7)--(2,6)--(0,6)--(0,5)--(2,5)--(2,4)--(0,4)--(0,3)--(2,3)--(2,2);
   
\draw[red,line width=2.5pt] (3,14)--(3,13);\draw[red,line width=2.5pt] (4,14)--(4,13);
   % \draw[red,line width=2pt](3,14)--(4,14);
   % \draw[red,line width=2pt](3,13)--(4,13);
\end{tikzpicture}
          }
        \caption{Reconnect Vertically}
        \label{small_example_3_3}
    \end{subfigure}
    \caption{An Example: Connect to Outer Grids}
    \label{small_example_3}
\end{figure}
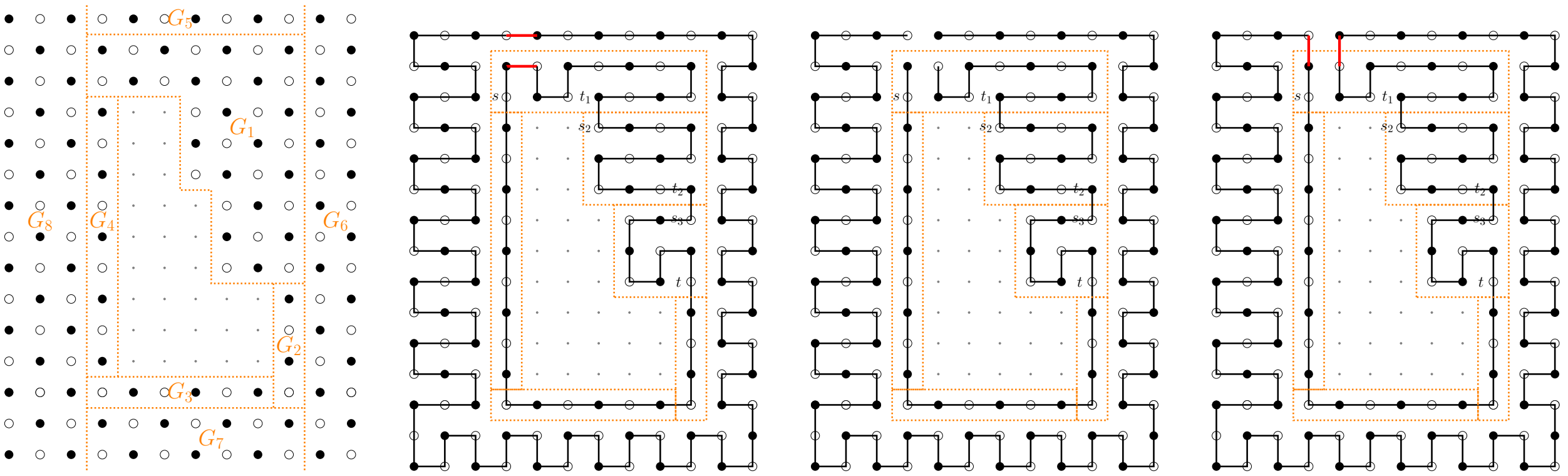

At this point, we have constructed a Hamiltonian cycle within the inner region 
$G_1 \cup G_2 \cup G_3 \cup G_4$, which forms a minimal rectangular grid containing 
the three-stacked right triangular hole. To obtain a Hamiltonian cycle for the 
entire grid graph, it remains to construct a Hamiltonian cycle for the outer region 
$G_5 \cup G_6 \cup G_7 \cup G_8$, as illustrated in Figure \ref{fig:outer}, and then 
connect the two cycles together.

The construction of Hamiltonian cycles in the outer grids follows from existing 
results on rectangular grid graphs with rectangular holes. In particular, prior work has shown 
that rectangular grid graphs with rectangular holes admit linear-time algorithms 
for constructing Hamiltonian paths and cycles under appropriate conditions 
\cite{rec_hole_even, rec_hole}. These results allow us to determine the Hamiltonian 
cycle structure of the outer region, since it can be viewed as a rectangular grid 
with a structured missing subgraph.

Furthermore, recent work on grid graphs with L-shaped holes adopts a divide-and-conquer 
approach: the graph is partitioned into several rectangular subgraphs, Hamiltonian 
cycles are constructed locally, and then combined using parallel edges 
\cite{Lhole_recent}. 
In particular, two Hamiltonian cycles can be merged into a 
single cycle by removing a pair of parallel edges and reconnecting them across the 
boundary, resulting in a larger Hamiltonian cycle. This merge 
operation provides a systematic way to connect disjoint cycles.

Following this approach, we select an edge from the inner cycle 
$G_1 \cup G_2 \cup G_3 \cup G_4$ and a corresponding parallel edge from the outer 
cycle $G_5 \cup G_6 \cup G_7 \cup G_8$. These edges must be adjacent: vertically 
aligned if they are horizontal edges, and horizontally aligned if they are vertical 
edges. By removing these two edges and reconnecting their endpoints across the 
boundary, the black edges shown in Figure \ref{fig:outer}, we merge the inner and outer cycles into a single Hamiltonian cycle 
covering the entire graph.

This completes the construction for the small hole case.

\paragraph{Example}
Figure \ref{small_example_3} shows the example of how to connect the inner Hamiltonian Cycle to the outer Grids. Given two Hamiltonian Cycles, one for the inner subgraph, one for the outer subgraph, we can find two parallel edges that are horizontally aligned, denoted red in Figure \ref{small_example_3_1}. We now remove these two edges and reconnect their endpoints vertically, as shown in Figure \ref{small_example_3_2} and \ref{small_example_3_3} respectively.

\subsection{The Large Hole Case}\label{sec:large-hole}
We organize the large-hole analysis by the boundary conditions $a<2$, $b=0$, $c_2=0$, or $d=0$. 

These are exactly the situations in which at least one of the four complementary boundary regions of the hole becomes too thin or disappears entirely. As a result, the usual Hamiltonian cycle/path construction is no longer fully flexible: at least one side of the graph cannot be used as an ordinary transit corridor.

In the regular cases, the construction typically allows the cycle to pass through one side of the graph, continue to another subgraph, and later reconnect through a different side. In contrast, in the large-hole cases, one side is constrained. Therefore, the construction must be modified so that the cycle does not rely on repeated traversal across that constrained side. Instead, one must first determine which side is limited, and then arrange the cycle so that the corresponding region is covered in a prescribed manner and merged with the rest of the graph through the remaining unconstrained sides.

\subsubsection{Sub Case I: $c_2=0$}
When $c_2=0$, the right complementary side is constrained. In particular, the right side can no longer serve as a full connecting corridor between the upper and lower parts of the graph. Thus, the usual construction---which passes through the upper-right part, continues along the right side, and then connects directly to the lower part---is no longer available.

Instead, the construction must avoid relying on the right side as a transit region. The upper-right (shaded area in Figure \ref{fig:general-setting}) remaining part must be covered and then merged back into the left/main portion of the graph, after which the cycle can continue toward the lower part. In other words, the traversal must return from the right toward the left, rather than using the right boundary as a direct channel from top to bottom.

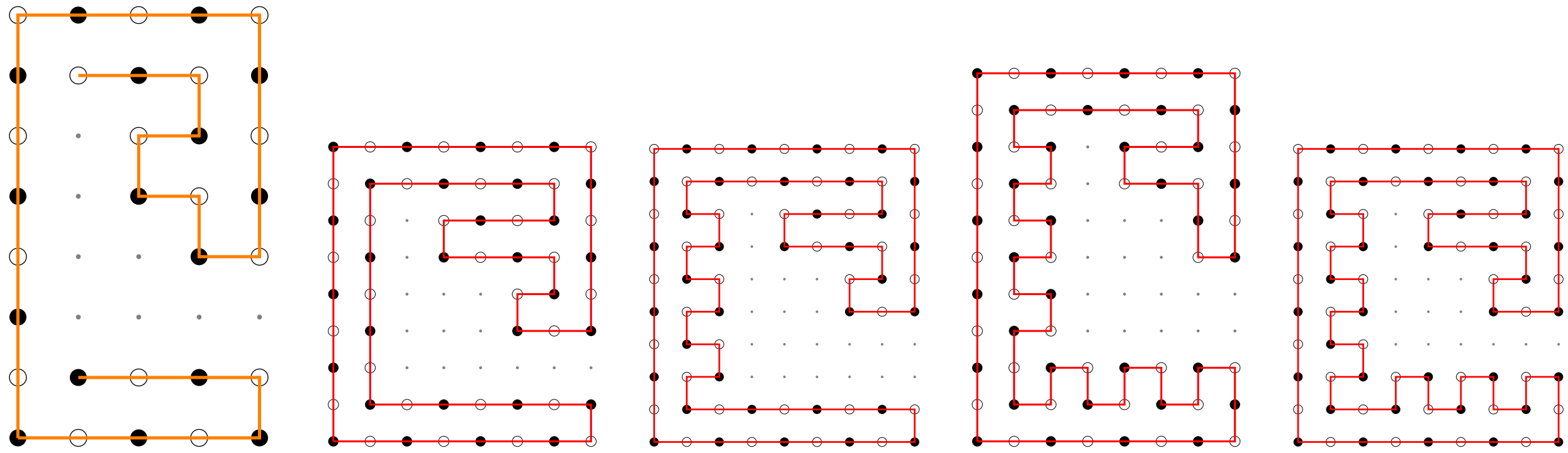
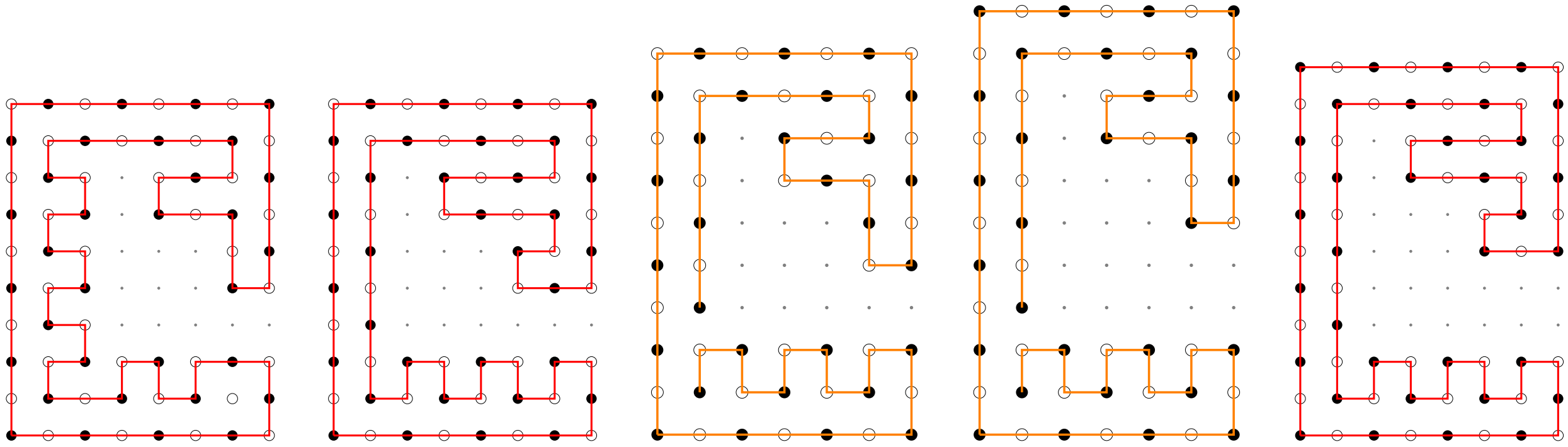
\begin{figure}[h!]
    \centering
    \begin{subfigure}[b]{0.18\textwidth}
        \centering
        \resizebox{\linewidth}{!}{
            \begin{tikzpicture}
  % Background gray dots for all grid points
  \foreach \x in {0,...,4}
    \foreach \y in {0,...,7}
      \fill[gray] (\x,\y) circle(0.04);

  % ── Parity-coloured nodes ──────────────────────────────────────
  % Rule: black if (x+y) even, white if (x+y) odd

  % All nodes using parity colouring

    \foreach \x in {0,...,4} {
    \foreach \y in {0,1,6,7} {
      \pgfmathparse{int(mod(\x+\y,2))==0 ? 1 : 0}
      \ifnum\pgfmathresult=1
        \fill[black] (\x,\y) circle(4pt);
      \else
        \draw[fill=white] (\x,\y) circle(4pt);
      \fi
    }
  }

  \foreach \x in {0,2,3,4} {
    \foreach \y in {4,5} {
      \pgfmathparse{int(mod(\x+\y,2))==0 ? 1 : 0}
      \ifnum\pgfmathresult=1
        \fill[black] (\x,\y) circle(4pt);
      \else
        \draw[fill=white] (\x,\y) circle(4pt);
      \fi
    }
  }

  \foreach \x in {0,3,4} {
    \foreach \y in {3} {
      \pgfmathparse{int(mod(\x+\y,2))==0 ? 1 : 0}
      \ifnum\pgfmathresult=1
        \fill[black] (\x,\y) circle(4pt);
      \else
        \draw[fill=white] (\x,\y) circle(4pt);
      \fi
    }
  }

  \foreach \x in {0} {
    \foreach \y in {2} {
      \pgfmathparse{int(mod(\x+\y,2))==0 ? 1 : 0}
      \ifnum\pgfmathresult=1
        \fill[black] (\x,\y) circle(4pt);
      \else
        \draw[fill=white] (\x,\y) circle(4pt);
      \fi
    }
  }

   \draw[orange, line width=1.5pt] (0,0)--(4,0)--(4,1)--(1,1);
   \draw[orange, line width=1.5pt] (0,0)--(0,6)--(0,7)--(4,7)--(4,3)--(3,3)--(3,4)--(2,4)--(2,5)--(3,5)--(3,6)--(1,6);

\end{tikzpicture}
          }
        \caption{}
        \label{largehole_1_1}
    \end{subfigure}
    \hfill
    \begin{subfigure}[b]{0.18\textwidth}
        \centering
        \resizebox{\linewidth}{!}{
            \begin{tikzpicture}
  % Background gray dots for all grid points
  \foreach \x in {0,...,7}
    \foreach \y in {0,...,8}
      \fill[gray] (\x,\y) circle(0.04);

  % ── Parity-coloured nodes ──────────────────────────────────────
  % Rule: black if (x+y) even, white if (x+y) odd

  % All nodes using parity colouring

    \foreach \x in {0,...,7} {
    \foreach \y in {0,1,7,8} {
      \pgfmathparse{int(mod(\x+\y,2))==0 ? 1 : 0}
      \ifnum\pgfmathresult=1
        \fill[black] (\x,\y) circle(4pt);
      \else
        \draw[fill=white] (\x,\y) circle(4pt);
      \fi
    }
  }

  \foreach \x in {0,1,3,4,5,6,7} {
    \foreach \y in {5,6} {
      \pgfmathparse{int(mod(\x+\y,2))==0 ? 1 : 0}
      \ifnum\pgfmathresult=1
        \fill[black] (\x,\y) circle(4pt);
      \else
        \draw[fill=white] (\x,\y) circle(4pt);
      \fi
    }
  }

  \foreach \x in {0,1,5,6,7} {
    \foreach \y in {3,4} {
      \pgfmathparse{int(mod(\x+\y,2))==0 ? 1 : 0}
      \ifnum\pgfmathresult=1
        \fill[black] (\x,\y) circle(4pt);
      \else
        \draw[fill=white] (\x,\y) circle(4pt);
      \fi
    }
  }

  \foreach \x in {0,1} {
    \foreach \y in {2} {
      \pgfmathparse{int(mod(\x+\y,2))==0 ? 1 : 0}
      \ifnum\pgfmathresult=1
        \fill[black] (\x,\y) circle(4pt);
      \else
        \draw[fill=white] (\x,\y) circle(4pt);
      \fi
    }
  }

   \draw[red, line width=1.5pt] (0,0)--(7,0)--(7,1)--(1,1)--(1,7)--(6,7)--(6,6)--(3,6)--(3,5)--(6,5)--(6,4)--(5,4)--(5,3)--(7,3)--(7,8)--(0,8)--(0,0);
   
\end{tikzpicture}
          }
        \caption{}
        \label{largehole_1_2}
    \end{subfigure}
    \hfill
    \begin{subfigure}[b]{0.18\textwidth}
        \centering
        \resizebox{\linewidth}{!}{
            \begin{tikzpicture}
  % Background gray dots for all grid points
  \foreach \x in {0,...,8}
    \foreach \y in {0,...,9}
      \fill[gray] (\x,\y) circle(0.04);

  % ── Parity-coloured nodes ──────────────────────────────────────
  % Rule: black if (x+y) even, white if (x+y) odd

  % All nodes using parity colouring

    \foreach \x in {0,...,8} {
    \foreach \y in {0,1,8,9} {
      \pgfmathparse{int(mod(\x+\y,2))==0 ? 1 : 0}
      \ifnum\pgfmathresult=1
        \fill[black] (\x,\y) circle(4pt);
      \else
        \draw[fill=white] (\x,\y) circle(4pt);
      \fi
    }
  }

  \foreach \x in {0,1,2,4,5,6,7,8} {
    \foreach \y in {6,7} {
      \pgfmathparse{int(mod(\x+\y,2))==0 ? 1 : 0}
      \ifnum\pgfmathresult=1
        \fill[black] (\x,\y) circle(4pt);
      \else
        \draw[fill=white] (\x,\y) circle(4pt);
      \fi
    }
  }

  \foreach \x in {0,1,2,6,7,8} {
    \foreach \y in {4,5} {
      \pgfmathparse{int(mod(\x+\y,2))==0 ? 1 : 0}
      \ifnum\pgfmathresult=1
        \fill[black] (\x,\y) circle(4pt);
      \else
        \draw[fill=white] (\x,\y) circle(4pt);
      \fi
    }
  }

  \foreach \x in {0,1,2} {
    \foreach \y in {2,3} {
      \pgfmathparse{int(mod(\x+\y,2))==0 ? 1 : 0}
      \ifnum\pgfmathresult=1
        \fill[black] (\x,\y) circle(4pt);
      \else
        \draw[fill=white] (\x,\y) circle(4pt);
      \fi
    }
  }

   \draw[red, line width=1.5pt] (0,0)--(8,0)--(8,1)--(1,1)--(1,2)--(2,2)--(2,3)--(1,3)--(1,4)--(2,4)--(2,5)--(1,5)--(1,6)--(2,6)--(2,7)--(1,7)--(1,8)--(7,8)--(7,7)--(4,7)--(4,6)--(7,6)--(7,5)--(6,5)--(6,4)--(8,4)--(8,9)--(0,9)--(0,0);
   
\end{tikzpicture}
          }
        \caption{}
        \label{largehole_1_3}
    \end{subfigure}
    \hfill
    \begin{subfigure}[b]{0.18\textwidth}
        \centering
        \resizebox{\linewidth}{!}{
            \begin{tikzpicture}
  % Background gray dots for all grid points
  \foreach \x in {0,...,7}
    \foreach \y in {0,...,10}
      \fill[gray] (\x,\y) circle(0.04);

  % ── Parity-coloured nodes ──────────────────────────────────────
  % Rule: black if (x+y) even, white if (x+y) odd

  % All nodes using parity colouring

    \foreach \x in {0,1,2} {
    \foreach \y in {0,...,10} {
      \pgfmathparse{int(mod(\x+\y,2))==0 ? 1 : 0}
      \ifnum\pgfmathresult=1
        \fill[black] (\x,\y) circle(4pt);
      \else
        \draw[fill=white] (\x,\y) circle(4pt);
      \fi
    }
  }

  \foreach \x in {3,4,5,6,7} {
    \foreach \y in {0,1,2,9,10} {
      \pgfmathparse{int(mod(\x+\y,2))==0 ? 1 : 0}
      \ifnum\pgfmathresult=1
        \fill[black] (\x,\y) circle(4pt);
      \else
        \draw[fill=white] (\x,\y) circle(4pt);
      \fi
    }
  }

  \foreach \x in {4,5,6,7} {
    \foreach \y in {7,8} {
      \pgfmathparse{int(mod(\x+\y,2))==0 ? 1 : 0}
      \ifnum\pgfmathresult=1
        \fill[black] (\x,\y) circle(4pt);
      \else
        \draw[fill=white] (\x,\y) circle(4pt);
      \fi
    }
  }

  \foreach \x in {6,7} {
    \foreach \y in {5,6} {
      \pgfmathparse{int(mod(\x+\y,2))==0 ? 1 : 0}
      \ifnum\pgfmathresult=1
        \fill[black] (\x,\y) circle(4pt);
      \else
        \draw[fill=white] (\x,\y) circle(4pt);
      \fi
    }
  }

   \draw[red, line width=1.5pt] (0,0)--(7,0)--(7,2)--(6,2)--(6,1)--(5,1)--(5,2)--(4,2)--(4,1)--(3,1)--(3,2)--(2,2)--(2,1)--(1,1)--(1,3)--(2,3)--(2,4)--(1,4)--(1,5)--(2,5)--(2,6)--(1,6)--(1,7)--(2,7)--(2,8)--(1,8)--(1,9)--(6,9)--(6,8)--(4,8)--(4,7)--(6,7)--(6,5)--(7,5)--(7,10)--(0,10)--(0,0);
   
\end{tikzpicture}
          }
        \caption{}
        \label{largehole_1_4}
    \end{subfigure}
    \hfill
    \begin{subfigure}[b]{0.18\textwidth}
        \centering
        \resizebox{\linewidth}{!}{
            \begin{tikzpicture}
  % Background gray dots for all grid points
  \foreach \x in {0,...,8}
    \foreach \y in {0,...,9}
      \fill[gray] (\x,\y) circle(0.04);

  % ── Parity-coloured nodes ──────────────────────────────────────
  % Rule: black if (x+y) even, white if (x+y) odd

  % All nodes using parity colouring

    \foreach \x in {0,...,8} {
    \foreach \y in {0,1,2,8,9} {
      \pgfmathparse{int(mod(\x+\y,2))==0 ? 1 : 0}
      \ifnum\pgfmathresult=1
        \fill[black] (\x,\y) circle(4pt);
      \else
        \draw[fill=white] (\x,\y) circle(4pt);
      \fi
    }
  }

  \foreach \x in {0,1,2,4,5,6,7,8} {
    \foreach \y in {6,7} {
      \pgfmathparse{int(mod(\x+\y,2))==0 ? 1 : 0}
      \ifnum\pgfmathresult=1
        \fill[black] (\x,\y) circle(4pt);
      \else
        \draw[fill=white] (\x,\y) circle(4pt);
      \fi
    }
  }

  \foreach \x in {0,1,2,6,7,8} {
    \foreach \y in {4,5} {
      \pgfmathparse{int(mod(\x+\y,2))==0 ? 1 : 0}
      \ifnum\pgfmathresult=1
        \fill[black] (\x,\y) circle(4pt);
      \else
        \draw[fill=white] (\x,\y) circle(4pt);
      \fi
    }
  }

  \foreach \x in {0,1,2} {
    \foreach \y in {3} {
      \pgfmathparse{int(mod(\x+\y,2))==0 ? 1 : 0}
      \ifnum\pgfmathresult=1
        \fill[black] (\x,\y) circle(4pt);
      \else
        \draw[fill=white] (\x,\y) circle(4pt);
      \fi
    }
  }

   \draw[red, line width=1.5pt] (0,0)--(8,0)--(8,2)--(7,2)--(7,1)--(6,1)--(6,2)--(5,2)--(5,1)--(4,1)--(4,2)--(3,2)--(3,1)--(1,1)--(1,2)--(2,2)--(2,3)--(1,3)--(1,4)--(2,4)--(2,5)--(1,5)--(1,6)--(2,6)--(2,7)--(1,7)--(1,8)--(7,8)--(7,7)--(4,7)--(4,6)--(7,6)--(7,5)--(6,5)--(6,4)--(8,4)--(8,9)--(0,9)--(0,0);
   
\end{tikzpicture}
          }
        \caption{}
        \label{largehole_1_5}
    \end{subfigure}
    \caption{Construct Hamiltonian Cycle for Sub Case I of Large Hole Case}
    \label{subcase_1_1}
\end{figure}

Under this condition, we identified the following sub-cases, which we can either prove that a Hamiltonian path doesn't exist or construct a Hamiltonian path. We divide the analysis into three main subcases, one extreme case and two regular cases based on the parity of other parameters, i.e. $a,b,d, h_0, h_1,m,$ and $n$. As shown below and in Figures~\ref{subcase_1_1} and~\ref{subcase_1_2}, when $a, h_0, h_1$ are even, the Hamiltonian path construction will be slightly different based on the parity of $b, d$ and $n-b$. The orange path indicates the nonexistence of a Hamiltonian Cycle under that specific subcase. A similarly pattern holds when $a, h_0$ or $h_1$ is odd, but we will not list it here due to the page limit.

\begin{table}[htbp]
\centering
\small
\begin{tabular}{@{}p{0.56\linewidth}p{0.36\linewidth}@{}}
\toprule
Condition (throughout, $c_2=0$) & Outcome or figure \\
\midrule
$a=1, b=1$ or $d=1$ & No Hamiltonian cycle (Figure~\ref{largehole_1_1}) \\
\midrule
\multicolumn{2}{@{}l}{The following rows assume $a,h_0,h_1$ are even.}\\
$b,d$ even & Figure~\ref{largehole_1_2} \\
$d$ odd, $b$ even, $n-b$ even & Figure~\ref{largehole_1_3} \\
$d$ odd, $b$ even, $n-b$ odd & No Hamiltonian cycle \\
$d,b$ odd, $n-b$ even, $m$ odd & Figure~\ref{largehole_1_4} \\
$d,b$ odd, $n-b$ even, $m$ even & No Hamiltonian cycle \\
$d,b$ odd, $n-b$ odd, $m$ odd & Figure~\ref{largehole_1_5} \\
$d,b$ odd, $n-b$ odd, $m$ even & No Hamiltonian cycle (Figure~\ref{largehole_1_6}) \\
$b$ odd, $d$ even, $n-b$ even, $m$ odd & No Hamiltonian cycle (Figure~\ref{largehole_1_9}) \\
$b$ odd, $d$ even, $n-b$ even, $m$ even & Figure~\ref{largehole_1_10} \\
$b$ odd, $d$ even, $n-b$ odd, $m$ odd & No Hamiltonian cycle (Figure~\ref{largehole_1_8}) \\
$b$ odd, $d$ even, $n-b$ odd, $m$ even & Figure~\ref{largehole_1_7} \\
\midrule
$a,h_0$, or $h_1$ odd & Detailed cases not included here \\
\bottomrule
\end{tabular}
\caption{Parity cases for $c_2=0$.}
\end{table}

\begin{figure}[h!]
    \centering

    \begin{subfigure}[b]{0.18\textwidth}
        \centering
        \resizebox{\linewidth}{!}{
            \begin{tikzpicture}
  % Background gray dots for all grid points
  \foreach \x in {0,...,7}
    \foreach \y in {0,...,9}
      \fill[gray] (\x,\y) circle(0.04);

  % ── Parity-coloured nodes ──────────────────────────────────────
  % Rule: black if (x+y) even, white if (x+y) odd

  % All nodes using parity colouring

    \foreach \x in {0,...,7} {
    \foreach \y in {0,1,2,8,9} {
      \pgfmathparse{int(mod(\x+\y,2))==0 ? 1 : 0}
      \ifnum\pgfmathresult=1
        \fill[black] (\x,\y) circle(4pt);
      \else
        \draw[fill=white] (\x,\y) circle(4pt);
      \fi
    }
  }

  \foreach \x in {0,1,2,4,5,6,7} {
    \foreach \y in {6,7} {
      \pgfmathparse{int(mod(\x+\y,2))==0 ? 1 : 0}
      \ifnum\pgfmathresult=1
        \fill[black] (\x,\y) circle(4pt);
      \else
        \draw[fill=white] (\x,\y) circle(4pt);
      \fi
    }
  }

  \foreach \x in {0,1,2,6,7} {
    \foreach \y in {4,5} {
      \pgfmathparse{int(mod(\x+\y,2))==0 ? 1 : 0}
      \ifnum\pgfmathresult=1
        \fill[black] (\x,\y) circle(4pt);
      \else
        \draw[fill=white] (\x,\y) circle(4pt);
      \fi
    }
  }

  \foreach \x in {0,1,2} {
    \foreach \y in {3} {
      \pgfmathparse{int(mod(\x+\y,2))==0 ? 1 : 0}
      \ifnum\pgfmathresult=1
        \fill[black] (\x,\y) circle(4pt);
      \else
        \draw[fill=white] (\x,\y) circle(4pt);
      \fi
    }
  }

   \draw[red, line width=1.5pt] (0,0)--(7,0)--(7,2)--(5,2)--(5,1)--(4,1)--(4,2)--(3,2)--(3,1)--(1,1)--(1,2)--(2,2)--(2,3)--(1,3)--(1,4)--(2,4)--(2,5)--(1,5)--(1,6)--(2,6)--(2,7)--(1,7)--(1,8)--(6,8)--(6,7)--(4,7)--(4,6)--(6,6)--(6,4)--(7,4)--(7,9)--(0,9)--(0,0);
   
\end{tikzpicture}
          }
        \caption{}
        \label{largehole_1_6}
    \end{subfigure}
    \hfill
    \begin{subfigure}[b]{0.18\textwidth}
        \centering
        \resizebox{\linewidth}{!}{
            \begin{tikzpicture}
  % Background gray dots for all grid points
  \foreach \x in {0,...,7}
    \foreach \y in {0,...,9}
      \fill[gray] (\x,\y) circle(0.04);

  % ── Parity-coloured nodes ──────────────────────────────────────
  % Rule: black if (x+y) even, white if (x+y) odd

  % All nodes using parity colouring

    \foreach \x in {0,...,7} {
    \foreach \y in {0,1,2,8,9} {
      \pgfmathparse{int(mod(\x+\y,2))==0 ? 1 : 0}
      \ifnum\pgfmathresult=1
        \fill[black] (\x,\y) circle(4pt);
      \else
        \draw[fill=white] (\x,\y) circle(4pt);
      \fi
    }
  }

  \foreach \x in {0,1,3,4,5,6,7} {
    \foreach \y in {6,7} {
      \pgfmathparse{int(mod(\x+\y,2))==0 ? 1 : 0}
      \ifnum\pgfmathresult=1
        \fill[black] (\x,\y) circle(4pt);
      \else
        \draw[fill=white] (\x,\y) circle(4pt);
      \fi
    }
  }

  \foreach \x in {0,1,5,6,7} {
    \foreach \y in {4,5} {
      \pgfmathparse{int(mod(\x+\y,2))==0 ? 1 : 0}
      \ifnum\pgfmathresult=1
        \fill[black] (\x,\y) circle(4pt);
      \else
        \draw[fill=white] (\x,\y) circle(4pt);
      \fi
    }
  }

  \foreach \x in {0,1} {
    \foreach \y in {3} {
      \pgfmathparse{int(mod(\x+\y,2))==0 ? 1 : 0}
      \ifnum\pgfmathresult=1
        \fill[black] (\x,\y) circle(4pt);
      \else
        \draw[fill=white] (\x,\y) circle(4pt);
      \fi
    }
  }

   \draw[red, line width=1.5pt] (0,0)--(7,0)--(7,2)--(6,2)--(6,1)--(5,1)--(5,2)--(4,2)--(4,1)--(3,1)--(3,2)--(2,2)--(2,1)--(1,1)--(1,8)--(6,8)--(6,7)--(3,7)--(3,6)--(6,6)--(6,5)--(5,5)--(5,4)--(7,4)--(7,9)--(0,9)--(0,0);
   
\end{tikzpicture}
          }
        \caption{}
        \label{largehole_1_7}
    \end{subfigure}
    \hfill
    \begin{subfigure}[b]{0.18\textwidth}
        \centering
        \resizebox{\linewidth}{!}{
            \begin{tikzpicture}
  % Background gray dots for all grid points
  \foreach \x in {0,...,6}
    \foreach \y in {0,...,9}
      \fill[gray] (\x,\y) circle(0.04);

  % ── Parity-colouorange nodes ──────────────────────────────────────
  % Rule: black if (x+y) even, white if (x+y) odd

  % All nodes using parity colouring

    \foreach \x in {0,...,6} {
    \foreach \y in {0,1,2,8,9} {
      \pgfmathparse{int(mod(\x+\y,2))==0 ? 1 : 0}
      \ifnum\pgfmathresult=1
        \fill[black] (\x,\y) circle(4pt);
      \else
        \draw[fill=white] (\x,\y) circle(4pt);
      \fi
    }
  }

  \foreach \x in {0,1,3,4,5,6} {
    \foreach \y in {6,7} {
      \pgfmathparse{int(mod(\x+\y,2))==0 ? 1 : 0}
      \ifnum\pgfmathresult=1
        \fill[black] (\x,\y) circle(4pt);
      \else
        \draw[fill=white] (\x,\y) circle(4pt);
      \fi
    }
  }

  \foreach \x in {0,1,5,6} {
    \foreach \y in {4,5} {
      \pgfmathparse{int(mod(\x+\y,2))==0 ? 1 : 0}
      \ifnum\pgfmathresult=1
        \fill[black] (\x,\y) circle(4pt);
      \else
        \draw[fill=white] (\x,\y) circle(4pt);
      \fi
    }
  }

  \foreach \x in {0,1} {
    \foreach \y in {3} {
      \pgfmathparse{int(mod(\x+\y,2))==0 ? 1 : 0}
      \ifnum\pgfmathresult=1
        \fill[black] (\x,\y) circle(4pt);
      \else
        \draw[fill=white] (\x,\y) circle(4pt);
      \fi
    }
  }

   \draw[orange, line width=1.5pt] (0,0)--(6,0)--(6,2)--(5,2)--(5,1)--(4,1)--(4,2)--(3,2)--(3,1)--(2,1)--(2,2)--(1,2)--(1,1);

   \draw[orange, line width=1.5pt] (1,3)--(1,8)--(5,8)--(5,7)--(3,7)--(3,6)--(5,6)--(5,4)--(6,4)--(6,9)--(0,9)--(0,0);
   
\end{tikzpicture}
          }
        \caption{}
        \label{largehole_1_8}
    \end{subfigure}
    \hfill
    \begin{subfigure}[b]{0.18\textwidth}
        \centering
        \resizebox{\linewidth}{!}{
            \begin{tikzpicture}
  % Background gray dots for all grid points
  \foreach \x in {0,...,6}
    \foreach \y in {0,...,10}
      \fill[gray] (\x,\y) circle(0.04);

  % ── Parity-coloured nodes ──────────────────────────────────────
  % Rule: black if (x+y) even, white if (x+y) odd

  % All nodes using parity colouring

    \foreach \x in {0,...,6} {
    \foreach \y in {0,1,2,9,10} {
      \pgfmathparse{int(mod(\x+\y,2))==0 ? 1 : 0}
      \ifnum\pgfmathresult=1
        \fill[black] (\x,\y) circle(4pt);
      \else
        \draw[fill=white] (\x,\y) circle(4pt);
      \fi
    }
  }

  \foreach \x in {0,1,3,4,5,6} {
    \foreach \y in {7,8} {
      \pgfmathparse{int(mod(\x+\y,2))==0 ? 1 : 0}
      \ifnum\pgfmathresult=1
        \fill[black] (\x,\y) circle(4pt);
      \else
        \draw[fill=white] (\x,\y) circle(4pt);
      \fi
    }
  }

  \foreach \x in {0,1,5,6} {
    \foreach \y in {5,6} {
      \pgfmathparse{int(mod(\x+\y,2))==0 ? 1 : 0}
      \ifnum\pgfmathresult=1
        \fill[black] (\x,\y) circle(4pt);
      \else
        \draw[fill=white] (\x,\y) circle(4pt);
      \fi
    }
  }

  \foreach \x in {0,1} {
    \foreach \y in {3,4} {
      \pgfmathparse{int(mod(\x+\y,2))==0 ? 1 : 0}
      \ifnum\pgfmathresult=1
        \fill[black] (\x,\y) circle(4pt);
      \else
        \draw[fill=white] (\x,\y) circle(4pt);
      \fi
    }
  }

   \draw[orange, line width=1.5pt] (0,0)--(6,0)--(6,2)--(5,2)--(5,1)--(4,1)--(4,2)--(3,2)--(3,1)--(2,1)--(2,2)--(1,2)--(1,1);

   \draw[orange, line width=1.5pt] (1,3)--(1,9)--(5,9)--(5,8)--(3,8)--(3,7)--(5,7)--(5,5)--(6,5)--(6,10)--(0,10)--(0,0);
   
\end{tikzpicture}
          }
        \caption{}
        \label{largehole_1_9}
    \end{subfigure}
    \hfill
    \begin{subfigure}[b]{0.18\textwidth}
        \centering
        \resizebox{\linewidth}{!}{
            \begin{tikzpicture}
  % Background gray dots for all grid points
  \foreach \x in {0,...,7}
    \foreach \y in {0,...,10}
      \fill[gray] (\x,\y) circle(0.04);

  % ── Parity-coloured nodes ──────────────────────────────────────
  % Rule: black if (x+y) even, white if (x+y) odd

  % All nodes using parity colouring

    \foreach \x in {0,...,7} {
    \foreach \y in {0,1,2,9,10} {
      \pgfmathparse{int(mod(\x+\y,2))==0 ? 1 : 0}
      \ifnum\pgfmathresult=1
        \fill[black] (\x,\y) circle(4pt);
      \else
        \draw[fill=white] (\x,\y) circle(4pt);
      \fi
    }
  }

  \foreach \x in {0,1,3,4,5,6,7} {
    \foreach \y in {7,8} {
      \pgfmathparse{int(mod(\x+\y,2))==0 ? 1 : 0}
      \ifnum\pgfmathresult=1
        \fill[black] (\x,\y) circle(4pt);
      \else
        \draw[fill=white] (\x,\y) circle(4pt);
      \fi
    }
  }

  \foreach \x in {0,1,5,6,7} {
    \foreach \y in {5,6} {
      \pgfmathparse{int(mod(\x+\y,2))==0 ? 1 : 0}
      \ifnum\pgfmathresult=1
        \fill[black] (\x,\y) circle(4pt);
      \else
        \draw[fill=white] (\x,\y) circle(4pt);
      \fi
    }
  }

  \foreach \x in {0,1} {
    \foreach \y in {3,4} {
      \pgfmathparse{int(mod(\x+\y,2))==0 ? 1 : 0}
      \ifnum\pgfmathresult=1
        \fill[black] (\x,\y) circle(4pt);
      \else
        \draw[fill=white] (\x,\y) circle(4pt);
      \fi
    }
  }

   \draw[red, line width=1.5pt] (0,0)--(7,0)--(7,2)--(6,2)--(6,1)--(5,1)--(5,2)--(4,2)--(4,1)--(3,1)--(3,2)--(2,2)--(2,1)--(1,1)--(1,9)--(6,9)--(6,8)--(3,8)--(3,7)--(6,7)--(6,6)--(5,6)--(5,5)--(7,5)--(7,10)--(0,10)--(0,0);
   
\end{tikzpicture}
          }
        \caption{}
        \label{largehole_1_10}
    \end{subfigure}
    \caption{Construct Hamiltonian Cycle for Sub Case I of Large Hole Case}
    \label{subcase_1_2}
\end{figure}

\subsubsection{Sub Case II: $a<2$}

When $a<2$, the upper complementary side is constrained. If $a=1$, the upper side is present but too thin to support the usual back-and-forth traversal. If $a=0$, the upper side disappears entirely. Hence the construction cannot rely on the upper part as a flexible passage region. In particular, the cycle cannot cross the upper side repeatedly; instead, the upper portion must be treated as a prescribed boundary segment and incorporated into the cycle in a controlled way. The detailed case analysis for this subcase remains
to be completed, although we expect its construction
to follow a similar approach to Sub Case I. As what is described in \textbf{Sub Case I}, each case of different parity combinations would make the way of constructing various, meanwhile, every case that has some parameters with the same parity, the Hamiltonian Path in that corresponding subgraph would be similar to each other.

\subsubsection{Sub Case III: $b=0$}
When $b=0$, the lower complementary side disappears. Therefore, the lower side cannot be used as a usual connecting corridor, and the cycle must be completed without relying on a bottom passage to reconnect different parts of the graph. 

\subsubsection{Sub Case IV: $d=0$}
When $d=0$, the left complementary side disappears, so the left side can no longer be used as a connecting corridor. Similarly, the cycle must be organized using the remaining upper, right, and lower parts only, with the left boundary playing no transitional role.

\paragraph{Remaining Case Analysis.}
So far, we have focused on the detailed case analysis for
Sub Case I ($c_2=0$), while Sub Cases II, III, and IV
remain to be completed. We expect these remaining cases
to follow similar construction strategies. In particular,
when $b=0$ or $d=0$, the situation is similar to
$c_2=0$, since one of the connecting regions disappears,
requiring the Hamiltonian cycle to be constructed through
the remaining regions. The main difference lies in the
location of the constrained boundary, which may require
adjustments to the traversal direction and parity conditions.
For Sub Case II ($a<2$), when $a=0$, the upper connecting
region disappears, leading to a similar restriction.
However, when $a=1$, a single row remains above the
hole, which may impose additional constraints on the
Hamiltonian cycle construction. Therefore, although
the general approach developed in Sub Case I may be
extended to the remaining subcases, further case analysis
is needed to establish the corresponding constructions
and nonexistence conditions.

\section{Rectangular-Decomposition Approach}
\label{sec:decomposition}

\subsection{Partition into Six Rectangular Subgraphs}

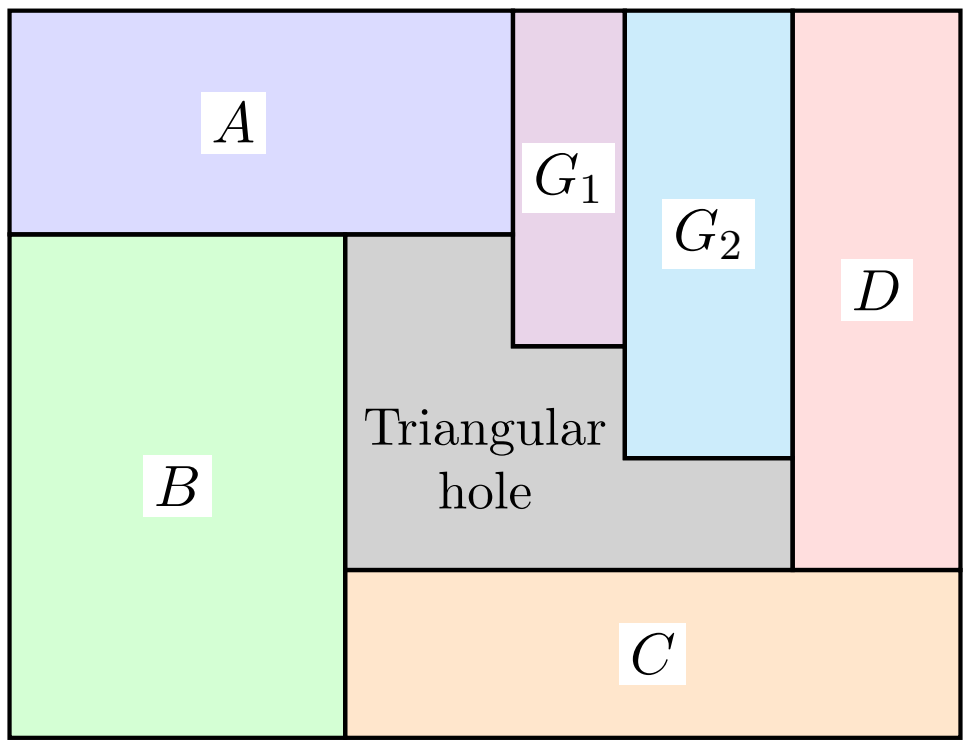
\begin{figure}[htbp]
\centering
\begin{tikzpicture}[x=0.37cm,y=-0.37cm]
  % Example: m=17, n=13; d=6; (c0,c1,c2)=(8,6,3);
  % a=4, (h0,h1,h2)=(2,2,2), b=3.
  \fill[blue!14]   (-0.5,-0.5) rectangle (8.5,3.5);
  \fill[green!17]  (-0.5,3.5) rectangle (5.5,12.5);
  \fill[orange!20] (5.5,9.5) rectangle (16.5,12.5);
  \fill[red!13]    (13.5,-0.5) rectangle (16.5,9.5);
  \fill[violet!17] (8.5,-0.5) rectangle (10.5,5.5);
  \fill[cyan!18]   (10.5,-0.5) rectangle (13.5,7.5);
  \fill[gray!35] (5.5,3.5) -- (8.5,3.5) -- (8.5,5.5)
    -- (10.5,5.5) -- (10.5,7.5) -- (13.5,7.5)
    -- (13.5,9.5) -- (5.5,9.5) -- cycle;

  % Draw only actual vertices and edges of the graph outside the hole.
  % \foreach \x in {0,...,16}{
  %   \foreach \y in {0,...,12}{
  %     \pgfmathtruncatemacro{\removed}{
  %       (\x>=6) && (\y>=4) && (\y<=9) &&
  %       (((\y<=5)&&(\x<=8)) ||
  %        ((\y>=6)&&(\y<=7)&&(\x<=10)) ||
  %        ((\y>=8)&&(\x<=13)))}
  %     \ifnum\removed=0
  %       \ifnum\x<16
  %         \pgfmathtruncatemacro{\rightremoved}{
  %           (\x+1>=6) && (\y>=4) && (\y<=9) &&
  %           (((\y<=5)&&(\x+1<=8)) ||
  %            ((\y>=6)&&(\y<=7)&&(\x+1<=10)) ||
  %            ((\y>=8)&&(\x+1<=13)))}
  %         \ifnum\rightremoved=0
  %           \draw[black!35,thin] (\x,\y) -- ++(1,0);
  %         \fi
  %       \fi
  %       \ifnum\y<12
  %         \pgfmathtruncatemacro{\belowremoved}{
  %           (\x>=6) && (\y+1>=4) && (\y+1<=9) &&
  %           (((\y+1<=5)&&(\x<=8)) ||
  %            ((\y+1>=6)&&(\y+1<=7)&&(\x<=10)) ||
  %            ((\y+1>=8)&&(\x<=13)))}
  %         \ifnum\belowremoved=0
  %           \draw[black!35,thin] (\x,\y) -- ++(0,1);
  %         \fi
  %       \fi
  %       \fill[black!65] (\x,\y) circle (0.035);
  %     \fi
  %   }
  % }
  \foreach \xa/\ya/\xb/\yb in {
    -0.5/-0.5/8.5/3.5, -0.5/3.5/5.5/12.5,
    5.5/9.5/16.5/12.5, 13.5/-0.5/16.5/9.5,
    8.5/-0.5/10.5/5.5, 10.5/-0.5/13.5/7.5}{
    \draw[thick] (\xa,\ya) rectangle (\xb,\yb);
  }
  \draw[thick] (5.5,3.5) -- (8.5,3.5) -- (8.5,5.5)
    -- (10.5,5.5) -- (10.5,7.5) -- (13.5,7.5)
    -- (13.5,9.5) -- (5.5,9.5) -- cycle;
  \node[fill=white,inner sep=2pt] at (3.5,1.5) {$A$};
  \node[fill=white,inner sep=2pt] at (2.5,8) {$B$};
  \node[fill=white,inner sep=2pt] at (11,11) {$C$};
  \node[fill=white,inner sep=2pt] at (15,4.5) {$D$};
  \node[fill=white,inner sep=2pt] at (9.5,2.5) {$G_1$};
  \node[fill=white,inner sep=2pt] at (12,3.5) {$G_2$};
  \node[align=center,font=\small] at (8,7.5) {Triangular\\hole};
\end{tikzpicture}
\caption{Partition into six rectangular subgraphs. The gray region is removed.}
\label{fig:six-region-partition}
\end{figure}

We partition the rectangular grid graph with a three-stacked
right-triangular hole into six rectangular subgraphs, denoted by
$A$, $B$, $C$, $D$, $G_1$, and $G_2$, as illustrated in
Figure~\ref{fig:six-region-partition}. The partition follows the
outer boundary and the three steps of the hole.

The upper-left subgraph $A$ extends from the left boundary of the
grid to the column containing the rightmost vertex of the top layer
of the hole. Its bottom row is one row above the top of the hole.
The subgraph $B$ lies along the left side of the hole. It begins
immediately below $A$, stops at the row containing the bottom of the hole,
and ends one column to the left of the hole's left edge.
The subgraph $C$ lies below the hole, beginning one row below its
bottom layer. It extends across the full width of the grid, from the left boundary
to the right boundary.
The subgraph $D$ occupies the right side: it extends from the top
boundary down to the row containing the bottom of the hole, and its
leftmost column is one column to the right of the hole's rightmost
vertex.

The remaining two subgraphs lie above the steps of the hole.
The subgraph $G_1$ is immediately to the right of $A$ and above
the portion of the second layer that extends beyond the first layer.
The subgraph $G_2$ lies between $G_1$ and $D$, above the portion
of the third layer that extends beyond the second layer.
Thus, the rectangles along the top boundary occur in the order
$A,G_1,G_2,D$ from left to right.

For a precise description, we use the coordinate convention from
Section~\ref{sec:problem}: $(0,0)$ is the top-left vertex, and the
$x$- and $y$-coordinates increase to the right and downward,
respectively. The top of the hole begins at row $a$, and its
three layers have heights $h_0,h_1,h_2$. The left boundary of the
hole is at column $d$, while the rightmost columns of the layers
are $m-c_0-1$, $m-c_1-1$, and $m-c_2-1$, respectively.
Thus, without introducing additional boundary variables, we obtain
the following partition:
\begin{center}
\begin{tabular}{ccl}
\toprule
Subgraph & Column range & Row range \\
\midrule
$A$   & $0\leq x<m-c_0$   & $0\leq y<a$ \\
$B$   & $0\leq x<d$       & $a\leq y<n$ \\
$C$   & $0\leq x<m$       & $a+h_0+h_1+h_2\leq y<n$ \\
$D$   & $m-c_2\leq x<m$   & $0\leq y<a+h_0+h_1+h_2$ \\
$G_1$ & $m-c_0\leq x<m-c_1$ & $0\leq y<a+h_0$ \\
$G_2$ & $m-c_1\leq x<m-c_2$ & $0\leq y<a+h_0+h_1$ \\
\bottomrule
\end{tabular}
\end{center}
Equivalently, their widths and heights are
\begin{equation}
\begin{aligned}
 A &: (m-c_0)\times a, &
 B &: d\times(h_0+h_1+h_2),\\
 C &: m\times b, &
 D &: c_2\times(a+h_0+h_1+h_2),\\
 G_1 &: (c_0-c_1)\times(a+h_0), &
 G_2 &: (c_1-c_2)\times(a+h_0+h_1).
\end{aligned}
\label{eq:six-region-dimensions}
\end{equation}
Here $n=a+h_0+h_1+h_2+b$, in accordance with the vertical
partition of the enclosing rectangle. We consider the nonempty
rectangles with the necessary minimum dimensions when applying
local Hamiltonian-cycle constructions.
These ranges assign every remaining vertex to exactly one subgraph.
Edges joining different subgraphs are retained as connections between
the rectangular components.

\subsection{Case Analysis by the Number of Odd-Sized Rectangles}
\label{sec:odd-rectangle-cases}

We now organize the construction according to the number of odd-sized
rectangles among $A$, $B$, $C$, $D$, $G_1$, $G_2$. A rectangle is odd-sized if its
number of vertices is odd, which only occurs if its width and height are both odd. Otherwise, it is even-sized.

Our objective is to construct a Hamiltonian cycle by repartitioning
the graph into even-sized rectangles and merging their local cycles.
The case analysis below gives explicit reductions for several parity
configurations, while identifying others that still require a
construction or compatibility argument.
Accordingly, throughout this case analysis, we assume that $G$ admits
a Hamiltonian cycle. In particular, $G$ has an even number of vertices.
This is a necessary condition, rather than a sufficient condition for
Hamiltonicity.

Since the six rectangles form a partition of the vertex set,
\[
 |V(G)|=|V(A)|+|V(B)|+|V(C)|+|V(D)|+|V(G_1)|+|V(G_2)|.
\]
Reducing this identity modulo two shows that the number of odd-sized
rectangles must be even. Therefore, the only possibilities are
$0$, $2$, $4$, and $6$ odd-sized rectangles.

\subsubsection{No Odd-Sized Rectangles}

If there are no odd-sized rectangles, all six components are even-sized.
Provided that each has width and height at least two, we construct a
Hamiltonian cycle within each rectangle using the proposed rectangular
construction. We then connect these local cycles through the
parallel-edge merging operation. The local cycles must be chosen so
that the required pairs of adjacent parallel edges are available at
each merge. Once these compatible merges are performed, the resulting
cycle covers every vertex of $G$ exactly once.

The remaining cases concern partitions containing odd-sized rectangles,
which cannot individually admit Hamiltonian cycles. Our approach is to
repartition appropriate regions into even-sized rectangles before
constructing and merging their local cycles.

\subsubsection{Case I: Exactly Two Odd-Sized Rectangles}

The construction depends on the relative positions of the two odd-sized
rectangles. The six components occur in the cyclic order
\[
 A,\ B,\ C,\ D,\ G_2,\ G_1,\ A.
\]
Here, adjacency refers to consecutive components in this order, which
share an interface in the six-region partition. There are three
possibilities: the two odd-sized rectangles are adjacent, separated by
one rectangle along a shortest route in this order, or separated by
two rectangles along either shortest route.

\paragraph{Case 1: Two adjacent odd-sized rectangular subgraphs.}

Suppose exactly two of the six rectangular subgraphs
$A$, $B$, $C$, $D$, $G_1$, and $G_2$ are odd-sized,
and these two subgraphs are adjacent.
There are six possible adjacent pairs:
$(A,B)$, $(B,C)$, $(C,D)$, $(D,G_2)$,
$(G_2,G_1)$, and $(G_1,A)$.

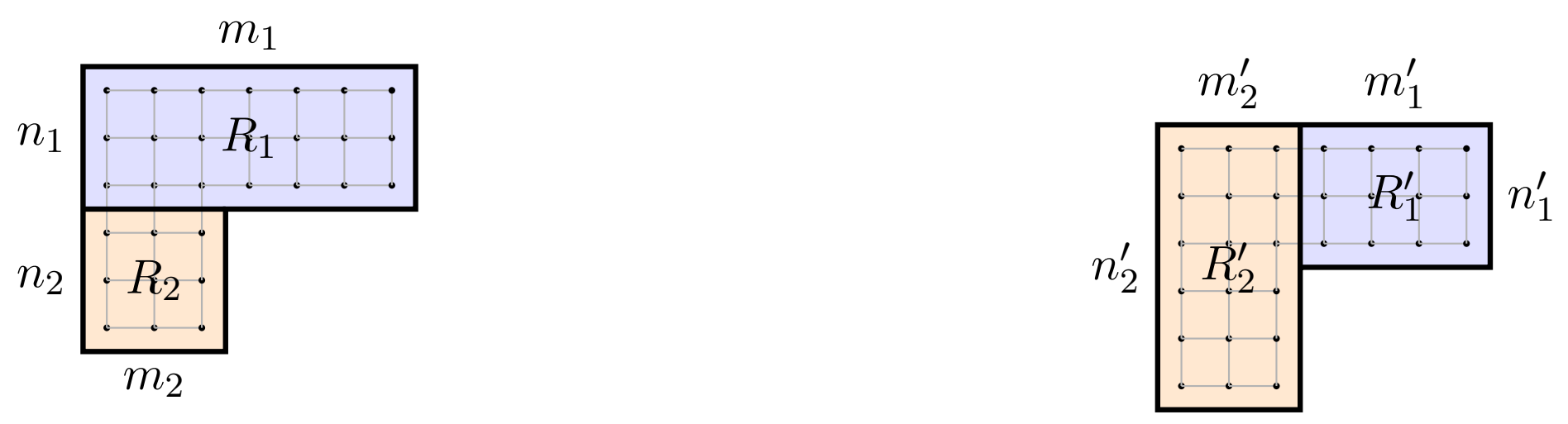
\begin{figure}[h]
    \centering
    \begin{subfigure}[b]{0.46\columnwidth}
        \centering
        \begin{tikzpicture}[scale=0.38]
            \def\Ma{7}
            \def\Na{3}
            \def\Mb{3}
            \def\Nb{3}

            % Original upper rectangle A
            \fill[blue!12]
                (0,0) rectangle (\Ma,\Na);

            % Original lower rectangle B
            \fill[orange!18]
                (0,-\Nb) rectangle (\Mb,0);

            % Grid edges and vertices
            \foreach \x in {0,...,6} {
                \foreach \y in {0,...,2} {
                    \fill (\x+0.5,\y+0.5)
                        circle (2pt);
                    \ifnum\x<6
                        \draw[gray!60]
                            (\x+0.5,\y+0.5) --
                            (\x+1.5,\y+0.5);
                    \fi
                    \ifnum\y<2
                        \draw[gray!60]
                            (\x+0.5,\y+0.5) --
                            (\x+0.5,\y+1.5);
                    \fi
                }
            }
            \foreach \x in {0,...,2} {
                \foreach \y in {-3,...,-1} {
                    \fill (\x+0.5,\y+0.5)
                        circle (2pt);
                    \ifnum\x<2
                        \draw[gray!60]
                            (\x+0.5,\y+0.5) --
                            (\x+1.5,\y+0.5);
                    \fi
                    \ifnum\y<-1
                        \draw[gray!60]
                            (\x+0.5,\y+0.5) --
                            (\x+0.5,\y+1.5);
                    \fi
                }
            }

            % Edges connecting A and B
            \foreach \x in {0,...,2} {
                \draw[gray!60]
                    (\x+0.5,-0.5) --
                    (\x+0.5,0.5);
            }

            % Original partition boundaries
            \draw[very thick]
                (0,-3)--(3,-3)--(3,0)--
                (7,0)--(7,3)--(0,3)--cycle;
            \draw[very thick] (0,0)--(3,0);

            % Labels
            \node at (3.5,1.5) {$R_1$};
            \node at (1.5,-1.5) {$R_2$};

            \node[above] at (3.5,3) {$m_1$};
            \node[left] at (0,1.5) {$n_1$};
            \node[left] at (0,-1.5) {$n_2$};
            \node[below] at (1.5,-3) {$m_2$};
        \end{tikzpicture}
        \caption{Original partition.}
        \label{fig:two-stack-ori}
    \end{subfigure}
    \hfill
    \begin{subfigure}[b]{0.46\columnwidth}
        \centering
        \begin{tikzpicture}[scale=0.38]
            \def\Ma{7}
            \def\Na{3}
            \def\Mb{3}
            \def\Nb{3}

            % New left rectangle
            \fill[orange!18]
                (0,-\Nb) rectangle
                (\Mb,\Na);

            % New right rectangle
            \fill[blue!12]
                (\Mb,0) rectangle
                (\Ma,\Na);

            % Grid edges and vertices
            \foreach \x in {0,...,6} {
                \foreach \y in {0,...,2} {
                    \fill (\x+0.5,\y+0.5)
                        circle (2pt);
                    \ifnum\x<6
                        \draw[gray!60]
                            (\x+0.5,\y+0.5) --
                            (\x+1.5,\y+0.5);
                    \fi
                    \ifnum\y<2
                        \draw[gray!60]
                            (\x+0.5,\y+0.5) --
                            (\x+0.5,\y+1.5);
                    \fi
                }
            }
            \foreach \x in {0,...,2} {
                \foreach \y in {-3,...,-1} {
                    \fill (\x+0.5,\y+0.5)
                        circle (2pt);
                    \ifnum\x<2
                        \draw[gray!60]
                            (\x+0.5,\y+0.5) --
                            (\x+1.5,\y+0.5);
                    \fi
                    \ifnum\y<-1
                        \draw[gray!60]
                            (\x+0.5,\y+0.5) --
                            (\x+0.5,\y+1.5);
                    \fi
                }
            }

            % Edges connecting the rows
            \foreach \x in {0,...,2} {
                \draw[gray!60]
                    (\x+0.5,-0.5) --
                    (\x+0.5,0.5);
            }

            % New partition boundaries
            \draw[very thick]
                (0,-3)--(3,-3)--(3,0)--
                (7,0)--(7,3)--(0,3)--cycle;
            \draw[very thick] (3,0)--(3,3);

            % Labels
            \node at (1.5,0) {$R_2'$};
            \node at (5,1.5) {$R_1'$};

            \node[above] at (1.5,3) {$m_2'$};
            \node[above] at (5,3) {$m_1'$};
            \node[left] at (0,0) {$n_2'$};
            \node[right] at (7,1.5) {$n_1'$};
        \end{tikzpicture}
        \caption{After repartitioning.}
        \label{fig:repartitioning}
    \end{subfigure}

    \caption{Repartitioning two adjacent odd-sized
    rectangular grid graphs into two even-sized
    rectangular grid graphs.}
    \label{fig:adjacent-odd-repartition}
\end{figure}

By rotating the corresponding pairs by
$0^\circ$, $90^\circ$ clockwise, $180^\circ$,
$0^\circ$, $90^\circ$ counterclockwise, and
$90^\circ$ counterclockwise, respectively,
each pair can be represented by the same
configuration: two rectangular grid graphs
sharing an aligned left boundary, where the
upper rectangle is strictly wider than the
lower rectangle, shown in Figure \ref{fig:two-stack-ori}.
Therefore, it suffices to consider this
representative case.

Let the upper and lower rectangular subgraphs
be $R_1(m_1,n_1)$ and $R_2(m_2,n_2)$,
respectively, where $m_1$, $n_1$, $m_2$, and $n_2$ are all odd, and $m_1 > m_2$. Since both rectangles have odd width and odd height, neither admits a Hamiltonian cycle individually.

We repartition $R_1 \cap R_2$ by extending the
right boundary of the bottom subgraphs($R_2$) upward through the top one($R_1$), as shown in Figure \ref{fig:repartitioning}.
This produces two new rectangular subgraphs,
denoted by $R_1'$ and $R_2'$, with dimensions
\[
\begin{aligned}
R_1' &(m_1', n_1'),
& m_1' &= m_1 - m_2,
& n_1' &= n_1,\\
R_2' &(m_2', n_2'),
& m_2' &= m_2,
& n_2' &= n_1+n_2.
\end{aligned}
\]

Since $m_1$ and $m_2$ are odd,
$m_1'=m_1-m_2$ is even.
Similarly, since $n_1$ and $n_2$ are odd,
$n_2'=n_1+n_2$ is even.
Thus, $R_1'$ has even width and $R_2'$ has even
height, implying that both are even-sized
rectangular grid graphs.

Consequently, a Hamiltonian cycle can be
constructed within each of $R_1'$ and $R_2'$.
By applying the previously established
cycle-merging theorem, these two local
Hamiltonian cycles can be connected to form
a Hamiltonian cycle in $R_1 \cap R_2$.
The resulting cycle can then be merged with
the Hamiltonian cycles of the remaining
even-sized rectangular subgraphs to obtain
a Hamiltonian cycle in the original grid graph.

\paragraph{Case 2: Two odd-sized rectangular subgraphs separated by one subgraph.}

Suppose exactly two of the six rectangular subgraphs
$A$, $B$, $C$, $D$, $G_1$, and $G_2$ are odd-sized,
and they are separated by exactly one even-sized
rectangular subgraph.
There are six possible pairs:
$(A,C)$, $(B,D)$, $(C,G_2)$, $(D,G_1)$,
$(G_2,A)$, and $(G_1,B)$.

Based on the relative positions of these subgraphs,
we distinguish two representative configurations.

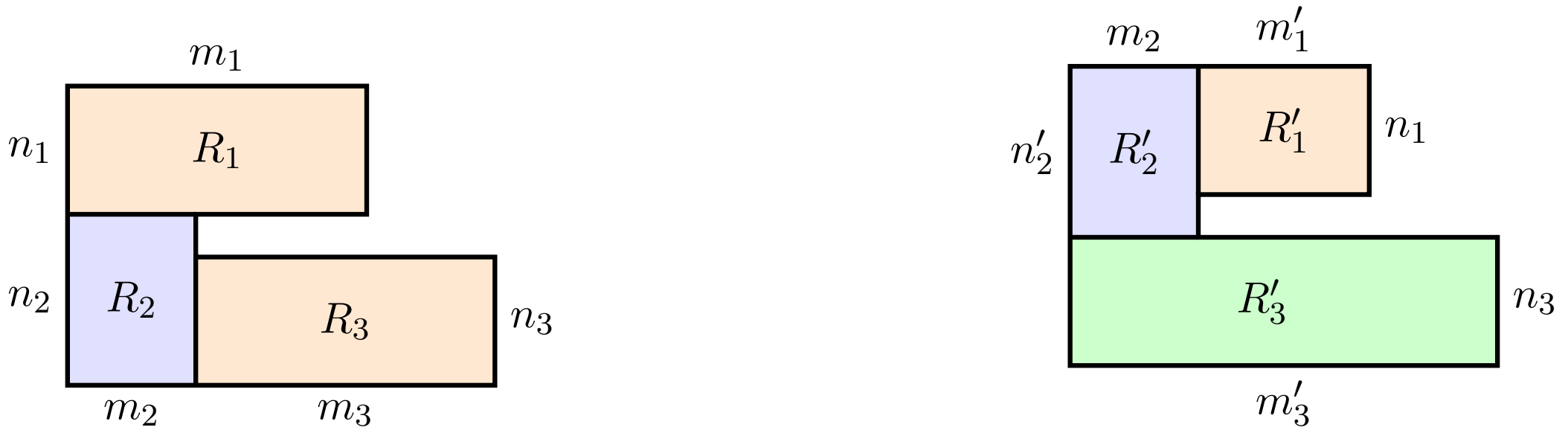
\begin{figure}[htbp]
    \centering
    \begin{subfigure}[b]{0.46\columnwidth}
        \centering
        \begin{tikzpicture}[scale=0.38]
            % R1: 7 x 3, R2: 3 x 4, R3: 7 x 3
            \fill[orange!18] (0,4) rectangle (7,7);
            \fill[blue!12] (0,0) rectangle (3,4);
            \fill[orange!18] (3,0) rectangle (10,3);

            \draw[very thick]
                (0,0)--(10,0)--(10,3)--(3,3)
                --(3,4)--(7,4)--(7,7)
                --(0,7)--cycle;
            \draw[very thick] (0,4)--(3,4);
            \draw[very thick] (3,0)--(3,3);

            \node at (3.5,5.5) {$R_1$};
            \node at (1.5,2) {$R_2$};
            \node at (6.5,1.5) {$R_3$};

            \node[above] at (3.5,7) {$m_1$};
            \node[left] at (0,5.5) {$n_1$};
            \node[left] at (0,2) {$n_2$};
            \node[below] at (1.5,0) {$m_2$};
            \node[below] at (6.5,0) {$m_3$};
            \node[right] at (10,1.5) {$n_3$};
        \end{tikzpicture}
        \caption{Original partition.}
        \label{fig:three-stack-ori}
    \end{subfigure}
    \hfill
    \begin{subfigure}[b]{0.46\columnwidth}
        \centering
        \begin{tikzpicture}[scale=0.38]
            % Illustrative geometric boundary reassignment
            \fill[orange!18] (3,4) rectangle (7,7);
            \fill[blue!12] (0,3) rectangle (3,7);
            \fill[green!20] (0,0) rectangle (10,3);

            \draw[very thick]
                (0,0)--(10,0)--(10,3)--(3,3)
                --(3,4)--(7,4)--(7,7)
                --(0,7)--cycle;
            \draw[very thick] (3,4)--(3,7);
            \draw[very thick] (0,3)--(3,3);

            \node at (5,5.5) {$R_1'$};
            \node at (1.5,5) {$R_2'$};
            \node at (4.5,1.5) {$R_3'$};
            
            \node[above] at (5,7) {$m_1'$};
            \node[right] at (7,5.5) {$n_1$};
            \node[left] at (0,5) {$n_2'$};
            \node[above] at (1.5,7) {$m_2$};
            \node[below] at (5,0) {$m_3'$};
            \node[right] at (10,1.5) {$n_3$};
        \end{tikzpicture}
        \caption{After boundary reassignment.}
        \label{fig:three-stack-repartition}
    \end{subfigure}

    \caption{Repartitioning three staggered rectangular
    subgraphs with two nonadjacent odd-sized subgraphs.}
    \label{fig:three-stack}
\end{figure}

\textbf{Configuration 1.}
The first configuration corresponds to $(A,C)$,
where $B$ lies between the two odd-sized subgraphs.
The configurations $(B,D)$, $(G_1,B)$, and $(C,G_2)$
can be reduced to this configuration by rotating
them by $180^\circ$, $90^\circ$ counterclockwise,
and $180^\circ$, respectively.
Therefore, it suffices to consider the
representative configuration shown in
Figure~\ref{fig:three-stack-ori}.

Let the upper, middle, and lower rectangular
subgraphs be $R_1(m_1,n_1)$, $R_2(m_2,n_2)$,
and $R_3(m_3,n_3)$, respectively.
The subgraphs $R_1$ and $R_2$ share an aligned
left boundary, while $R_2$ and $R_3$ share
an aligned bottom boundary.
Furthermore, the right edge of $R_2$ is
adjacent to the left edge of $R_3$ along
their overlapping portion.
We have
\[
m_1>m_2,\qquad m_2+m_3>m_1.
\]
Since $R_1$ and $R_3$ are odd-sized,
$m_1,n_1,m_3,n_3$ are all odd.
The repartitioning depends on the parity
of $m_2$ and $n_2$.

\emph{(i) $m_2$ is odd and $n_2$ is even.}

We first apply the repartitioning operation
introduced in Case 1 to $R_1$ and $R_2$,
extending the right boundary of $R_2$
upward through $R_1$.
We then extend the bottom boundary of $R_2$
rightward through $R_3$.
As shown in Figure~\ref{fig:three-stack-repartition},
this produces three new rectangular subgraphs,
denoted by $R_1'$, $R_2'$, and $R_3'$,
with dimensions
\[
\begin{aligned}
R_1' &(m_1',n_1'),
& m_1' &= m_1-m_2,
& n_1' &= n_1,\\
R_2' &(m_2',n_2'),
& m_2' &= m_2,
& n_2' &= n_2+n_1-n_3,\\
R_3' &(m_3',n_3'),
& m_3' &= m_3+m_2,
& n_3' &= n_3.
\end{aligned}
\]

All three new regions after boundary reassignment are even size: $m_1$ and $m_2$ are odd, so $m_1'=m_1-m_2$ is even; $n_1$ and $n_3$ are odd, $n_2$ is even, so $n_2'=n_2+n_1-n_3$ is even; $m_2$ and $m_3$ are odd, so $m_3'=m_3+m_2$ is even.

Followed by Algorithm \ref{alg:even-rectangle-cycle}, a Hamiltonian cycle can be constructed within each of $R_1'$, $R_2'$, and $R_3'$. By Theorem \ref{thm:connect-two-cycles}, these three local Hamiltonian cycles can be connected one by one to form a Hamiltonian cycle. The connected Hamiltonian cycle can then merge with the Hamiltonian cycles of the remaining three regions.

% \textcolor{red}{
% TODO: Verify the precise repartitioning dimensions
% for the staggered configuration and establish
% the necessary nonemptiness conditions.
% }

\emph{(ii) $m_2$ is even and $n_2$ is odd.}

\begin{figure}[htbp]
\centering

% Common outer boundary for all five panels
\def\OuterShape{%
    \draw[very thick]
        (0,0)--(13,0)--(13,3)--(6,3)
        --(6,5)--(7,5)--(7,8)--(0,8)--cycle;
}

% ==========================================
% (a) Original partition
% ==========================================
\begin{subfigure}[t]{0.31\columnwidth}
    \centering
    \begin{tikzpicture}[x=0.26cm,y=0.26cm]
        \fill[orange!18] (0,5) rectangle (7,8);
        \fill[blue!12] (0,0) rectangle (6,5);
        \fill[orange!18] (6,0) rectangle (13,3);

        \OuterShape
        \draw[thick] (0,5)--(6,5);
        \draw[thick] (6,0)--(6,3);

        \node at (3.5,6.5) {$R_1$};
        \node at (3,2.5) {$R_2$};
        \node at (9.5,1.5) {$R_3$};
    \end{tikzpicture}
    \caption{Original partition.}
    \label{fig:eo-original}
\end{subfigure}
\hfill
% ==========================================
% (b) Vertical split of R2
% ==========================================
\begin{subfigure}[t]{0.31\columnwidth}
    \centering
    \begin{tikzpicture}[x=0.26cm,y=0.26cm]
        \fill[orange!18] (0,5) rectangle (7,8);
        \fill[blue!12] (0,0) rectangle (3,5);
        \fill[green!20] (3,0) rectangle (6,5);
        \fill[orange!18] (6,0) rectangle (13,3);

        \OuterShape
        \draw[thick] (0,5)--(6,5);
        \draw[thick] (6,0)--(6,3);
        \draw[thick] (3,0)--(3,5);

        \node at (3.5,6.5) {$R_1$};
        \node[font=\scriptsize] at (1.5,2.5)
            {$R_{2L}$};
        \node[font=\scriptsize] at (4.5,2.5)
            {$R_{2R}$};
        \node at (9.5,1.5) {$R_3$};
    \end{tikzpicture}
    \caption{Vertical separation of $R_2$.}
    \label{fig:eo-vertical}
\end{subfigure}
\hfill
% ==========================================
% (c) Boundary reassignment of R2R and R3
% ==========================================
\begin{subfigure}[t]{0.31\columnwidth}
    \centering
    \begin{tikzpicture}[x=0.26cm,y=0.26cm]
        \fill[orange!18] (0,5) rectangle (7,8);
        \fill[blue!12] (0,0) rectangle (3,5);
        \fill[green!20] (3,3) rectangle (6,5);
        \fill[orange!18] (3,0) rectangle (13,3);

        \OuterShape
        \draw[thick] (0,5)--(6,5);
        \draw[thick] (3,0)--(3,5);
        \draw[thick] (3,3)--(6,3);

        \node at (3.5,6.5) {$R_1$};
        \node[font=\scriptsize] at (1.5,2.5)
            {$R_{2L}$};
        \node[font=\scriptsize] at (4.5,4)
            {$R_{2R}'$};
        \node at (8,1.5) {$R_3'$};
    \end{tikzpicture}
    \caption{Reassignment of $R_{2R}$ and $R_3$.}
    \label{fig:eo-lower-reassignment}
\end{subfigure}

\par\medskip

% ==========================================
% (d) Horizontal split of R2L
% ==========================================
\begin{subfigure}[t]{0.39\columnwidth}
    \centering
    \begin{tikzpicture}[x=0.30cm,y=0.30cm]
        \fill[orange!18] (0,5) rectangle (7,8);
        \fill[blue!12] (0,2) rectangle (3,5);
        \fill[violet!18] (0,0) rectangle (3,2);
        \fill[green!20] (3,3) rectangle (6,5);
        \fill[orange!18] (3,0) rectangle (13,3);

        \OuterShape
        \draw[thick] (0,5)--(6,5);
        \draw[thick] (3,0)--(3,5);
        \draw[thick] (3,3)--(6,3);
        \draw[thick] (0,2)--(3,2);

        \node at (3.5,6.5) {$R_1$};
        \node[font=\scriptsize] at (1.5,3.5)
            {$R_{2LU}$};
        \node[font=\scriptsize] at (1.5,1)
            {$R_{2LB}$};
        \node[font=\scriptsize] at (4.5,4)
            {$R_{2R}'$};
        \node at (8,1.5) {$R_3'$};
    \end{tikzpicture}
    \caption{Horizontal separation of $R_{2L}$.}
    \label{fig:eo-horizontal}
\end{subfigure}
\hspace{0.03\columnwidth}
% ==========================================
% (e) Final boundary reassignment
% ==========================================
\begin{subfigure}[t]{0.39\columnwidth}
    \centering
    \begin{tikzpicture}[x=0.30cm,y=0.30cm]
        \fill[orange!18] (3,5) rectangle (7,8);
        \fill[blue!12] (0,2) rectangle (3,8);
        \fill[violet!18] (0,0) rectangle (3,2);
        \fill[green!20] (3,3) rectangle (6,5);
        \fill[orange!18] (3,0) rectangle (13,3);

        \OuterShape
        \draw[thick] (3,0)--(3,8);
        \draw[thick] (0,2)--(3,2);
        \draw[thick] (3,3)--(6,3);
        \draw[thick] (3,5)--(6,5);

        \node at (5,6.5) {$R_1'$};
        \node[font=\scriptsize] at (1.5,5)
            {$R_{2LU}'$};
        \node[font=\scriptsize] at (1.5,1)
            {$R_{2LB}$};
        \node[font=\scriptsize] at (4.5,4)
            {$R_{2R}'$};
        \node at (8,1.5) {$R_3'$};
    \end{tikzpicture}
    \caption{Final partition into five even-sized rectangles.}
    \label{fig:eo-final}
\end{subfigure}

\caption{Repartitioning three staggered rectangular
subgraphs when $m_2$ is even and $n_2$ is odd.
The middle rectangle is first separated vertically
into two odd-width rectangles. After boundary
reassignment with $R_3$, the left rectangle is
separated horizontally and its upper portion
is combined with part of $R_1$.
The resulting five rectangular subgraphs are
all even-sized.}
\label{fig:three-stack-even-odd}
\end{figure}
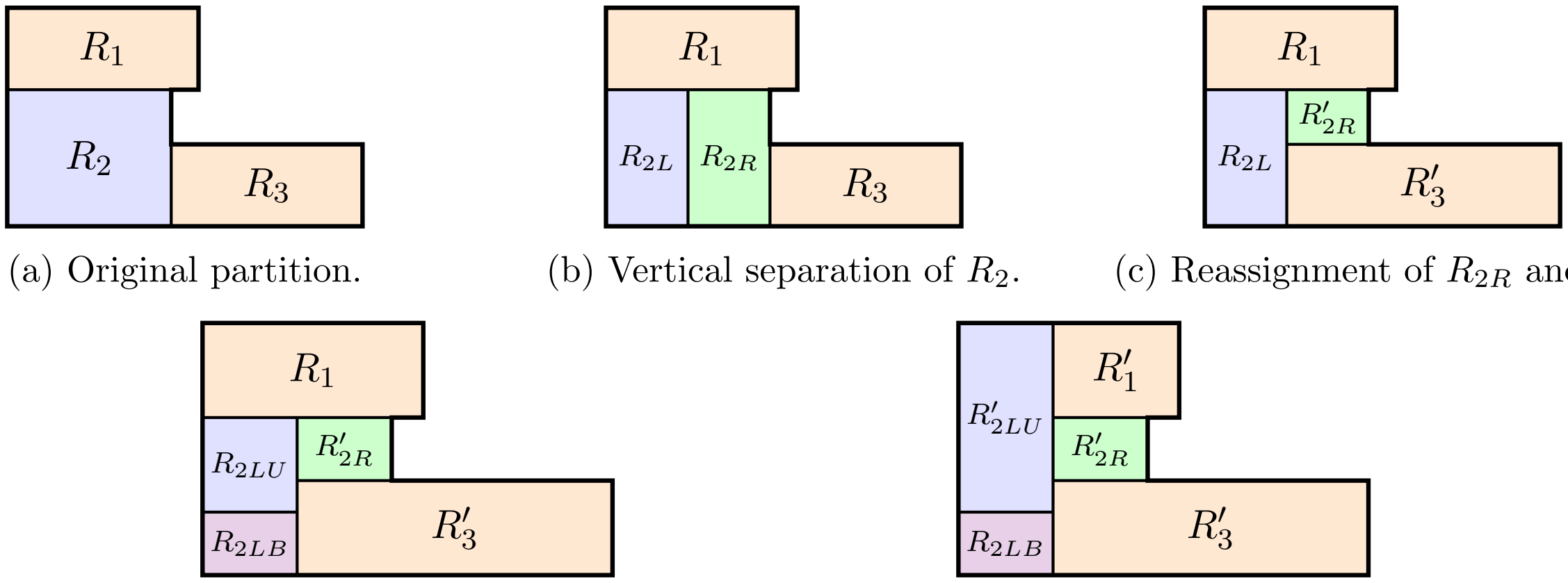

In this case, we repartition the three rectangular
subgraphs into five even-sized rectangular subgraphs.
We first separate $R_2$ vertically into two
odd-width rectangles, then apply boundary reassignment
to $R_{2R}$ and $R_3$. Afterward, we separate
$R_{2L}$ horizontally and apply another boundary
reassignment to its upper part and $R_1$, as shown
in Figure~\ref{fig:three-stack-even-odd}.

Specifically, we first divide $R_2$ vertically
into two rectangular subgraphs, denoted by
$R_{2L}$ and $R_{2R}$, where
\[
m_{2L}+m_{2R}=m_2,
\]
and both $m_{2L}$ and $m_{2R}$ are odd.
Since $n_2$ is odd, both $R_{2L}$ and $R_{2R}$
are odd-sized rectangular subgraphs.

We then extend the upper boundary of $R_3$
leftward through $R_{2R}$. This produces two
new rectangular subgraphs, denoted by
$R_{2R}'$ and $R_3'$. Next, we separate
$R_{2L}$ horizontally into an upper part
$R_{2LU}$ and a bottom part $R_{2LB}$,
such that
\[
n_{2LU}+n_{2LB}=n_2,
\]
where $n_{2LU}$ is odd and $n_{2LB}$ is even.

Finally, we apply the boundary reassignment
introduced in Case 1 to $R_1$ and $R_{2LU}$,
extending the right boundary of $R_{2LU}$
upward through $R_1$. This produces
$R_1'$ and $R_{2LU}'$.

As shown in Figure~\ref{fig:eo-final},
the resulting five rectangular subgraphs
are $R_1'$, $R_{2R}'$, $R_3'$,
$R_{2LB}$, and $R_{2LU}'$,
with dimensions
\[
\begin{aligned}
R_1' &(m_1',n_1'),
& m_1' &= m_1-m_{2L},
& n_1' &= n_1,\\
R_{2R}' &(m_{2R}',n_{2R}'),
& m_{2R}' &= m_{2R},
& n_{2R}' &= n_2-n_3,\\
R_3' &(m_3',n_3'),
& m_3' &= m_3+m_{2R},
& n_3' &= n_3,\\
R_{2LB} &(m_{2LB},n_{2LB}),
& m_{2LB} &= m_{2L},
& n_{2LB} &= n_2-n_{2LU},\\
R_{2LU}' &(m_{2LU}',n_{2LU}'),
& m_{2LU}' &= m_{2L},
& n_{2LU}' &= n_1+n_{2LU}.
\end{aligned}
\]

All five resulting rectangular subgraphs are
even-sized. Since $m_1$ and $m_{2L}$ are odd,
$m_1'=m_1-m_{2L}$ is even. Similarly,
since $n_2$ and $n_3$ are odd,
$n_{2R}'=n_2-n_3$ is even.
Since $m_3$ and $m_{2R}$ are odd,
$m_3'=m_3+m_{2R}$ is even.
By construction, $n_{2LB}$ is even.
Finally, since $n_1$ and $n_{2LU}$ are odd,
$n_{2LU}'=n_1+n_{2LU}$ is even.

Therefore, all five resulting rectangular
subgraphs are even-sized.

Provided that each resulting rectangle has
width and height at least two, a Hamiltonian
cycle can be constructed within each of
$R_1'$, $R_{2R}'$, $R_3'$, $R_{2LB}$,
and $R_{2LU}'$ by
Algorithm~\ref{alg:even-rectangle-cycle}.
By applying Theorem~\ref{thm:connect-two-cycles}
whenever the required compatible boundary edges
are available, these five local Hamiltonian
cycles can be connected successively to form
a Hamiltonian cycle in the union of the three
original subgraphs. The resulting cycle can
then be merged with the Hamiltonian cycles
of the remaining three regions to obtain
a Hamiltonian cycle in the original grid graph.

\emph{(iii) $m_2$ and $n_2$ are both even.}

We claim that this configuration does not
admit a Hamiltonian cycle because the numbers
of black and white vertices are unequal.
A proof based on the bipartite coloring
of the grid graph will be provided.

\medskip

\begin{figure}[htbp]
    \centering
    \begin{subfigure}[b]{0.46\columnwidth}
        \centering
        \begin{tikzpicture}[scale=0.42]
            % R1: 3 x 3, R2: 2 x 5, R3: 3 x 7
            \fill[orange!18] (0,4) rectangle (3,7);
            \fill[blue!12] (3,2) rectangle (5,7);
            \fill[orange!18] (5,0) rectangle (8,7);

            \draw[very thick]
                (0,4)--(0,7)--(8,7)--(8,0)
                --(5,0)--(5,2)--(3,2)
                --(3,4)--cycle;
            \draw[very thick] (3,4)--(3,7);
            \draw[very thick] (5,2)--(5,7);

            \node at (1.5,5.5) {$R_1$};
            \node at (4,4.5) {$R_2$};
            \node at (6.5,3.5) {$R_3$};

            \node[above] at (1.5,7) {$m_1$};
            \node[above] at (4,7) {$m_2$};
            \node[above] at (6.5,7) {$m_3$};

            \node[left] at (0,5.5) {$n_1$};
            \node[below] at (4,2) {$n_2$};
            \node[right] at (8,3.5) {$n_3$};
        \end{tikzpicture}
        \caption{Original partition.}
        \label{fig:three-side-ori}
    \end{subfigure}
    \hfill
    \begin{subfigure}[b]{0.46\columnwidth}
        \centering
        \begin{tikzpicture}[scale=0.42]
            \fill[orange!18] (0,4) rectangle (8,7);
            \fill[blue!12] (3,2) rectangle (5,4);
            \fill[green!20] (5,0) rectangle (8,4);

            \draw[very thick]
                (0,4)--(0,7)--(8,7)--(8,0)
                --(5,0)--(5,2)--(3,2)
                --(3,4)--cycle;
            \draw[very thick] (3,4)--(8,4);
            \draw[very thick] (5,2)--(5,4);

            \node at (4,5.5) {$R_1'$};
            \node at (4,3) {$R_2'$};
            \node at (6.5,2) {$R_3'$};

            \node[above] at (4,7) {$m_1'$};
            \node[right] at (8,5.5) {$n_1'$};
            \node[below] at (4,2) {$m_2'$};
            \node[left] at (3,3) {$n_2'$};
            \node[below] at (6.5,0) {$m_3'$};
            \node[right] at (8,2) {$n_3'$};
        \end{tikzpicture}
        \caption{After repartitioning.}
        \label{fig:three-side-repartition}
    \end{subfigure}

    \caption{Repartitioning three horizontally adjacent
    rectangular subgraphs into three even-sized
    rectangular subgraphs.}
    \label{fig:three-side}
\end{figure}
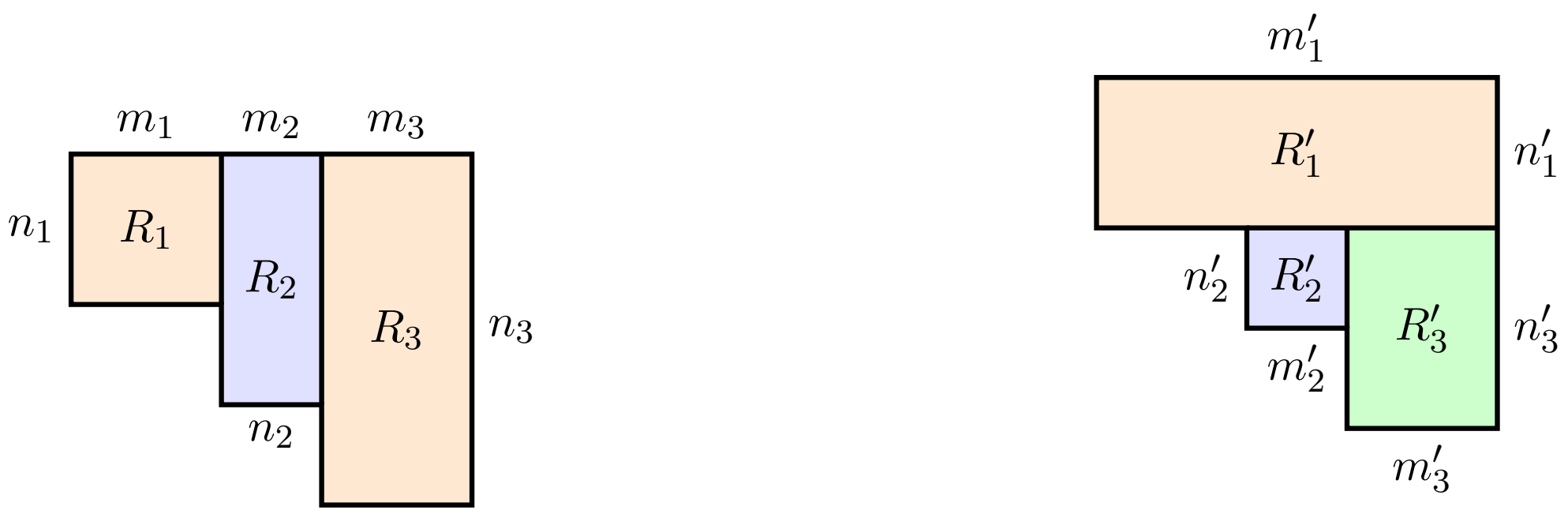

\textbf{Configuration 2.}

The second configuration corresponds to $(D,G_1)$,
where $G_2$ lies between the two odd-sized subgraphs.
The configuration $(G_2,A)$ can be reduced to
this configuration by reflection.
Therefore, it suffices to consider the
representative configuration shown in
Figure~\ref{fig:three-side-ori}.

Let the three rectangular subgraphs,
from left to right, be $R_1(m_1,n_1)$,
$R_2(m_2,n_2)$, and $R_3(m_3,n_3)$.
They share an aligned top boundary,
and their heights satisfy
\[
n_1<n_2<n_3.
\]
The right boundary of $R_1$ is adjacent
to the upper portion of the left boundary
of $R_2$, while the right boundary of
$R_2$ is adjacent to the upper portion
of the left boundary of $R_3$.

Since $R_1$ and $R_3$ are odd-sized,
$m_1,n_1,m_3,n_3$ are all odd.
We distinguish the following cases
according to the parity of $R_2$.

\emph{(i) $m_2$ is even.}

In this case, $R_2$ is either even by odd
or even by even.
We repartition the three subgraphs by
extending the bottom boundary of $R_1$
rightward through $R_2$ and $R_3$,
as shown in Figure~\ref{fig:three-side-repartition}.

This produces three new rectangular subgraphs,
denoted by $R_1'$, $R_2'$, and $R_3'$,
with dimensions
\[
\begin{aligned}
R_1' &(m_1',n_1'),
& m_1' &= m_1+m_2+m_3,
& n_1' &= n_1,\\
R_2' &(m_2',n_2'),
& m_2' &= m_2,
& n_2' &= n_2-n_1,\\
R_3' &(m_3',n_3'),
& m_3' &= m_3,
& n_3' &= n_3-n_1.
\end{aligned}
\]

Since $m_1$ and $m_3$ are odd and
$m_2$ is even,
$m_1'=m_1+m_2+m_3$ is even.
Thus, $R_1'$ is even-sized.

Since $m_2'=m_2$ is even,
$R_2'$ is also even-sized,
regardless of the parity of $n_2'$.

Finally, since $n_1$ and $n_3$ are odd,
$n_3'=n_3-n_1$ is even.
Thus, $R_3'$ is even-sized.

Consequently, all three resulting
rectangular subgraphs are even-sized.

\emph{(ii) $m_2$ is odd and $n_2$ is even.}

This case can be handled using a repartitioning
strategy similar to that introduced in
Configuration 1(ii). However, the detailed
construction and correctness analysis have not
yet been completed and are left for future work.

For the completed cases, a Hamiltonian cycle
can be constructed within each resulting
even-sized rectangular subgraph.
By applying the previously established
cycle-merging theorem, these local cycles
can be connected to form a Hamiltonian cycle
in the union of the three subgraphs.
The resulting cycle can then be merged
with the Hamiltonian cycles of the remaining
even-sized rectangular subgraphs to obtain
a Hamiltonian cycle in the original grid graph.

\subsubsection{Case II: Exactly Four Odd-Sized Rectangles}

Suppose exactly four of the six rectangular subgraphs
$A,B,C,D,G_1,G_2$ are odd-sized. Recall that these subgraphs occur in
the cyclic adjacency order
\[
 A,\ B,\ C,\ D,\ G_2,\ G_1,\ A.
\]
We first show that at least two of the odd-sized rectangles must be
adjacent. Consider the three disjoint pairs of consecutive rectangles
\[
 \{A,B\},\qquad \{C,D\},\qquad \{G_2,G_1\}.
\]
If no two odd-sized rectangles were adjacent, each of these pairs
could contain at most one odd-sized rectangle. Hence there would be
at most three odd-sized rectangles, contradicting the assumption
that there are four. Therefore, at least one adjacent odd-sized pair
must exist.

Choose such a pair and apply the adjacent-odd repartitioning
introduced in Case I, whenever its geometric and minimum-dimension
conditions are satisfied. The chosen pair is then replaced by
even-sized rectangular subgraphs while preserving the underlying
vertex set. The two remaining odd-sized rectangles can be examined
using the constructions from Case I: if they are adjacent, use the
adjacent-pair repartitioning; if they are separated, use the
appropriate three-rectangle repartitioning according to their
relative positions and the parity of the intervening rectangle.

This yields a reduction strategy from four odd-sized rectangles to
the two-odd configurations already considered. However, after the
first boundary adjustment, the second pair must be analyzed in the
updated partition rather than in the original six-region diagram.
In particular, it remains necessary to verify that the first cut
does not invalidate the second repartitioning, that all resulting
rectangles have dimensions at least two, and that their local
Hamiltonian cycles admit a compatible sequence of edge exchanges.
Thus, the adjacency argument guarantees a candidate first pair,
but does not by itself establish a Hamiltonian cycle for every
four-odd configuration.

\subsubsection{Case III: Six Odd-Sized Rectangles}

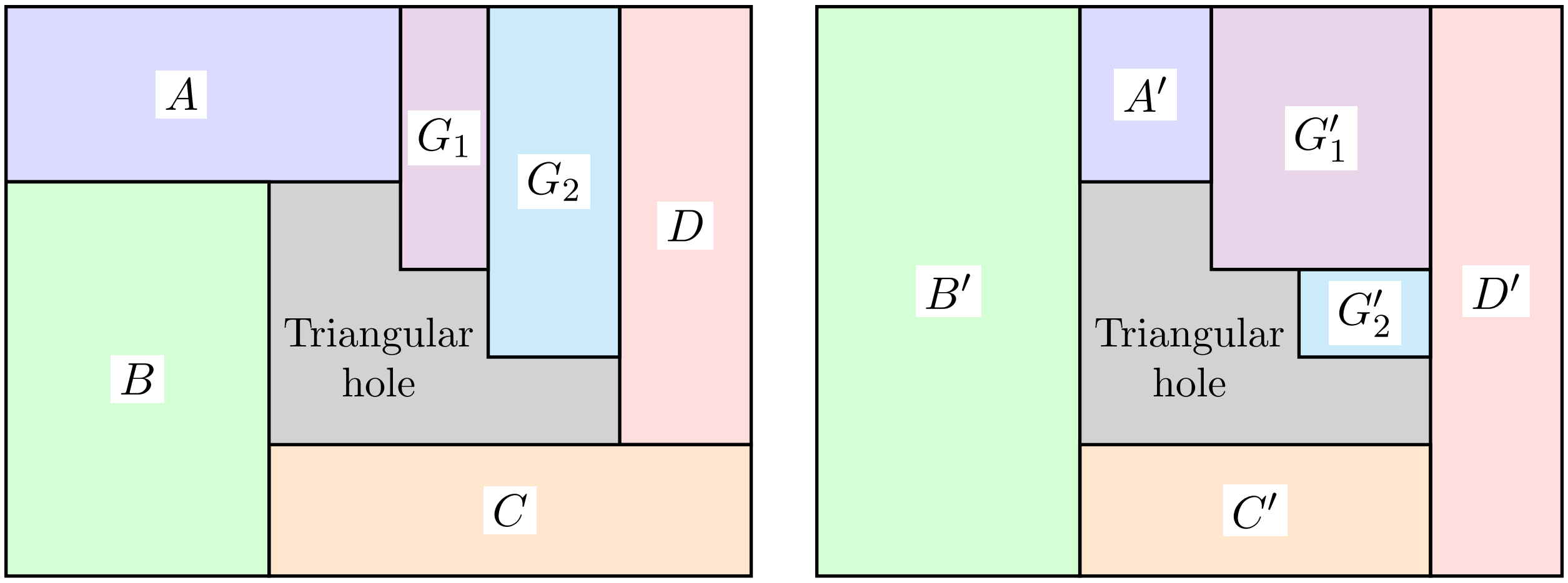
\begin{figure*}[t]
\centering

% ==========================================
% (a) Original partition
% ==========================================
\begin{subfigure}[t]{0.48\textwidth}
\centering
\resizebox{\linewidth}{!}{%
\begin{tikzpicture}[x=0.37cm,y=-0.37cm]

  % Original six rectangles
  \fill[blue!14]   (-0.5,-0.5) rectangle (8.5,3.5);
  \fill[green!17]  (-0.5,3.5) rectangle (5.5,12.5);
  \fill[orange!20] (5.5,9.5) rectangle (16.5,12.5);
  \fill[red!13]    (13.5,-0.5) rectangle (16.5,9.5);
  \fill[violet!17] (8.5,-0.5) rectangle (10.5,5.5);
  \fill[cyan!18]   (10.5,-0.5) rectangle (13.5,7.5);

  % Triangular hole
  \fill[gray!35]
    (5.5,3.5)--(8.5,3.5)--(8.5,5.5)
    --(10.5,5.5)--(10.5,7.5)
    --(13.5,7.5)--(13.5,9.5)
    --(5.5,9.5)--cycle;

  % Original boundaries
  \foreach \xa/\ya/\xb/\yb in {
    -0.5/-0.5/8.5/3.5,
    -0.5/3.5/5.5/12.5,
    5.5/9.5/16.5/12.5,
    13.5/-0.5/16.5/9.5,
    8.5/-0.5/10.5/5.5,
    10.5/-0.5/13.5/7.5}{
    \draw[thick] (\xa,\ya) rectangle (\xb,\yb);
  }

  % Labels
  \node[fill=white,inner sep=2pt] at (3.5,1.5) {$A$};
  \node[fill=white,inner sep=2pt] at (2.5,8) {$B$};
  \node[fill=white,inner sep=2pt] at (11,11) {$C$};
  \node[fill=white,inner sep=2pt] at (15,4.5) {$D$};
  \node[fill=white,inner sep=2pt] at (9.5,2.5) {$G_1$};
  \node[fill=white,inner sep=2pt] at (12,3.5) {$G_2$};

  \node[align=center,font=\small]
    at (8,7.5) {Triangular\\hole};

\end{tikzpicture}%
}
\caption{Original six-region partition.}
\label{fig:six-partition-before}
\end{subfigure}
\hfill
% ==========================================
% (b) After three pairwise reassignments
% ==========================================
\begin{subfigure}[t]{0.48\textwidth}
\centering
\resizebox{\linewidth}{!}{%
\begin{tikzpicture}[x=0.37cm,y=-0.37cm]

  % Pair (A,B):
  % A': 3 x 4; B': 6 x 13
  \fill[blue!14]   (5.5,-0.5) rectangle (8.5,3.5);
  \fill[green!17]  (-0.5,-0.5) rectangle (5.5,12.5);

  % Pair (C,D):
  % C': 8 x 3; D': 3 x 13
  \fill[orange!20] (5.5,9.5) rectangle (13.5,12.5);
  \fill[red!13]    (13.5,-0.5) rectangle (16.5,12.5);

  % Pair (G1,G2):
  % G1': 5 x 6; G2': 3 x 2
  \fill[violet!17] (8.5,-0.5) rectangle (13.5,5.5);
  \fill[cyan!18]   (10.5,5.5) rectangle (13.5,7.5);

  % Triangular hole (unchanged)
  \fill[gray!35]
    (5.5,3.5)--(8.5,3.5)--(8.5,5.5)
    --(10.5,5.5)--(10.5,7.5)
    --(13.5,7.5)--(13.5,9.5)
    --(5.5,9.5)--cycle;

  % New rectangular boundaries
  \foreach \xa/\ya/\xb/\yb in {
    5.5/-0.5/8.5/3.5,
    -0.5/-0.5/5.5/12.5,
    5.5/9.5/13.5/12.5,
    13.5/-0.5/16.5/12.5,
    8.5/-0.5/13.5/5.5,
    10.5/5.5/13.5/7.5}{
    \draw[thick] (\xa,\ya) rectangle (\xb,\yb);
  }

  % New region labels
  \node[fill=white,inner sep=2pt] at (7,1.5) {$A'$};
  \node[fill=white,inner sep=2pt] at (2.5,6) {$B'$};
  \node[fill=white,inner sep=2pt] at (9.5,11) {$C'$};
  \node[fill=white,inner sep=2pt] at (15,6) {$D'$};
  \node[fill=white,inner sep=2pt] at (11,2.5) {$G_1'$};
  \node[fill=white,inner sep=2pt] at (12,6.5) {$G_2'$};

  \node[align=center,font=\small]
    at (8,7.5) {Triangular\\hole};

\end{tikzpicture}%
}
\caption{After the three pairwise boundary reassignments.}
\label{fig:six-partition-after}
\end{subfigure}

\caption{Repartitioning the six rectangular subgraphs
through three pairwise boundary adjustments:
$(A,B)$, $(C,D)$, and $(G_1,G_2)$.
The original partition is shown in (a), while
(b) illustrates the resulting rectangular partition.
The triangular hole and the set of retained vertices
remain unchanged.}
\label{fig:six-region-reassignment}

\end{figure*}

Suppose all six rectangular subgraphs
$A,B,C,D,G_1,G_2$ are odd-sized.
We consider the three pairs of adjacent rectangular
subgraphs
\[
(A,B), \qquad (C,D), \qquad (G_1,G_2).
\]
For each pair, we apply the boundary reassignment
introduced in Case I.

Specifically, for the pair $(A,B)$, we extend the
right boundary of $B$ upward through $A$.
For the pair $(C,D)$, we extend the left boundary
of $D$ downward through $C$.
Finally, for $(G_1,G_2)$, we extend the bottom
boundary of $G_1$ rightward through $G_2$.

Figure~\ref{fig:six-region-reassignment}
illustrates the original partition and the partition
obtained after these three boundary adjustments.
The operations preserve the vertex set of the
original graph while changing the dimensions
of the rectangular components.

When both rectangles in each pair are odd-sized
and the required geometric conditions hold,
the adjacent-pair repartitioning produces
even-sized rectangles. Hamiltonian cycles can
then be constructed locally, provided that
each resulting rectangle has both dimensions
at least two, and merged through compatible
boundary edges.

\section{Discussion}\label{sec:discussion}
\section{Discussion}\label{sec:discussion}

The geometry-driven construction distinguishes between
small-hole and large-hole cases. The small-hole procedure constructs
a path in the reversed triangular region, closes it through the bounding
rectangle, and merges the inner and outer cycles through parallel edges.
The large-hole analysis considers the parity cases for $c_2=0$
and describes the constrained upper, lower, and left sides.
The remaining detailed large-hole cases and the nonadjacent
repartitioning cases require further analysis to establish completeness.

Furthermore, the rectangular-decomposition approach relies on
repartitioning the original rectangular subgraphs into smaller
rectangular regions. Since each resulting region must have width
and height at least two to admit a Hamiltonian cycle, additional
analysis is needed to determine the minimum allowable values
of the relevant parameters under which the proposed
repartitioning operations remain valid. In particular, when
the original regions are too narrow or too short, some resulting
subgraphs may not satisfy the minimum dimension requirements.
These boundary cases require separate analysis or alternative
constructions to establish the completeness of the approach.

\bibliographystyle{plain}
\bibliography{reference}

% Source ambiguities retained for author verification; no mathematical repair inferred:
% The reversed-triangle algorithm uses m without a consistent definition and
% mixes local endpoint coordinates with the fixed s=(0,1) statement.
% It gives m_3=w_2-w_1+1 in the dimensions but m_3=w_2-w_0 later.
% The extreme large-hole condition retains the source wording a=1,b=1 or d=1.
\end{document}